\documentstyle[aps,pre,eqsecnum,tighten]{revtex}

\begin{document}
\bibliographystyle{plain}

\title{Discrete Breathers}
\author{S. Flach}
\address{Max-Planck-Institut f\"ur Physik komplexer Systeme,
Bayreuther Str.40 H.16, D-01187 Dresden, Germany}
\author{C. R. Willis}
\address{Department of Physics, Boston University, Boston, MA 02215,
USA}
\date{\today}
\maketitle
\begin{abstract}
Nonlinear classical Hamiltonian lattices exhibit generic solutions
in the form of discrete breathers. These solutions are time-periodic
and (typically exponentially) localized in space. The lattices
exhibit discrete translational symmetry. Discrete breathers are not
confined to certain lattice dimensions. Necessary ingredients for
their occurence are the existence of upper bounds on the phonon
spectrum (of small fluctuations around the groundstate)
of the system as well as the nonlinearity in the differential
equations. We will present
existence proofs, formulate necessary existence conditions,
and discuss structural stability of discrete breathers. The following
 results
will be also discussed:
the creation of breathers through tangent bifurcation of
band edge plane waves; dynamical stability; details of the
spatial decay; numerical methods of obtaining breathers; interaction
of breathers with phonons and electrons; movability; influence
of the lattice dimension on discrete breather properties; quantum
lattices - quantum breathers.

Finally we will formulate a new conceptual aproach capable
of predicting whether discrete breather exist for a given system or not,
without actually solving for the breather. We discuss potential applications
in lattice dynamics of solids (especially molecular crystals), selective
bond excitations in large molecules, dynamical properties of coupled
arrays of Josephson junctions, and localization of electromagnetic
waves in photonic crystals with nonlinear response.
\end{abstract}
\noindent
pacs: 03.20.+i , 63.20.Pw , 63.20.Ry
\\
keywords: lattices, breathers, localization
\\
\\
\\
To be published in Physics Reports

\newpage

\tableofcontents

\newpage

\noindent
LIST OF FREQUENTLY USED ABBREVATIONS
\\
\\
\\
\begin{tabbing}
ODE \hspace*{1cm}\= ordinary differential equation \\
PDE \> partial differential equation \\
DB \> discrete breather \\
NLS \> nonlinear Schr\"odinger equation \\
DNLS \> discrete nonlinear Schr\"odinger equation \\
AL \> Ablowitz-Ladik (lattice) \\
RWA \> rotating wave approximation \\
FPU \> Fermi-Pasta-Ulam (model) \\
KG \> Klein-Gordon (model) \\
sG \> sine-Gordon (model) \\
PM \> Poincare map \\
PBC \> periodic boundary conditions
\end{tabbing}

\newpage

\vspace*{5cm}
\noindent
Dedicated to Peter Fulde on the occasion of his 60th birthday.

\newpage

\section{Introduction} \label{s1}

Discreteness is at the very heart of almost any many-particle
system, if the processes under discussion involve length scales
of the order of (or not too large compared to) the interparticle 
distances. Condensed Matter physics is a particular well-known
area of research where discreteness plays an important role, and
the same could also be said of topics in chemistry or biology.
Discreteness is here typically meant with respect to space
(particles versus fields). Consequently one deals with Hamiltonians
which are discrete in space but define a continuous phase space flow
with respect to time. 

In this review we will discuss time-periodic localized excitations in 
discrete systems. The localization is due to the discreteness
combined with the nonlinearity of the system, as opposed
to the well known disorder induced localization. The reader is
encouraged to check the subsection 'Summary' of the conclusions
to this report for a full list of the results discussed. Some of
the results discussed here can be found in \cite{sp95},\cite{sf96-2}%
,\cite{sa96}.

\subsection{General remarks} \label{s1-1}

Typically the dynamics of a many-particle system can be regarded
as a discretization of certain continuous systems, that is the
classical coupled ordinary differential equations (ODE) of the
many-particle system can be regarded as discretized versions of
partial differential equations (PDE). The replacement of the
spatial derivatives by spatial differences
in the equations of motion implies the reduction of the symmetry
of the Hamiltonian.
In general lowering the symmetry means enriching the class of solutions
because less restrictions are imposed. Of course one also looses solutions
when lowering the symmetry - namely the ones which were generated
by the higher symmetry. Let us discuss two simple examples of linear
equations (PDE or ODE). Assuming that the PDE has a continuous translational
symmetry, we can expect a linear spectrum (i.e. a dispersion
relation) $\omega(q)$ which is
unbounded. The discrete version related to the PDE will
yield a linear spectrum which is i) bounded and ii) might
have additional gaps depending on the discretization procedure. The 
finite upper (or lower)
bound of the linear spectrum is intimately connected
with the interparticle distance (or the discretization step of the
space continuum). Replacing e.g. the wave equation $U_{,tt}-U_{,xx}=0$ 
by $U(l)_{,tt} - (U(l+1) + U(l-1) - 2U(l))=0$ with $l=0,\pm1,\pm2,...$ 
means replacing
the linear spectrum $\omega^2(q)=q^2$ by $\omega^2=2(1-{\rm cos}(q))$.
Let us now add a translational symmetry 
breaking inhomogeneity
to the system in the form $(1+\epsilon \delta (x))U_{,tt}-U_{,xx}=0$ for
the continuum and $(1+\epsilon \delta_{l,0})
U(l)_{,tt} - (U(l+1) + U(l-1) - 2U(l))=0$ for the lattice. For $\epsilon < 0$ 
it is well-known that a spatially localized and time-periodic solution (eigenmode)
exists for the lattice version. 
This happens because the linear spectrum has a finite bound, and 
eigenmodes with frequencies outside the linear spectrum can not
be spatially extended. However the corresponding PDE does not posess
a bound state, simply because the linear spectrum occupies the whole       
real axis. Indeed, substituting $U(x,t) = u(x){\rm e}^{i\omega t} $
we obtain $-u_{,xx} + \omega^2 (-1 + \epsilon \delta(x))u(x)=0$ which is the 
time-independent Schroedinger equation of a particle in the potential
$V(x)=\omega^2(-1+\epsilon \delta(x))$. Since the corresponding 
eigenenergy is zero,
obviously no bound state exists - rather an extended state is found.
The pattern of thoughts which leads to the last result without performing
all calculations is: i) if we discretize a PDE then the linear spectrum
will be bounded and might even lead to additional gaps 
in the PDE linear spectrum (think of electronic band structures in solids);
ii) thus we can by proper modifications of the discretized equations
produce spatially localized but time-periodic states, whose frequencies
are located in the new gaps opened due to discreteness.

\subsection{Nonlinearity and Discrete Breathers} \label{s1-2}

As long as we stick to linear equations, we have to break the 
discrete translational symmetry of the discretized equations in order
to obtain spatially localized modes, i.e. we have to add 'impurities'. 
In this review we will 
show that one can retain the discrete symmetry and still obtain
localized excitations - by adding nonlinear terms to the equations.
Certainly nonlinear terms are present in any application of many-particle
dynamics. The thought pattern capable of predicting the existence
of localized excitations will be much more elaborate than for the linear
case. We will see that due to the nonlinearities higher harmonics of
the excitation's frequency have to be considered too (i.e. their positions
with respect to the linear spectrum). Consequently nonlinearity is crucial
with respect to the difference between continuum and discreteness. Indeed,
the above demonstrated linear examples can not change the fact that one
can add terms to the PDE which cause the opening of gaps (and still 
preserve the continuous symmetry). Then nothing hinders us from adding
symmetry-perturbing terms which are capable of producing localized or
bound states. However in the case of a nonlinear PDE the fact that the
linear spectrum is still unbounded typically forbids the existence
of bound states (because the mentioned higher harmonics of the bound
state's frequency will always resonate with the linear spectrum).
Here is the big advantage of a discrete system, because the finite
upper bound of the linear spectrum still allows for frequencies whose
entire higher harmonics might lie outside the linear spectrum.

Thus we are going to investigate solutions
of Hamiltonian lattice equations (e.g. describing the classical dynamics
of a crystal lattice) which are periodic in time and localized in space.
Different names of these objects have appeared in the literature -
Intrinsic Localized Modes (ILM), Self-localized Anharmonic Modes (SLAM),
Nonlinear Localized Excitations (NLE), Discrete Breathers (DB), etc.
The first three entries of this exotic list are perhaps an attempt
to separate the appearance of these objects from some well-known but
only weakly related solutions of PDEs - namely breathers. The fourth
entry of the list tries to create the difference by the attribute 'discrete'.
We choose the entry 'discrete breather' (although
we used NLE before). The main reason is that the objects are indeed
breathers living on a lattice, so it is sort of confusing to create
new names. Perhaps 'lattice breathers' would be better? There is still some
dynamics in that issue, so we will have to wait and see which entry will
be dominant.

\subsection{Some Properties of breathers in Hamiltonian Field Equations}
\label{s1-3}

Breather solutions are well-known from the study of the sine-Gordon (sG) PDE. 
The sG equation for a real-valued field $\Psi(x,t)$ is a particular
example out of the class of nonlinear Klein-Gordon (KG) equations
\begin{equation}
\Psi_{,tt} = C \Psi_{,xx} - F(\Psi) \label{1-1}
\end{equation}
with the choice $F_{sG}(z)={\rm sin}z$.
The breather solution is given by
\begin{equation}
\Psi_b(x,t)=4{\rm tan}^{-1}\left[ \frac{m}{\omega}\frac{{\rm sin}(\omega t)}
{{\rm ch}(mx)}\right] \;,\; \omega = \sqrt{1-m^2}. \label{1-2}
\end{equation}
It represents a field which is periodically oscillating in time and
decays exponentially in space as the distance from the center $x=0$
is increased (see Fig.1). 
%\begin{figure}
%\psfig{figure=/home/flach/papers.DIR/report.DIR/figures.DIR/fig1.ps}
%\caption{$\Psi_b(x,t)$ from (\ref{1-2}) versus $x$ for 26 different
%times equally spaced and covering one breather period and $m=0.5$.}
%\end{figure}
As shown stationary breather solutions form one-parameter families
of solutions.

Most probably these breather solutions are {\sl nongeneric}. 
Birnir \cite{bb94},\cite{bb94-2} and 
Denzler \cite{jd93},\cite{jd95} showed
that sG breathers are
{\sl isolated}, i.e. the solutions survive only under a finite number
of perturbations $\delta (z)$ of $F_{sG}(z) \rightarrow F_{sG}(z)+\delta (z)$.
Years of searching for breathers in $\phi^4$ systems ($F(z)=-z+z^3$)
were terminated by the nonexistence proof of breathers by Segur and
Kruskal \cite{sk87}. Related results have been derived by Kichenassamy 
\cite{sk91}. Geicke \cite{jg94} has calculated the decay 
rate of $\phi^4$-'breathers'.
So it appears that breathers are nongeneric (structurally unstable)
and thus rare objects in PDEs.
The reason
for that lies in the fact that a decomposition of (\ref{1-2}) into a
Fourier series with respect to time yields higher harmonics with frequencies
$k\omega$ ($k$ is the Fourier number). These frequencies resonate with
the linear spectrum of (\ref{1-1}) $\omega_q = \sqrt{Cq^2+F'(z=0)}$
($q$ - wave number), if $k$ is larger than a given number which depends on
the choice of $\omega$. Consequently the corresponding separatrix manifolds
associated with (\ref{1-1}) are of finite dimension. 
The intersection of finite-dimensional manifolds in an infinite-dimensional
phase space is comparable to a miracle - translated this implies an infinite
number of symmetries. Indeed the sG equation is integrable!
Together with the
infinity of the dimension of  the corresponding
phase space the structural instability of sG breathers
follows immediately \cite{ekns84}. It is interesting to note that those
arguments are obtained by linearising the equation (\ref{1-1}) and then
analyzing the ODEs for the Fourier coefficient functions. However a 
test of solution (\ref{1-2}) shows that the linearization is not adequate -
simply because (\ref{1-2}) has nonzero higher Fourier components (with respect
to time) and those components would not spatially decay in the linearized
version of (\ref{1-1}). We will discuss similar effects in the lattice case.

Boyd \cite{jpb90} introduced a very interesting concept of the
'nanopteron' (still another exotic name). The nanopteron concept
is the following: assume that the PDE does not allow for a breather.
Then it can still allow for a solution which is locally close to a breather,
but which does not decay to zero in space - rather it decays to some nonzero
but small (compared with the center) amplitude. The essential difference
between the breather and the nanopteron is that breathers are solutions of 
finite energies, whereas nanopterons are solutions of infinite energies.
The nanopteron construction essentially uses the fact that there can be
solutions periodic in time with frequencies outside the linear spectrum.
Because higher harmonics resonate with the linear spectrum, the solutions do
not decay in space to zero, 
but nevertheless keep their inhomogeneous spatial properties.

It is worthwhile mentioning another field equation which supports
breather solutions - the Nonlinear Schr\"odinger Equation (NLS)
in $1+1$ dimensions:
\begin{equation}
{\rm -i} \Psi_{,t} = \Psi_{,xx} + 2|\Psi |^2 \Psi\;\;. \label{1-3}
\end{equation}
This equation is also integrable (!) (note that the field is complex-valued)
and posesses soliton solutions \cite{degm82}. One of them
is nothing but a breather solution. The breather solution is obtained
by substituting $\Psi(x,t)=\phi(x){\rm e}^{{\rm i}\omega t}$. The
time-independent field $\phi$ is assumed to be real valued and satisfies
the equation
\begin{equation}
\phi_{,xx} = \omega \phi - 2\phi^3 \;\;. \label{1-4}       
\end{equation}
For $\omega > 0$ this equation allows for homoclinic orbits
defined by $\phi(0)= \sqrt{\omega}$, $\phi_x(0)=0$. 
These homoclinic orbits yield breather solutions of the original
NLS equation (\ref{1-3}).
It is interesting
to note that in contrast to the sG case perturbations of the NLS equation
sometimes do not destroy the breather solutions. However they will not 
destroy the breather solution
typically only if the perturbations preserve the conservation law
$B = \int |\Psi|^2 {\rm d}x$. This is intimately connected to the {\sl
nongeneric} structure of the NLS equation itself - namely that the 
nonlinearity appears only in the amplitudes, but not in the phases
(of the complex field $\Psi$).

We have found only one example of a PDE in 2+1 dimensions which allows
for numerically obtained breather-like objects \cite{arm91}. 
Also we want to mention
the efforts to obtain breather solutions for Einstein-Maxwell equations
of the electro-vacuum, and for equations describing one-dimensional optical
media with Kerr nonlinearities capable of self-induced transparency
\cite{rtm89},\cite{emm89}. 
Altogether the efforts of finding breathers in PDEs have been discouraging.
It is hard to believe that isolated, nongeneric and structurally unstable
objects are capable of describing phenomena in nature.

\subsection{The Discrete World} \label{s1-4}

Although there has been interest in the relationship between
the continuum and discreteness for a considerable amount of time
many of the results present here are quite new.

One possible way of testing the difference between PDEs and their
discretized versions with respect to breathers would be to find
a lattice model with discrete breathers and then to test whether
they are structurally stable or not. Indeed we can write down one
such lattice model - the Ablowitz-Ladik (AL) lattice \cite{al76}
\begin{equation}
i\Psi(l)_{,t} + C(\Psi(l+1)+\Psi(l-1)-2\Psi(l))+(\Psi(l+1)+\Psi(l-1))
|\Psi(l)|^2=0\;\;\;. \label{1-5}   
\end{equation}
The 
AL lattice is a discretized version of the NLS equation, and the AL lattice
is integrable. It allows for breather solutions. The structural stability
analysis of AL breathers has not been done yet. Also any structural stability
analysis will not yield too much insight into why breathers might exist
for many different lattice realizations. Thus we choose a different way
of analysis. At the end of this analysis we will obtain a 
conceptual approach which allows
us to predict the existence/nonexistence of breathers for
lattices without actually solving for the breather solutions.

\section{First approximate results}
\label{s2}

In this section we will introduce the models, 
discuss multiple scale expansions and
rotating wave approximations as well as some numerical studies
which have been used in order to describe discrete breathers.

Let us consider a $d$-dimensional hypercubic lattice with $N$ sites.
Each site is labeled by a $d$-dimensional vector $l \in Z^d$.
Assign to each lattice site a vector $X_l \in R^f$, where $f$
is the number of components and is to be finite.
The evolution of the system is given by the set of equations
\begin{equation}
\dot{X}_l = F_l(\{X_m\})\;\;,\;\;F_l(\{0\})=0\;\;. \label{2-1}
\end{equation}
It will be important for the discussion to agree upon
a general form of the expansion of $F$ in the variables $X$, which
is usually a Taylor expansion:
\begin{equation}
F \sim  \sum_{\mu=1}^{\infty} f_{\mu} X^{\mu}\;\;.  \label{2-2}
\end{equation}
In fact the expansion coefficients in (\ref{2-2}) are tensors reflecting
the true topology of the interaction incorporated in the general
form (\ref{2-1}).
For finite number of sites $N$ we impose periodic boundary
conditions. Then the discrete translational invariance of the system
implies that any solution of (\ref{2-1}) generates new solutions
under translation of all $l$ by the same vector $L \in Z^d$
: $l \rightarrow l+L$. According to (\ref{2-1}) $X_l=0$ is a solution
for all times, hereafter coined {\sl groundstate} solution (GS).
Finally the evolution defined by (\ref{2-1}) is required to have
an integral of motion
(energy) $E(\{X_l\}) \geq 0$, $E(\{0\})=0$ with $\dot{E}=0$.
In the following we will briefly discuss various perturbation theories
applied to specifications of (\ref{2-1}).

\subsection{The beginning}
\label{s21}

In preparing this report we came across a remarkable paper by
Kosevich and Kovalev \cite{kk74} (hereafter referred to as KK) which seems to
be the first direct attempt to solve for discrete breathers. Furthermore
the authors of this early work addressed a whole set
of questions connected with the existence of discrete breathers,
which will be discussed in detail in coming sections - e.g. the
connection of the existence of discrete breathers with instabilities
of band edge plane waves. 
One of the models studied is given by equation (1) in \cite{kk74}: 
\begin{equation}
\ddot{u}_n + b^2(2u_n - u_{n+1} - u_{n-1}) - f(u_n) = 0\;\;\;. \label{2-3}
\end{equation}
This is a discretized version of the Klein-Gordon equation in 1+1 dimensions
(and is a special case of (\ref{2-1}) with $X_n$ being a two-component
vector with the components $u_n$ and $\dot{u}_n$). The function $f(z)$
is set to $f(z)=-z+\alpha z^2 + \beta z^3$. Linearization of (\ref{2-3})
around the stable static solution $u_n(t)=0$ yields the linear spectrum
(i.e. dispersion relation) for plane waves $\omega^2_q=
1+2b^2(1-{\rm cos}q)$.
Note that the linear spectrum has i) a finite upper bound - this is
important with respect to higher harmonics of a breather solution, and
ii) a lower nonzero bound - this allows for breather frequencies
below the linear spectrum (sometimes called gap modes). KK consider solutions
of (\ref{2-3}) with frequencies below and close to the lower band edge
of the linear spectrum $\omega = 1$. Intuitively those solutions should
be of large wavelength, thus KK argue in favor of considering the
PDE $u_{,tt}-u_{,xx}-f(u)=0$. Breather solutions are then derived using
multiple scale expansions.
In lowest order the solution reads (equation (8) in \cite{kk74}) 
\begin{equation}
u(x,t)=\left( \frac{8}{3\beta}\right)^{1/2} \frac{\sqrt{1-\omega^2}}{{\rm ch}
\sqrt{1-\omega^2}x} {\rm cos} \omega t \;\;\;. \label{2-3a}
\end{equation}
However 13 years later Segur and Kruskal have
demonstrated that those asymptotic expansions are not correct \cite{sk87}.
Still the analysis of KK is impressive, especially the idea that also
solutions with frequencies larger than and close to the linear spectrum 
can be considered in a continuum limit after proper substitutions. 
However the continuum approximation is clearly not appropriate for 
the discussion of breather solutions, because the breather solution will
inevitably have higher harmonics with respect to a Fourier series in time,
and those higher harmonics will ultimately resonate with the unbounded linear
spectrum of the PDE. However the starting point was (\ref{2-3}) whose
linear spectrum is bounded from above.

\subsection{Multiple scale expansions}
\label{s22}

Further progress using multiple time scales was
achieved by Remoissenet \cite{mr86} (see also \cite{ysk93-pla},%
\cite{bp91}). This
semi-discrete multiple-scale approach keeps the discreteness yet still
uses the frequency distance from the band edge of the linear spectrum 
as a small parameter. Typically one looks for solutions in the form
\begin{equation}
u_n = \phi_n + (\psi_n {\rm e}^{it} + cc) + (\xi_n {\rm e}^{2it} + cc)
+ ...\;\;, \label{2-4}
\end{equation}
where the functions $\phi_n,\psi_n,\xi_n,...$ are small in amplitude
and slowly varying in time. Using $\epsilon=1-\omega$ as a small parameter
(where $\omega$ is some frequency of the solutions we are looking for)
and assuming $\psi_n \sim \sqrt{\epsilon}$, $\phi_n \sim \epsilon$, 
$\xi_n \sim \epsilon$, $d/dt \sim \epsilon$ one finally arrives at
the celebrated discrete nonlinear Schroedinger equation (DNLS)
(we do not care about the explicit form of the coefficients, so only
the general form of the equation is given):
\begin{equation}
i\dot{\psi}_n + (\psi_{n-1} + \psi_{n+1}-2\psi_n) + 
\kappa | \psi_n|^2\psi_n = 0\;\;\;. \label{2-5}
\end{equation}
This is a remarkable equation which has been studied in great detail
and has been used to describe many different phenomena (sometimes
this equation is called discrete self-trapping equation 
\cite{els85})\footnote{For a list of publications on the DNLS
click on web page: http://www.ma.hw.ac.uk/$^{\sim}$chris/dst/.}.
Here we want only to discuss its connection to discrete breathers.
Let us assume $\psi_n = A_n {\rm e}^{i\epsilon t}$ with $A_n$ being
real-valued and time-independent, then we obtain
\begin{equation}
-(\epsilon+2)A_n+A_{n-1}+A_{n+1} +\kappa A^3_n=0\;\;\;. \label{2-6}
\end{equation}
Equation (\ref{2-6}) defines a two-dimensional discrete map
(knowing two neighbouring amplitudes allows for the computation of the
next amplitude). We are looking for solutions of this map which are
finite in some region of the chain and decay to zero if we iterate the
map to plus or minus infinity. Solutions of that type are homoclinic
orbits and arise from the intersection of the invariant manifolds
of a given hyperbolic fixed point. Thus we have to check whether
the fixed point solution $A_n=0$ is a hyperbolic one, and if yes, whether
the invariant stable and unstable manifolds intersect. The first question
is easy to solve. For that one linearizes (\ref{2-6}) around $A_n=0$,
obtains the matrix defining the linear two-dimensional map and calculates
its eigenvalues. The eigenvalues are $\lambda=1+\epsilon/2
\pm \sqrt{\epsilon + (\epsilon/2)^2}$. For $\epsilon > 0$
the fixed point $A_n=0$ is hyperbolic and thus two invariant one-dimensional
manifolds exist, which originate at the fixed point. The second
task - showing that these manifolds intersect - is harder to achieve. However
reliable numerical methods exist, and Hennig has shown the intersection
of the invariant manifolds \cite{dh96}. Strictly speaking equation (\ref{2-6})
has to be analysed in the limit of small $\epsilon \ll 1$, due to
the multiple scale expansions. In that case the amplitudes will be small too,
and the variation on the lattice is slow compared to the lattice spacing,
so that one could replace the differences in (\ref{2-6})  by differentials.
That would bring us back to KK's result (\ref{2-3a}). But we will know that
the persistence of the manifold intersection in the discrete system
allows for an approximate analytical solution of (\ref{2-6}) by (\ref{2-3a}).
The DNLS (\ref{2-5}) is an interesting model by itself, i.e. without tying
it to the multiple-scale expansion. We will come back to it in the later
course of this work.  

Let us discuss the multiple-scale expansions in general. 
First of all when stopping
at the first nontrivial order of the expansion (\ref{2-5}) we did
not advance much as compared to the more crude replacement of the
difference equations by differential ones as done by KK. Indeed the whole
problem of higher harmonics of the breather's frequency resonating (or not)
with the linear spectrum appears only in higher orders of the expansion.
Bang and Peyrard \cite{bp95} proceeded to higher orders of the expansion
and found several complications. The interested reader will find details
in \cite{bp95}. Here we want to mention two principal problems with those
expansions. The multiple-scale expansions attempt to make a perturbation
approach to the linear problem which is perturbed with nonlinear terms. 
However the nonlinear terms in fact destroy the integrability of the 
linear problem. So what the perturbation approach faces is a nonintegrable
perturbation of an integrable problem. This has been quite a topic in
mathematics of this century \cite{via89}. 
One type of mathematical approach focus
on analytical continuation of solutions of the integrable case into
the nonintegrable one (e.g. the famous Kolmogorov-Arnold-Moser theorem
continues tori, but only for finite dimensional systems \cite{via89}).
Another type of approach tries to establish new phase space properties
of the nonintegrable case, still using the distance to the integrable
system as a small parameter. This second type of approach 
is very delicate, and the
attempt to find discrete breathers as described above belongs to the
second type - simply because no breathers exist in the linear (integrable)
case (indeed, if we choose $\epsilon=0$ which corresponds to the
linear limit, the fixed point $A_n=0$ in (\ref{2-6})
is not hyperbolic anymore). However the first
type of approach is useful for describing discrete breathers
as we will see. It can be used either to analytically continue
discrete breather solutions from some trivial integrable limit, or it
can be used to analytically continue plane wave solutions of the linear
integrable limit and study their stability properties. Since only
periodic orbits will be continued, the continuation can be carried out
equally for a finite-dimensional as for an infinite-dimensional phase
space! 

\subsection{Rotating wave approximation}
\label{s23}

Another approximate method frequently used in the literature is the 
rotating wave approximation (RWA) 
\cite{tks88},\cite{st88},\cite{st90-jpsj-3},\cite{thor91},\cite{st92-jpsj-2}%
,\cite{sw93}. 
This approach is directly
searching for discrete breather solutions as solutions harmonic in time.
The solution is parametrized using a Fourier series expansion
\begin{equation}
u_n(t)=\sum_{k=-\infty}^{k=+\infty}A_{nk}{\rm e}^{ik\omega t}\;\;\;.
\label{2-7}
\end{equation}
Then one inserts the  ansatz into the
original equations of motion, collects  terms with
equal harmonics $k\cdot \omega$, $k=0,\pm1,\pm2,...$
and neglects all coefficients in the Fourier representation
 for $|k| \geq k_0$ for some $k_0$. Usually one puts
$k_0=2$. In that case the RWA is very close to the mentioned multiple-scale
expansions. Indeed the RWA is nothing but a perturbation theory
of weakly nonlinear oscillations for small amplitudes. 
This can be immediately obtained for system (\ref{2-3}) with $\alpha=0$
(cf. text below (\ref{2-3}). Then $A_{n0}=0$ for all $n$ (in fact
$A_{n,2m}=0$ due to the additional symmetry of the Hamiltonian 
being invariant under change of sign of the variables). We find
\begin{equation}
-\omega^2 A_{n,1} + b^2(2A_{n,1} - A_{n+1,1} - A_{n-1,1})
+ A_{n,1} - 6 \beta A_{n,1}^3 = 0 \;\;\;. \label{2-7a}
\end{equation}
The similarity to equation (\ref{2-6}) is evident.
However the
RWA has been frequently used for quite strong nonlinearities. 
For instance for the trivial case $b=0$ (independent sites) we
solve (\ref{2-7a}) for any site and find $A_{n,1} = 0$ or $A_{n,1}=
\omega / \sqrt{6\beta}$. We can choose any of the two solutions at
any lattice site and thus create an infinite number of solutions
characterized by a binary sequence. The position of an element
of the sequence is the lattice site, and the value of the element encodes
either of the two possible solutions for $A_{n,1}$. Using Aubry's 
'antiintegrability' approach \cite{aa90} 
it is possible to prove continuation of
those solutions into the interacting $b \neq 0$ case. If we would choose
e.g. a sequence $(...0001000...)$ the continuation into the interacting
case would correspond to a solution which is still spatially localized -
and could be called a discrete breather. The richness of the possible
internal structure of localized solutions is evident due to the arbitrariness
in the choice of the coding sequence - e.g. why not take $(...000101011000...)$?

The reason
why the RWA can give quite reasonable estimations even when being far
away from the usual perturbation region of amplitudes has been explained
in \cite{fwo94} by studying a single nonlinear oscillator. 
We apply the RWA to the one particle problem
\begin{equation}
\ddot{Q} = -V'(Q) \;\;\;. \label{2-7b}
\end{equation}
The potential $V(Q)$ should provide bound motion, at least
for the cases under consideration. Since this problem
is integrable, one can (at least numerically) calculate
the period of the periodic solution (\ref{2-7b})
and thus the fundamental frequency.
E.g. for the class of potentials
\begin{equation}
V(Q) = \frac{1}{2m}Q^{2m}   \label{2-7c}
\end{equation}
the solution for the frequency can be found analytically
as a function of the energy $E$:
\begin{equation}
\omega = \frac{1}{2}\sqrt{2\pi}(2m)^{1-1/2m} \cdot
\frac{\Gamma (1/2m + 1/2)}{\Gamma(1/2m)}\cdot E^{1/2 - 1/2m} \;\;.
\label{2-7d}
\end{equation}
Here $\Gamma(x)$ is the Gamma-function. Applying RWA with
$k_o=2$ one derives the approximation
\begin{equation}
\omega_1 = 2^{1-m}\cdot (2m)^{1/2-1/2m}\cdot
\sqrt{\frac{(2m-1)!}{m!(m-1)!}}\cdot E^{1/2-1/2m}\;\;. \label{2-7e}
\end{equation}
First we note the remarkable coincidence between the exact
result (\ref{2-7d}) and the approximation (\ref{2-7e})
with respect to the energy dependence, which was achieved
within the RWA using the exact relation between the amplitude
and the energy as it follows from (\ref{2-7c}).
Moreover comparing the prefactors for e.g. $m=2$ in (\ref{2-7c})
yields 1.1981 and 1.2247 for the exact result (\ref{2-7d})
and the approximation (\ref{2-7e}), respectively. That means that
the simplest RWA in the case of (\ref{2-7c}) already gives
an error of less then 2.3\% !
The errors accumulate however for larger
$m$. 
In the limit $m \rightarrow \infty$ the potential (\ref{2-7c})
becomes box-like shaped, and the approximation of the dynamics in this
potential by a harmonic function in time is obviously wrong.
In general the RWA fails whenever the true solution has either
weakly decaying Fourier components or a maximum of $A_{kn}$ with respect
to $k$ for some value of $k > 1$. We will see that this can happen to discrete
breathers. In principle the RWA can be extended by increasing $k_0$
(in \cite{fwo94} this is demonstrated using a potential with barriers
instead of (\ref{2-7c}), for which the $k_0=2$ RWA fails when the energy
of the solution is close to the barrier) , and
indeed the RWA then goes over into what is called numerical calculation
of discrete breathers up to machine precision. The typical case $k_0=2$
however is again missing the problem of resonances of higher harmonics
with the linear spectrum. Thus the most reliable results can be achieved
with RWA when the frequency of the discrete breathers is located above the
linear spectrum. Numerical solutions of the RWA equations have been
obtained e.g. using the Gauss procedure \cite{dw96}.

\subsection{Numerical evidence}
\label{s24}

Finally we want to shortly mention a number of papers demonstrating
the existence of breather-like objects through numerical integrations
of the equations of motion. This is achieved using either Runge-Kutta
methods or Verlet (leap-frog) algorithms \cite{numrec}. The results
are typically presented as some snapshots of the amplitude or energy
distribution in the lattice during the actual integration 
\cite{thor90},\cite{bkp90-prb},\cite{bkp90-ssc},\cite{bkr90-jetpl},\cite{%
bkr90-pla},\cite{kr90},\cite{sak90},\cite{bk91},\cite{thom90},\cite{%
thor91},\cite{ht92-pla},\cite{ht92-jpsj-1},\cite{ht92-jpsj-2},\cite{%
kh93},\cite{ats93},\cite{ma92},\cite{dpw92},\cite{hsx93},\cite{zwl93}.
The findings strongly suggest that there exist solutions of the
lattice dynamics which appear to be similar to breather solutions. 
Moreover Burlakov, Kisilev and Pyrkov \cite{bkp90-prb} have numerically
obtained breatherlike structures for a two-dimensional lattice,
which implies that discrete breathers are supported  by lattice
dimensions one and two at least. Thus although most of the analysis
in the literature has been done for one-dimensional lattices, it is very
important to keep in mind that discrete breather existence is not
restricted to one-dimensional systems.
For some systems these solutions even move - despite the discreteness
of the system. All those studies of course do not answer the question as to
whether the observed objects correspond to exact solutions or not.
Still it is tempting to conjecture the existence of discrete breathers,
because the variety of systems analyzed numerically makes the opposite
statement difficult to believe.

\section{The local ansatz}
\label{s3}

In this chapter we will discuss numerical efforts of characterizing
the breather-like objects which appear in many simulations (cf. preceeding
section). These studies have been performed for one- and two-dimensional
lattices. 
These studies generate a 'local integrability' picture, which has been
useful in proceeding with the understanding of the properties of discrete
breathers.
For a complete discussion see Refs. 
\cite{fw93},\cite{fwo94},\cite{fkw94}. We will finally discuss the
possibility of wavelet analysis of breather-like objects \cite{kh93}.
Note that a number of sophisticated numerical studies aimed more at
the 'interaction' of discrete breathers with other objects - like discrete
breathers again, or impurities etc - will be discussed later in the appropriate
chapters of this review.

\subsection{A numerical experiment}
\label{s31}

Consider a Hamiltonian which yields a more general form of equation (\ref{2-3}):
\begin{equation}
H=\sum_l \left(\frac{1}{2} \dot{u}_l^2 + V(u_l) + \Phi(u_l - u_{l-1})\right) \;\;\;.
\label{3-1a}
\end{equation}
We specify the potential terms
in (\ref{3-1a}) in form of an expansion around this groundstate:
\begin{eqnarray}
V(z)= \sum_{k=2}^{\infty} \frac{1}{k}v_k z^k \;\;, \label{3-1b} \\
\Phi(z) = \sum_{k=2}^{\infty} \frac{1}{k}\phi_k z^k \;\;. \label{3-1c}
\end{eqnarray}
A specific realization is $v_2=1$, $v_3=-1$, $v_4=1/4$, $\phi_2=0.1$,
with all other expansion coefficients being zero.
The number of lattice sites $N$ is 
$N=3000$. Periodic boundary conditions are used.
Choose an initial condition for
a numerical integration which is spatially localized: all particles at
rest in their groundstate position, and one (central) particle displaced
from the groundstate position by some amplitude , having zero velocity.
Define the discrete energy density
\begin{equation}
e_l = \frac{1}{2}\dot{u}_l^2 + V(u_l) +\frac{1}{4}\phi_2\cdot\left[ (u_l-u_{l-1})^2
+ (u_l - u_{l+1})^2 \right] \;\;\;. \label{3-1}
\end{equation}
The sum over all local energy densities gives the total conserved
energy. If DBs are excited, the initial local energy burst should
mainly stay within the DB. Thus defining
\begin{equation}
e_{(2m+1)}= \sum_{-m}^{m}e_l \label{3-2}
\end{equation}
and exciting the local energy burst at lattice site $l=0$
by choosing a proper value of $m$ in (\ref{3-1}) we will control
the time dependence of $e_{(2m+1)}$. If this function doesn't
(or slowly enough) decay to zero, the existence of a breather-like object
can be confirmed. The term 'slowly enough' has to be specified
with respect to the typical group velocities of small amplitude
phonons. This sets the time scale we are interested in:
\begin{equation}
t \gg \frac{\sqrt{2+2\phi_2}}{2\phi_2}=7.416 \;\;\;. \label{3-3}
\end{equation}
In Fig.2 (Fig.2 of \cite{fwo94})
%\begin{figure}
%\caption{}
%\end{figure}
we show the time dependence of $e_{(5)}$ for
an initial condition $u_0(t=0)=2.3456$, $u_{l\neq 0}=0$,
$\dot{u}_l=0$.
Clearly a localized object is found. After
a short time period of the order of 100 time units nearly constant values
of $e_{(5)}$ are observed. The breather-like object is stable over a long period
of time with some weak indication of energy radiation. The energy distribution
within the object is shown in the inset of Fig.2. Essentially three
lattice sites are involved in the motion, so we find a rather localized
solution. 
While the central
particle performs large amplitude oscillations,
the nearest neighbours oscillate with small amplitudes. All oscillations
take place around
the groundstate.

\subsection{The internal dynamics}
\label{s32}

We show a $\sc Fourier$ transformation
of the motion of the central particle in the breather-like object 
(Fig.2 of \cite{fw93}) in Fig.3
%\begin{figure}
%\caption{}
%\end{figure}
. We see clearly that
there are two frequencies determining the motion of the central
particle $\omega_1 = 0.822\;,\;\;\omega_2 = 1.34$. All peak positions in
Fig.3 can be explained through linear combinations of these two
frequencies (similar observations were found recently in \cite{dvw96}. 
To proceed in the understanding of the phenomenon, we
plot in the inset in Fig.3 the {\sc Fourier} transformation of the motion
of the nearest neighbour(s) to the central particle. As
expected, we not only observe the two frequency spectrum,
but surprisingly the peak with the highest intensity is
not at $\omega_1$ as for the central particle, but at
$\omega_2$. It looks like every particle has its major
frequency.
Because of the symmetry
of the initial condition the two nearest neighbours move
in phase. Thus we are left with an effective 2 degree of freedom
problem (cf. inset in Fig.2).

\subsection{The reduced problem}
\label{s33}

Now it is a small step to recognize, that we might be confronted
with a kind of integrability phenomenon. Indeed, fixing the
rest of the particles at their groundstate positions reduces
the dynamical problem to a two degree of freedom system, which
might be integrable in parts of its phase space:
\begin{eqnarray}
\ddot{u}_0= -V'(u_0) - 2\phi_2(u_0 - u_{\pm 1}) \;\;, \label{3-4a} \\
\ddot{u}_{\pm 1} = -V'(u_{\pm 1}) - \phi_2 ( u_{\pm 1}
- u_0) \;\;. \label{3-4b}
\end{eqnarray}
We will call these types of few degree of freedom systems {\sl reduced
problems}.
In \cite{fw93}
a {\sc Poincare}
map between the trajectory and the subspace
$\left\{ \dot{u}_0,u_0,u_{\pm 1}=0,\dot{u}_{\pm 1} > 0\right\}$ has been
performed for the reduced problem, where the initial conditions correspond
to those of the breather-like trajectory of the extended lattice.
The same map has been then performed in the extended lattice itself, and
the two results were compared.
Not only was the existence of
regular motion on a twodimensional torus found in both cases, 
but the 
tori intersections for the reduced and full problems were practically
identical \cite{fw93}. 
Thus we arrive at two conclusions: i) the breather-like object
corresponds to a trajectory in the phase space of the full system
which is for the times observed practically embedded on a two-dimensional
torus manifold, thus being quasiperiodic in time;
ii) the breather-like object can be reproduced
within a reduced problem, where all particles but the central one
and its two neighbours are
fixed at their groundstate positions,
thereby reducing the number of relevant degrees of freedom.

With these properties in mind, it is clear, that there
have to appear two frequencies in the Fourier spectrum.
If the phase space flow of the reduced problem is 
regular in some parts of the phase space,
there should appear two actions $I_n$, $n=1,2$ as
functions of the original variables,
so that the Hamiltonian of the reduced problem can be
expressed through the two action variables only, and these
actions become integrals of motion. The corresponding
two frequencies
\begin{equation}
\omega_n = \frac{\partial H}{\partial I_n}    \label{3-5}
\end{equation}
determine the motion of system on the surface of the torus.
Obviously all linear combinations of multiples of these frequencies
appear in the {\sc Fourier} spectrum of the original particle
displacements. That is exactly what we observe.
The conclusions from above imply another
consequence - namely, that a spatially asymmetric breather-like
object (with respect to
the central particle) should also exist, i.e. that the
two nearest neighbours perform not-in-phase motions, even with
different amplitudes. That would mean, that in the language of
actions we lift a degeneracy by choosing asymmetric initial
conditions and have to expect
three instead of two fundamental frequencies, i.e. the frequency
$\omega _2$ splits into two frequencies $\omega _2 \neq \omega _3$.
This has been checked in \cite{fwo94} in a simulation with an asymmetric
initial condition, which differs from the previous symmetric
initial condition by additionally choosing $u_1(t=0) \neq 0$.
Indeed it was found i) that the local asymmetry is conserved throughout
the evolution of the system, and as the {\sc Fourier} spectrum of
the central particle motion and the two nearest neighbours
motions show, there are three frequencies:
$\omega_1 = 0.83$, $\omega_2=1.32$ and $\omega_3=1.35$ \cite{fw93}.

\subsection{A correspondence conjecture} 
\label{s34}

Intuitively it is evident, that none of the observed
frequencies describing the dynamics of the local entity should resonate
with the linear spectrum, since one expects radiation then, which would
violate the assumption that the object stays local without essential change.
In truth the conditions are much stricter, which we will discuss 
later. Since the reduced problem defined above can not be expected
to be integrable, we expect the typical phase space pattern
of a nonintegrable Hamiltonian system with two degrees of freedom
of having
some regular islands filled with nearly regular motion (tori) embedded
in a 'sea' of chaotic trajectories.
Chaotic trajectories have continuous (as opposed to discrete) Fourier
spectra (with respect to time), and so we should always expect that
parts of this spectrum overlap with the linear spectrum of the 
infinite lattice. Thus chaotic trajectories of the reduced problem 
appear not as candidates for breather-like entities. The regular islands
have to be checked with respect to their set of frequencies. If those frequencies
are located outside the linear spectrum of the infinite lattice, we
can expect localization - i.e. that a trajectory with the same initial
conditions if launched in the lattice will essentially form a localized
object. Islands which do not fulfill this nonresonance criterion should
be rejected as candidates for localized objects. Thus we arrive at
a selection rule for initial conditions in the lattice by studying the
low-dimensional dynamics of a reduced problem! This conjecture has
been successfully tested in \cite{fwo94}. In Fig.4 
%\begin{figure}
%\caption{}
%\end{figure}
we show a representative
Poincare map of the reduced problem. In Fig.5 
%\begin{figure}
%\caption{}
%\end{figure}
the time dependence of
the above defined local energy $e_{(5)}(t)$ is  shown for different
initial conditions which correspond to different trajectories of the
reduced problem. Clearly the initial conditions of regular islands
1,2 of the reduced problem yield localized patterns in the lattice, whereas
regular island 3 and the chaotic trajectory if launched into the lattice
lead to a fast decay of the local energy due to strong radiation of
plane waves. It is interesting to note that the energy decay of these
objects stops around $e_{(5)}=0.35$. In \cite{fwo94} it was noted
that the fraction of chaotic trajectories in the reduced problem is
practically vanishing for energies below that value. This aspect is not
yet fully understood.

Another observation which comes from this systematic analysis is
that the fixed points in the Poincare map of the reduced problem
(in the middle of the regular islands in Fig.4) correspond to periodic
orbits. A careful analysis of the decay properties as shown in Fig.5
has shown that all objects were slightly radiating - but some stronger
and some less. The objects corresponding to the periodic orbits of
the regular islands 1,2 of the reduced problem showed the weakest decay.
Thus we arrive at the suggestion that time-periodic local objects could
be free of any radiation - i.e. be exact solutions of the equations
of motion of the lattice! It makes then sense to go beyond the
approximations of discrete breathers as described in the previous
section and to look for a way of understanding why discrete breathers
can be exact solutions of the dynamical equations - provided they
are periodic in time. Further the question arises, why their 
quasiperiodic extensions appear to decay - i.e. why do quasiperiodic 
discrete breathers seem not to exist. We can also ask: suppose
quasiperiodic DBs do not exist - what are then their patterns of decay;
what about their life-times; what about moving DBs (certainly they 
can not be represented as time-peiodic solutions)? Much of what follows
will be finding answers to the questions raised here. 

Two things have to be said before we can proceed. First the linear
spectrum of the model used for the numerical results here is
optical-like, with a ratio of the band width to the gap of about
1/10. However this does not imply that the discrete breathers 
are merely due to some weakness of the interaction. First of all
we will see later, that the degree of localization of DBs is not
simply related to some bandwidth of the linear spectrum. Secondly
we know that DBs exist for systems without on-site potentials $V(z)$
but with nonlinear interactions. Third an estimation of the energy part
stored in the interaction of the DB object presented here yields
a value of 0.4. Compare that to the full energy $E\approx 0.7$.
Roughly half of the energy is stored in the interaction! that shows
how misleading the intuition can be. Another important addendum is that
of course the local ansatz was tested for different model realizations
(for details see \cite{fwo94}). Apparently the whole scenario is
rather independent of the choice of the model!

\subsection{Two-dimensional lattice}
\label{s35}

We left one question to be answered, i.e. - what is the 
impact of the dimension of the lattice? The answer seems to be - not much.
This must be so because the DBs can be described by local
few-degree-of-freedom systems (reduced problem). Then the only impact
the dimension of the lattice will have is to increase the number
of nearest neighbours, which implies simply some rescaling of the 
parameters of the reduced problem. To see whether that happens, 
the local ansatz was carried out in a two-dimensional analog of (\ref{2-6}).
The interested reader will find details in \cite{fkw94}. Here
we shorten the story by stating that practically the whole local ansatz
can be carried through in the two-dimensional lattice. An analog
of Fig.2 for the two-dimensional case is shown in Fig.6
%\begin{figure}
%\caption{}
%\end{figure}
where the energy distribution in a discrete breather solution is
shown, and the inset displays the time dependence of a local energy
similar to $e_{(5)}(t)$. 

The success of this step $d=1 \rightarrow d=2$ implies that nothing
terrible happens if we proceed to $d=3$ - which brings us to the possibility
of obtaining discrete breathers in the generic physical case (which
is $d=3$). Certainly it becomes worthwhile to continue our studies
now, since the expectations of finding some new interesting physical
phenomenon increase. Note how far we proceeded as compared to the
PDE studies mentioned in the introduction.

\subsection{Wavelet analysis}
\label{s36}

Before proceeding with the study of discrete breathers, it
is worthwhile to mention a numerical technique which is capable of 
efficiently tracking discrete breathers. 
If we ask for an efficient encoding of a spatially localized entity
(forget about the time-dependence for a moment), then it is certainly
not a spatial Fourier series, because many normal modes will be
'excited'. A simple basis is the lattice itself (a local basis),
but only as long as the discrete breather is localized on a few lattice
sites (call it a 'microscopic' breather). 
A 'macroscopic' breather could be then described by a set of normal
modes (whose number should decrease with increasing size of the breather).
What about a `mesoscopic` breather? The method of wavelets could be a
good candidate to fill this window. As shown by Hori \cite{kh93}
it takes only a few wavelet modes to encode a breather which is
neither localized on a few sites nor too extended. This can be very
useful when monitoring certain patterns of a numerical integration.

\section{Localization properties of discrete breathers}
\label{s4}

A natural way to proceed is to use an ansatz for the analytic
form of a discrete breather and to test what restrictions apply
to the parameters of the ansatz. Up to now we have seen that
discrete breathers are most probably regularly evolving in time,
so we will use a quasiperiodic function of time. This will bring
us to necessary existence conditions for discrete breathers, will
explain why in general discrete breathers can be only periodic in time,
but not quasiperiodic, and will allow us to estimate the decay
of Fourier coefficients (which are the parameters of the ansatz) in
space and in $k$-space. For details see \cite{sf94},\cite{sf95-pre-2}.

\subsection{The ansatz for a breather solution}
\label{s41}

As we have seen in the previous section the breather-like objects
evolve regularly in time, so that we choose an ansatz which
yields a quasiperiodic evolution of all variables:
\begin{equation}
u_l(t)=\sum_{k_1,k_2,...,k_n=-\infty}^{+\infty}
A_{lk_1k_2...k_n}e^{i(k_1\omega_1+k_2\omega_2+...+k_n\omega_n)t}\;\;.
\label{4-1}
\end{equation}
We will use the notations
\[
\vec{\omega}=(\omega_1,...,\omega_n)\;,\;
\vec{k}=(k_1,...,k_n)\;\;.
\]
The localization property of (\ref{4-1}) is defined by the boundary
condition
\begin{equation}
A_{lk_1k_2...k_n}\mid_{l \rightarrow \pm \infty} \rightarrow 0 \;\;\;.
\label{4-2}
\end{equation}
Since the $u_l(t)$ are real-valued we can reduce the number of independent
Fourier coefficients $A_{lk_1k_2...k_n}$ by demanding 
$A_{lk_1k_2...k_m...k_n}=A_{lk_1k_2...-k_m...k_n}$ for any $m$. Inserting
(\ref{4-1}) into the equations of motion for $u_l(t)$ and eliminating
$t$ by sorting exponentials with same exponents we will arrive at
some infinite number of coupled algebraic equations for the Fourier
coefficients $A$, where the control parameters are the model parameters
of the Hamiltonian and the frequencies $\omega_1 ... \omega_n$. These equations
will have a very complicated form due to the nonlinearities in the
equations of motion.
Because we consider nearest neighbour interaction (\ref{3-1a})
we can formally write down the resulting set of equations
as a nonlinear map:
\begin{equation}
M_{l+1,\vec{k}}=F(\{ M_{l,\vec{k'}} \} ,
\{ M_{l-1,\vec{k''}} \} ) \;\;\;. \label{4-3}
\end{equation}
Here we introduced a function $M_{l,\vec{k}}$ which is defined on
a discrete $n$-dimensional lattice. The lattice is given by
all combinations of $\{k_1,k_2,...,k_n\}$ where each integer $k_{n'}$
varies from $-\infty$ to $+\infty$. We have
\begin{equation}
M_{l,\vec{k}}=A_{lk_1k_2...k_n}\;\;\;. \label{4-4}
\end{equation}
Let us study (\ref{4-3}) in the tails of the DB, i.e. for
$l \rightarrow \pm \infty$ where (\ref{4-2}) holds by assumption. In the
generic case $v_2$ and $\phi_2$ from (\ref{3-1b}),(\ref{3-1c}) will be
nonzero. Then we can linearize the mapping (\ref{4-3}):
\begin{equation}
M_{l+1,\vec{k}}=(\kappa_{\vec{k}}(\vec{\omega})+2)
M_{l,\vec{k}}-M_{l-1,\vec{k}}
\;\;\;. \label{4-5}
\end{equation}
Here we have introduced another function on the $n$-dimensional
discrete space which is given by
\begin{equation}
\kappa_{\vec{k}}(\vec{\omega})=\frac{v_2-(\vec{k}\cdot \vec{\omega})^2}
{\phi_2}\;\;\;. \label{4-6}
\end{equation}
Equation (\ref{4-5}) is linear and thus every component of $M$ in the
$n$-dimensional discrete space decouples in this equation from
all other components. In fact (\ref{4-5}) is a linear two-dimensional
map, and $M_{l,\vec{k}}=0$ is a fixed point of this map (as it is for
the original nonlinear map (\ref{4-3})). It is characterized by the matrix $G$:
\begin{equation}
G =
\left(
\begin{array}{lr}
1 & \kappa_{\vec{k}}(\vec{\omega})  \\
1 & 1+\kappa_{\vec{k}}(\vec{\omega})
\end{array}
\right)\;\;\;. \label{4-7}
\end{equation}
For the eigenvalues of $G$ we find
\begin{eqnarray}
\lambda_{1,2} & = & 1+\frac{\kappa_{\vec{k}}(\vec{\omega})}
{2}\pm \sqrt{(1+\frac{\kappa_{\vec{k}}(\vec{\omega})}{2})^2 - 1}\;\;,
\label{4-8} \\
\lambda_1 \lambda_2 &  =  & 1 \;\;. \label{4-9}
\end{eqnarray}
We can consider three cases:
\begin{equation}
a)\;\;\kappa_{\vec{k}}(\vec{\omega}) > 0 \;\;: \;\;\;
0 < \lambda_2 < 1 \label{4-10}
\end{equation}
i.e. $\lambda_2$ is real. Especially $\lambda_2(\kappa_{\vec{k}}(\vec{\omega})
\rightarrow 0
)\rightarrow 1$ and $\lambda_2(\kappa_{\vec{k}}(\vec{\omega})
\rightarrow \infty)
\rightarrow 0$.
\begin{equation}
b) \;\; \kappa_{\vec{k}}(\vec{\omega}) < -4 \;\;:
-1 < \lambda_1 < 0 \label{4-11}
\end{equation}
i.e. $\lambda_1$ is real. Especially $\lambda_1(\kappa_{\vec{k}}(\vec{\omega})
\rightarrow -
\infty) \rightarrow 0$ and $\lambda_1(\kappa_{\vec{k}}(\vec{\omega})
\rightarrow -4)
\rightarrow -1$.
\begin{equation}
c) \;\;-4 \leq \kappa_{\vec{k}}(\vec{\omega}) \leq 0 \;\;:
\;\;\; \left|\lambda_1\right|
= \left|\lambda_2\right| = 1 \label{4-12}
\end{equation}
i.e. $\lambda_{1,2}$ are complex conjugated numbers on the unit circle.
Consequently in cases a) and b) the fixed point of the
mapping is a hyperbolic one, i.e. there exists exactly one direction
(eigenvector) in which the fixed point can be asymptotically reached
after an infinite number of iterations. In case c) the fixed point is
a marginally stable elliptic point, i.e. starting from any direction
the fixed point can be never reached after an infinite number of steps,
instead the mapping will produce a (deformed) circle around the
fixed point. Thus we find that case c) (\ref{4-12}) contradicts the
localization condition (\ref{4-2}).

\subsection{Existence conditions for time-periodic discrete breathers}
\label{s42}

Let us consider $n=1$. Then the DB solution is periodic
(cf. (\ref{4-1})). Equation (\ref{4-6}) can be simplified to
\begin{equation}
\kappa_{\vec{k}}(\vec{\omega})=\frac{v_2-k_1^2\omega_1^2}{\phi_2}
\;\;\;. \label{4-12a}
\end{equation}
The frequencies $\omega_q$ for small-amplitude plane waves around
the considered groundstate of (\ref{3-1a}) (where $q$ is the wave number)
are related to the parameters $v_2$ and $\phi_2$ by
\begin{equation}
v_2 \leq \omega_q^2 \leq v_2 + 4 \phi_2 \;\;\;. \label{4-13}
\end{equation}
Then it follows that case c) given in (\ref{4-12}) is identical with
\begin{equation}
c) \;\;\; k_1^2\omega_1^2 = \omega_q^2 \;\;\;. \label{4-14}
\end{equation}
We find that a time-periodic discrete breather 
can not exist if any multiple of its fundamental frequency
equals any frequency of the linear spectrum. The reason is that we can not satisfy
(\ref{4-14}) and (\ref{4-2}) simultaneously because of (\ref{4-12}).
One can interpret (\ref{4-14}) as a definition of nonexistence bands on
the $\omega_1$-frequency axis for time-periodic discrete breathers. 
Introducing a
normalized frequency $\tilde{\omega}_1=\omega_1/\sqrt{v_2}$
and normalized interaction $\tilde{\phi}_2=\phi_2/v_2$
(only if $v_2 \neq 0$) those bands are given by
\begin{equation}
\frac{1}{k_1^2} \leq \tilde{\omega}_1^2 \leq \frac{1+4\tilde{\phi}_2}{k_1^2}
\;\;\;. \label{4-15}
\end{equation}
For frequencies $\omega_1$ below the linear spectrum $\omega_q$ these
nonexistence bands increase in width with increasing 
$\tilde{\phi}_2$ and completely
overlap at $\tilde{\phi}_2=3/4$. Thus time-periodic DBs with frequencies
below the linear spectrum should only exist if the band width of the
linear spectrum is narrow enough. On the contrary no
frequency restrictions
seem to apply for the existence of single frequency DBs if the frequency
is located above the linear spectrum.

Now let us make some statements about spatial correlations
of phases of periodic DBs
if they exist. If the frequency of the DB is above the linear spectrum
then it follows $\kappa_{\vec{k}}(\vec{\omega}) < -4 $ for all $k_1$.
This corresponds to case b) in (\ref{4-11}). Then we have $-1 < \lambda_1
< 0$. Consequently $-1 < M_{l+1,\vec{k}}/M_{l,\vec{k}} < 0$
for all $l$ in the tail of the DB. Thus we expect a coherent out-of-phase
type of the motion of neighbouring particles in the tails of the DB.
If the frequency of the periodic DB is below the linear
spectrum ($v_2 > 0$)
things become more complicated. Namely there will always exist a certain
finite integer $k_c$ such that for $k_1 < k_c$ it follows
$\kappa_{\vec{k}}(\vec{
\omega})> 0$ which corresponds to case a) in (\ref{4-10}).
The corresponding Fourier components would yield in-phase type
of motion in the tails of the DB. However for all $k_1>k_c$
the case b) in (\ref{4-11}) applies. Those Fourier components would
yield out-of-phase motion. 
Thus we have to expect a complicated mixture of in- and out-of-phase
type of motion.

\subsection{Quasiperiodic discrete breathers}
\label{s43}

Let us consider $n=2$. Then
case c) in (\ref{4-12}) applies if
\begin{equation}
v_2 \leq (k_1\omega_1+k_2\omega_2)^2 \leq v_2+4\phi_2
\;\;\;. \label{4-16}
\end{equation}
Now it is possible to show that there exists an infinite number
of pairs of the integers $(k_1,k_2)$ such that (\ref{4-16})
is satisfied if the ratio $\omega_1/\omega_2$ is irrational
and $v_2 \geq 0$ and $\phi_2 > 0$ (cf. Appendix in \cite{sf94}).
Thus we can not expect the existence of two-frequency
DBs. The argument for $n \geq 3$ is also straightforward and
yields the same result. 

Some comments are appropriate at this stage. If we consider
the breather solution (\ref{1-2}) of the sG PDE, then its decomposition
into Fourier components (which are smooth functions of $x$) can be
also done. But then why does the sG breather exists at all in the
presence of an infinite number of resonances of multiples of
the breather's frequency with the linear spectrum? This undoubtly
is a subtle question. We will try to give a qualitative explanation.
Writing down the differential equations for the Fourier amplitudes
with all nonlinearities yields for each amplitude a differential
equation with homogeneous and inhomogeneous terms. Say we linearize
the equations in the tails of the breather solution. Suppose we did that
for a Fourier component not in resonance with the linear spectrum.
Then the linearized equations give us a one-parameter manifold
of solutions contracting to zero if integrated to infinity, with exponential
falloff. Did we miss something by neglecting the nonlinear inhomogeneous
terms? It could be, that the inhomogeneous terms decay slower than
the homogeneous solution. Then we know that the inhomogeneous part will
be the leading one if integrating far away from the center. Still we
are not restricted in the choice of the homogeneous solution - we still
have a one-parameter family of solutions. Take now a Fourier component
which resonates with the linear spectrum. Then the linearized equation
yields oscillating solutions. The only way to avoid that is to choose
the homogenenous solution to be zero - thus loosing the one-parameter
manifold of solutions. If now the inhomogeneous nonlinear parts omitted
by the linearization will show up with exponential decay, the 
full solution will decay in space too - but it ceases to be a one-parameter
manifold of solutions, it is just one fixed solution. 
This is precisely what happens to the sG breather solution (a nice
calculational exercise!).
Suppose we
change the Hamiltonian slightly - i.e. we change the control parameters
defining the differential equation or the difference equations discussed
in this review - then any solution should stay a solution, simply
change slightly. But if we had a resonance, this will have a dramatic
effect - we will in general slightly change the homogenenous solution
(from zero to nonzero) and thus loose the breather, instead obtaining
a breather-like object which is however not fully decaying to zero
in the tails! This is nothing else but Boyd's nanopteron \cite{jpb90},
which Kivshar and Turitsyn coin soliton on a standing carrier
wave \cite{kt92}.

This should make clear that any resonance is deadly for the breather -
either periodic or quasi-periodic in time. Exceptions can occur, but
will be nongeneric. So is the sG breather for PDEs, so are quasi-periodic
DBs (or time-periodic DBs with resonances). 
Another way of explaining it is that in the case of resonance(s) the
additional requirement of putting the mentioned homogeneous solutions
to zero makes the set of equations for the Fourier components overdetermined,
and thus solvable only for special cases (careful: this type of argument
is applicable if one considers only solutions which do decay in the tails
to zero!).

Cai, Bishop and Gr$\o$nbech-Jensen \cite{cbg95} have demonstrated that
quasiperiodic DBs exist for the Ablowitz-Ladik lattice. This is one
of the mentioned exceptions. Not only is the AL lattice integrable
(note that there is no direct connection between integrability and
the existence of breathers) but it is very restrictive with respect
to the equations for the Fourier amplitudes. We have checked that the
equations for the Fourier amplitudes allow for solutions when
all Fourier amplitudes with say negative Fourier numbers $k$ are zero
(the variables here are complex). Consequently this puts a tremendous
constraint on the combinations of $(k_1,...,k_n)$ for which resonances
with the linear spectrum can appear - in fact it is easy to see
that for the choosen solution in \cite{cbg95} no resonance
appears at all!

\subsection{The spatial decay of DBs}
\label{s44}

From now on discrete breather implies a time-periodic spatially
localized solution. Let us assume that such a solution is given,
and analyze its localization properties. For that we will 
first linearize the equations of motion in the tails and solve
for each Fourier component. The result is an exponential decay
given by the eigenvalues (\ref{4-8}):
\begin{equation}
A_{kl} \sim  ({\rm sgn}(\lambda_k))^l{\rm e}^{ln\mid
\lambda_k \mid l} \;\;, \label{4-17}
\end{equation}
where the eigenvalue choosen is the one with absolute value lower
than one. Note that (\ref{4-8}) depends only on the
linear parameters defining the linear spectrum. Consequently
all one has to know about the system in order to predict the spatial
decay are the frequency of the DB and the position of the linear spectrum
defined by $(v_2,\phi_2)$.
In Fig.7
%\begin{figure}
%\caption{}
%\end{figure}
the dependence of the Fourier amplitudes $A_{kl}$ on $l$ is shown
for a solution obtained in \cite{sf94} (we will discuss the method
of obtaining these solutions later). An exponential decay
of the $A_{kl}$ is obtained, with $k$-dependent exponents.
Comparing the numerically obtained exponents with the prediction
of the linearization (\ref{4-17}) we find excellent agreement in
Fig.8.
%\begin{figure}
%\caption{}
%\end{figure}
Another prediction we can formulate is, that if the
DB frequency $\omega_1$ is restricted to be in the nonzero
gap of the linear spectrum, then there exists a nonzero
value of $\omega^{(m)}_1$ such that the exponential decay in the
DB tails will be weaker for all other frequencies $\omega_1\neq \omega^{(m)}_1$
(still belonging to the gap). Let us explain why this statement
follows from the previous considerations.
First the frequency $\omega_1$ has to be larger than the
linear spectrum width - else one (or more) of multiples of $\omega_1$
will always lie in the linear spectrum. Secondly if $\omega_1$
is slightly below the lower edge of the linear spectrum, then the 
spatial decay of the $k=1$ component
will be very weak. Lowering $\omega_1$ we increase the decay exponent
for $k=1$,
but since $2\omega_1$ comes closer to the upper band edge
of the linear spectrum, there
will be a certain frequency when the decay in the tails will be governed
by the second harmonic rather than the first harmonic of $\omega_1$.
A calculation of $min(1-|\lambda(\kappa_k)|)$ with $|\lambda(\kappa_k)| \leq 1$
for different $\omega_1$ has been performed in \cite{sf94} (Fig.2
of \cite{sf94}). 
A comparison of the positions of the observed maxima with numerical
simulations which yielded whole families of DBs for different frequencies
clearly demonstrated the validity of the prediction - the DBs with frequencies
in the gap can not be infinitely strongly localized, thus there has to
be a DB solution that shows up with the strongest spatial localization.
Note that for DBs with frequencies above the linear spectrum nothing similar
occurs - the  Fourier components with larger $k$ are always stronger
localized than those with smaller $k$ (an exception is the $k=0$ component,
which we will discuss later).

\subsection{Nonlinear corrections}
\label{s45}

As already discussed, the linearization procedure fails whenever
nonlinear inhomogeneous terms decay more weakly in space than the
linear homogeneous solution. This will  happen typically whenever
any of the multiples of the DB's frequency comes close to the
linear spectrum. Note that this includes the limit when the
DB's frequency is itself very close to the linear spectrum -
a limit where several perturbation theories are usually applied.
Consequently the findings of this subsection are of relevance also
for DB solutions with frequencies above the linear spectrum.

Let us define the function $d(k)={\rm ln} (|\lambda_k |)$, which
is by definition always negative. Consider $k \geq 0$. Then $d(k)$ has
a single maximum for some $k=k_m$ and is monotonically decreasing with
further increasing $k$. The linearization procedure is violated
for a given $k_0$ if there exists a sequence of $k_1,k_2,...,k_{\alpha -1}$
(where either $v_{\alpha} \neq 0$ or $\phi_{\alpha} \neq 0$)
such that
\begin{equation}
d(k_0) \leq (d(k_1) + d(k_2) + ...+d(k_{\alpha-1})) \label{4-18}
\end{equation}
holds together with the condition 
\[
k_0=\pm k_1 \pm k_2 \pm ... \pm k_{\alpha -1} \;\;.
\]
It follows that the linearization is always correct for $k_m$.
In \cite{sf95-pre-2} it was shown that the number of other $k$ for which
the linearization does not hold is always finite for finite distance
of $\omega_1$ from the linear spectrum. A numerical test has verified
the correctness of the above statement, and in \cite{sf95-pre-2} quantitative
agreement has been found (one can predict the corrected exponents of
the spatial decay of the affected components by finding the leading
inhomogeneous term) between numerical DB solutions and the 
theoretical prediction. 

\subsection{The decay in $k$-space}
\label{s46}

Because higher harmonics are at large distance from the linear
spectrum , the corresponding Fourier components decay much faster
in space. That also implies that the decay in $k$ is increased when
increasing the distance from the breather's center \cite{sf95-pre-2}.
Using the linearization procedure Flach obtained for large $k$ 
\cite{sf95-pre-2}
\begin{equation}
|A_{kl}| \sim k^{-2|l|} s(k) \;\;, \label{4-19}
\end{equation}
where $l$ measures the distance from the DB center and $s(k)$ is some
site-independent function (typically decreasing exponentially with increasing
$k$ \cite{sf95-pre-2}). 
Thus hopping from one lattice site to the neighbour site
increases the decay in $k$-space by multiplying the original decay
function with a power $1/k^2$! The numerical DB solution analysis
yields perfect agreement to machine precision (Figs.7,8 in 
\cite{sf95-pre-2}).

\subsection{The case $v_2=0$}
\label{s47}

So far we discussed cases when $\phi_2 \neq 0$ and $v_2 \neq 0$.
The linear spectrum is in that case optical-like, i.e. it has
a nonzero lower bound and of course a finite upper bound.
Let us see what to expect when this is not the case anymore.

First of all in order to keep the groundstate solution
of the system we have to demand $v_3=0$ then without loss
of generality. The same applies to the later discussed case
$\phi_2=0$ of course.

The linear spectrum still exists, but it is acoustic-like -
i.e. the lower bound is zero. Nothing spectacular will happen
then for all Fourier components with $k \neq 0$. However the
$k=0$ component (the dc component) will resonate with the 
lower bound of the linear spectrum! We are saved if this dc component
is zero by some symmetry reason - e.g. for all $\Phi(z)=\Phi(-z)$
and $V(z)=V(-z)$
$A_{kl}=0$ for all even $k$, including the dc term. 
What if we
change the interaction potential and loose this symmetry? 
Numerical simulations \cite{bks93},%
\cite{ck93},\cite{hsx93},\cite{kbs93},\cite{%
kbs94-prb},\cite{kbs94-jl},\cite{kbs94-pla},\cite{sp94} have shown that
for one-dimensional systems of that type (with the additional
restriction $V(z)=0$ which can be called a Fermi-Pasta-Ulam (FPU)
lattice)  a breather-kink state forms, i.e. the Fourier components
with $k \neq 0$ decay in space to zero 
whereas the dc component shows up with a kink-like
structure. 
Assuming that these numerically found objects correspond to
solutions with finite energy, we can immediately obtain the
spatial variation of the $k=0$ component. 
Because far away from the solution
the dc displacements can be viewed as a strain field caused by a 
point-like source (in full analogy with the coresponding Maxwell equation
for the electric field, see \cite{LLVII}) it follows that the
decay can be calculated using Gauss's theorem. For a $d$-dimensional
system the decay of the dc component should be then $ \sim 1/r^{d-1}$
\cite{sf96}. So for a one-dimensional lattice we indeed obtain no decay
at all, which corresponds to the mentioned numerical findings of a
kink structure. In two- and three-dimensional lattices the prediction
would be an algebraic decay $1/r$ and $1/r^2$ correspondingly. That could
affect the exponential decay of the $k\neq 0$ components by changing
the exponential decay to an exponential times an algebraic decay.
Numerical tests are needed in order to check these possibilities.

Clearly if $v_{\mu} \neq 0$ with $\mu\geq 4$ then even for $d=1$
a kink-like structure would be impossible together with the decay
of all other Fourier components. Solving the equations for the 
dc component and taking the smallest $\mu$ for which  $v_{\mu} \neq 0$
we can predict $A_{0l} \sim l^{-2/(\mu-1)}$ for $d=1$. Since $\mu \geq 4$ here
this decay is either $1/(l^{2/3})$ or weaker. 
Altogether the issue of the dc component resonating with the linear
spectrum is an issue which requires further investigation.

\subsection{Systems without a linear spectrum}
\label{s48}

Consider $\phi_2=0$.
This implies that the linear spectrum is highly degenerated and
given by one value $\sqrt{v_2}$. We do not have problems with resonances
here (provided $v_2 \neq 0$ or $V(z),\Phi(z)$ are even functions), 
so DBs will exist. Moreover the decay of the Fourier components
will be faster than exponential in the tails. Sorting the equations
for the Fourier components we will obtain in lowest order
$\omega_1^2k^2 A_{kl} \sim A_{k,l-1}^{\mu-1}$ where $\mu$ is the
smallest even integer for which $\phi_{\mu} \neq 0$ (note that
the interaction has to be nonzero - otherwise we solve the trivial
problem of uncoupled particles). The resulting decay is
\begin{equation}
A_{kl} \sim {\rm e}^{(\mu-1)^l a_k} \label{4-20}
\end{equation}
where $a_k$ is some number which depends on $k$. The result (\ref{4-20})
is an exponential decay with the variable being not the distance itself
but rather an exponential function of the distance again! A nice example
to test this is the FPU lattice in one dimension with $\phi_{2m} \neq 0$
and all other control parameters of the potentials vanishing. The
amplitudes obtained e.g. in \cite{sf94} or in \cite{ma96} 
fit the predicted spatial decay.

It is not clear at present what happens if $v_2=0$ and the potential
functions are not even. Clearly these problems are waiting for an answer.

\subsection{Higher lattice dimensions}
\label{s49}

In this section we studied DBs for one-dimensional systems. The
generalization to higher lattice dimensions is straightforward.
Instead of solving linear maps one has to introduce Green's functions
\cite{st92-jpsj-2},\cite{dpw92}. The results will be essentially
the same. Especially quasiperiodic DBs will again turn out to be
nongeneric, and time-periodic DBs will exist if all multiples
of their frequency do not resonate with the linear spectrum.

\section{Existence proofs} 
\label{s5}

In this section we will discuss existence proofs for discrete breathers -
not for isolated model systems (like the Ablowitz-Ladik one) but for
classes of systems. The central part in this section will be the
continuation of periodic orbits - starting with the limiting case
of uncoupled particles (e.g. $\Phi(z)=0$) and then continuing to
finite interaction. This approach has been coined antiintegrability
method by Aubry \cite{sa94}, which is connected to the coding sequence
structure which defines a certain periodic orbit of the noninteracting
many-particle system. The proof has been performed by MacKay and Aubry
\cite{ma94},
and with some modifications also by Bambusi \cite{db96}. Further we will
discuss the proof of existence of DBs for systems with homogeneous
potentials by Flach \cite{sf95-pre-1}, 
which works especially for systems with $V(z)=0$ where
the antiintegrability method is not applicable directly. 
Finally we will discuss some related results obtained in the mathematical
literature. 
Only the essential ideas of the proofs and the conditions for their
application are given here. 
For a more complete discussion see Refs. 
\cite{ma94},\cite{db96},\cite{sf95-pre-1},\cite{sa96}.

\subsection{The antiintegrability approach}
\label{s51}

The concept of antiintegrability was initially introduced for the 
Standard map \cite{aa90}. This concept turned out to be a simple and
powerful tool for finding solutions of nonlinear systems. It is 
in essence the continuation of zeroes of an operator $F$ from
a solvable limit. The nonlinearity is necessary in order
to be sure that the eigenvalues of a certain Newton operator (a differential
of the operator $F$) are well behaved for the solution to be continued.
The antiintegrability method allows for a more efficient and numerically
easier and reliable calculation of chaotic trajectories of nonlinear
maps than other methods do. It can be easily extended to coupled maps.   

Consider the model (\ref{3-1a}) with parametrized interaction 
$\Phi(z)=\epsilon \Psi(z)$. Consider the case $\epsilon=0$. Then
the equations of motion are trivial since they decouple in space.
All we have to do is to solve for each individual oscillator
moving in its potential $V(z) \neq 0$. Note that we can not discuss
$V(z)=0$ since then the particles are unbounded, which makes
the following impossible. Each local solution can be then given 
in the form 
\begin{equation}
u_n(t)=g(\omega_n t + \alpha_n)\;\;, \label{5-1}
\end{equation}
where the function $g(t)$ is periodic in time with period $2\pi$.
The frequency $\omega_n$ is a unique function of the action $I_n$
of the $n$th oscillator (not that the reverse is not true - in general
each value of $\omega_n$ can correspond to several isolated values
of the action). 

The general idea of the proof is: consider a time-periodic
solution of the whole lattice for $\epsilon=0$  - e.g. one where all oscillators
are at rest but one oscillator is moving (encoded with its values
of the frequency $\omega_n$ and phase $\alpha_n$). Then it is possible
to find a periodic orbit of the lattice with $\epsilon \neq 0$ but
close to zero such, that the new periodic solution will have the 
same frequency and be close to the old periodic solution.  In other
words we can continue the periodic orbit of the noninteracting system
to finite interactions. It can be shown then that these periodic orbits
will be exponentially localized around the chosen center $n$, and
thus are discrete breathers.

The method starts with considering $F(u;\epsilon)=(F_1(u;\epsilon),
F_2(u;\epsilon),...)$ with 
\begin{equation}
F_n(u;\epsilon) = \ddot{u}_n + \frac{\partial H}{\partial u_n}\;\;.
\label{5-2}
\end{equation}
Note that $H$ parametrically depends on $\epsilon$. 
We are looking for zeroes of $F(u;\epsilon)$ (i.e. each component
of $F$ must vanish) since the zeroes of $F(u;\epsilon)$ solve the
equations of motion. The goal is to apply the implicit function theorem
(see e.g. \cite{jd69}). For that we need a subspace of
all functions $u=\{u_n(t)\}$ which yields isolated zeroes of $F$.
This can be done by restricting the consideration to time-periodic
functions. Still the frequency and the phase at each lattice
site could be choosen in different ways, which creates smooth
manifolds of zeroes of $F$. To guarantee that $F$ has only isolated
zeroes for $\epsilon=0$, MacKay and Aubry
demand that the frequency and phases are fixed \cite{ma94}
by staying in the subspace of all periodic functions $u=\{u_n(t)\}$ with:
\begin{equation}
u_n(t)=u_n(t+2\pi/\omega_b) \label{5-3}
\end{equation}
and
\begin{equation}
u_n(t)=u_n(-t)\;\;. \label{5-4}
\end{equation}
Here (\ref{5-4}) is allowed because the Hamiltonian equations of motion
are time reversible.
Now the zeroes of $F(u;\epsilon)$ are isolated for $\epsilon=0$.
Application of the implicit function theorem is then possible.
One is looking for a continuation of $u(\epsilon)$ such that
$F(u(\epsilon),\epsilon)=0$. The main problem is to show that
$\partial_u F$, the derivative with respect to $u$, is invertible
at the initial solution (for $\epsilon=0$). The Newton operator
$\partial_u F(u(\epsilon);\epsilon)$ has to fulfill the equation
\begin{equation}
\partial_u F(u(\epsilon);\epsilon) \frac{{\rm d} u(\epsilon)}{
{\rm d} \epsilon} + \partial _{\epsilon}  F(u(\epsilon);\epsilon)
=0\;\;, \label{5-5}
\end{equation}
which is a differential form of $F=0$. 
Continuation to $\epsilon \neq 0$ can be performed provided
the nonresonance condition $k \omega_b \neq \sqrt{v_2}$ and the
nonlinearity condition $V(z)-(v_2/2) z^2 \neq 0$ hold. It can be continued
until the Newton operator ceases to be invertible - which corresponds
to a collision of Floquet multipliers at $+1$ in the dynamical stability
analysis of the discrete breather orbit (see below).
The exponential degree of localization of the discrete breather
is proven in \cite{ma94} which studies the properties of the inverse
of the Newton operator.

The remarkable thing about this proof is that nothing has to be
said neither about the lattice dimension nor about the size of the system,
which can be infinite. Thus indeed discrete breathers exist e.g. in
two- and three-dimensional lattices. Another interesting fact is
that MacKay and Aubry can also continue 'multisite' breathers - 
i.e. solutions which are still exponentially localized in space but
whose internal structure is far more complicated than the simple
structure of the 'one-site' breathers. 

Bambusi has recently provided a similar proof with the difference
that the energy of the solution is fixed (continuation for constant
energy rather than for constant frequency). 

Using the antiintegrability approach MacKay and Aubry prove the
existence of discrete breathers in the discrete nonlinear Schroedinger
equation \cite{ma94}. 
This is especially interesting because the DNLS structure
is close to different models of interacting classical spins, so
that one can expect classical spin breathers to exist (numerically
these solutions have been found in \cite{wmb95},\cite{tkt93}
using RWA).

Aubry \cite{sa96} and Perfetti \cite{pp96} demonstrate
the existence of 'rotobreathers' for Hamiltonians which are
periodic in the displacements (angles). These solutions correspond
to a finite number of rotators rotating, and the rest of the rotators
librating with exponentially decreasing amplitude as the distance to
the center is increased. Numerical evidence of rotobreathers is
reported in \cite{tp96}.

\subsection{The homoclinic orbit approach}
\label{s52}

Consider a system with homogenenous potentials $V(z)=v_{2m}z^{2m}/(2m)$
and $\Phi(z)=\phi_{2m}z^{2m}/(2m)$. The equations of motion become
\begin{equation}
\ddot{u}_l = -v_{2m} u^{2m-1}_l - 
(u_l-u_{l-1})^{2m-1}-(u_l-u_{l+1})^{2m-1}\;\;. \label{5-5b}
\end{equation}
Note that we can always use $\phi_{2m}=1$ by rescaling time.
Kivshar \cite{ysk93-pre} 
noted that one can look for time-periodic solutions
of (\ref{5-5b}) in the form
\begin{equation}
u_l(t)=A_l G(t) \label{5-6}
\end{equation}
which separates time and space. Inserting (\ref{5-6}) into (\ref{5-5b})
we obtain a differential equation $\ddot{G}(t)=-\kappa G^{2m-1}(t)$
with a separation parameter $\kappa > 0$ and a two-dimensional
map for the amplitudes $A_l$:
\begin{equation}
\kappa A_l= v_{2m} A_l^{2m-1} + 
(A_l-A_{l-1})^{2m-1}+(A_l-A_{l+1})^{2m-1}\;.\label{5-7}
\end{equation}
Since the equation for the masterfunction $G(t)$ yields time-periodic
solutions, we have to show that homoclinic orbits exist for the discrete
map (\ref{5-7}) and will thus prove the existence of discrete breathers.
This has been done by Flach in \cite{sf95-pre-1} for $v_{2m}=0$, i.e.
for a system which does not fall into the class considered by MacKay
and Aubry in the previous subsection. The proof can be easily extended
to incorporate $v_{2m} \neq 0$. Let us sketch the main ideas. First of all
we have no restriction on $\kappa$ other than $\kappa > 0$.
Secondly we consider  $f_l=|A_l|$ with
\begin{equation}
\kappa f_l = v_{2m} f_l^{2m-1} +
(f_l+f_{l-1})^{2m-1}+(f_l + f_{l+1})^{2m-1}\;. \label{5-8}
\end{equation}
Equation (\ref{5-8}) has a hyperbolic fixed point (0,0) and an
elliptic fixed point $(f,f)$ for $ \kappa_{fp2}=(v_{2m}+2^{2m})f^{2m-2}$.
 
We start with a set of two nearest neighbour values $f_{l-1}$ and
$f_l$ and compute $f_{l+1}$ using (\ref{5-8}). Requiring
$0 < f_{l+1} < f_l$ we obtain an allowed region for $\kappa$, where
the lower bound of this region corresponds to $f_{l+1} = 0$. 
Repeating this procedure with increasing $l$ we find that
the lower bound on $\kappa$ is monotonically increasing with $l$.
Making sure that at the beginning of the map the lower bound was below
$\kappa_{fp2}$ we thus arrive at the conclusion that it will always
stay below $\kappa_{fp2}$, but it can not asymptotically reach it either for
$l \rightarrow \infty$, so we will find a $\kappa$ for which the
the iteration leads to a decay to zero. By proper choice of the initial
pair of neighbouring values and using the inversion symmetry of (\ref{5-8})
we can complete the proof.

The existence of homoclinic orbits of (\ref{5-8}) also implies
the existence of intersection points of the stable and unstable invariant
manifolds of the hyperbolic fixed point $(0,0)$. In \cite{sf95-pre-1} the
existence of horseshoe patterns was demonstrated. Then it is evident
that the invariant manifolds will allow for a much richer set of homoclinic
orbits than say the one-site breathers considered. To our understanding
these rich intersection patterns of the invariant manifolds directly 
compare to the possibility of continuing multi-site breathers using
the antiintegrability approach.

\section{Numerical methods obtaining discrete breathers}
\label{s6}

In this section we will briefly review some numerical methods
which allow to calculate a discrete breather solution to
machine precision. Thus we will not discuss here approximate
methods which are not aimed at a precise calculation (like
e.g. letting some initial localized conditions to evolve and
to radiate away what is not belonging to the solution, or
using from the beginning an approximation like RWA - thus
limiting oneself much above machine precision). At the end of 
each part we will list the advantages and disadvantages of
each method.

The choice of relevant variables separates
these methods into two classes. One class corresponds to choosing
initial conditions of the Hamiltonian system. Defining the
frequency of the DB one numerically integrates the initial conditions
over one DB period, and obtains a map of the phase space
into itself. The fixed points of this map are periodic solutions
with the given period. Methods operating with these variables
can be efficient in programming, because one needs typically two
variables per lattice site. The numerical integration over
the DB period can be performed very fast using standard integration
routines - $10^4$ lattice sites is well in the range of perfomance of
a typical workstation or PC. The disadvantage of this method
is that we create numerical errors already at the stage of computing
the map. However the efficiency of integration routines practically
eliminates these problems.
The second class of methods works with the Fourier coefficients
$A_{kl}$. These methods have the advantage that one is searching
for solutions without leaving the subclass of periodic solutions
(defined by the Fourier series). The disadvantage of this method
is that the number of variables is strictly speaking infinite
(because one needs an infinite number of Fourier coefficients per
lattice site). In practice we will have then to define a threshold
value for the Fourier number $k_t$ such that we neglect all Fourier
coefficients with $k > k_t$. This can be usually done because
any smooth periodic function with finite period will show up
with a decay of its Fourier coefficients for large $k$. So one has
to check after finding a solution whether the cutting value $k_t$ was
justified. This is especially important because the nonlinearity
in the original equations of motion produces infinite-range
interactions in Fourier space! Still we will be left with more than two variables per lattice
site - typically 10-30. Also the explicit form of the equations
of the Fourier coefficients can be complicated, such that we will
have to use integration routines in order to find the Fourier
coefficients of a certain function - which in addition creates
the necessity of integrations and again increases the numerical errors.

The methods to be discussed differ not only with respect to the
chosen variables. Once we have the variables, we can use
Newton's method, the method of steepest descent, or other nonlinear maps.
The first two of them rely on continuing from a limiting model case
where the solution can be approximated to good precision, some other
methods do not. For an extensive discussion and technical details
see Refs. \cite{cp90},\cite{sf95-pre-2},\cite{ma96},\cite{sa96},%
\cite{bms95-prl},\cite{bms95-prb}.

\subsection{Newton method}
\label{s61}

Recall the Newton method for the problem $f(x)=0$
which is solved for some $x=s$. We choose a good
approximation for the true solution $x_{(0)}$ and expand
$f(x)=f(x_{(0)}) + f'(x_{(0)})(x-x_{(0)})$. We thus obtain the iteration
procedure $x_{(n+1)} = x_{(n)} - f(x_{(n)}) / f'(x_{(n)})$. Here $f'(x)$ is
the first derivative of $f$ with respect to $x$. The method converges
if $x_{(0)}$ is sufficiently close to $s$ and if we do not encounter
a zero in $f'(x)$ on our iteration route.

\subsubsection{Phase space variables}
\label{s611}

Consider the numerical integration of the equations of
motion for the lattice over a given period $T$ 
\begin{equation}
u_n(T) = I_n(\{ u\};T)
\label{6-1}
\end{equation}
where the integration procedure $I$ depends parametrically on $T$
and on the initial conditions $\{u\}$. The initial
velocities can be chosen to be zero $\dot{u}(0)=0$. 
A periodic orbit of
period $T$ will then lead to
\begin{equation} 
f_n=0 \;\;,\;\;f_n=u_n - I_n(\{ u\};T) \;\;. \label{6-2}
\end{equation}
This is so because the kinetic energy of the system is
a positive definite quadratic form of the velocities.
Integrating over $T$ and finding all positions to be the same
then implies that the kinetic energy is again zero - and thus
the velocities are the same as the initial ones (one can easily
abandon these restrictions if searching for solutions of systems
without either time-reversal symmetry or some other nontrivial
properties).
Define 
$\vec{u} = (u_1,u_2,...,u_n,...)$,
$\vec{f} = (f_1,f_2,...f_n,...)$ and the matrix $M$ with
elements $M_{lm} = \partial f_l / \partial u_m$. Then the 
Newton method generates the map 
\begin{equation}
\vec{u}_{(n+1)} = \vec{u}_{(n)} - 
M^{-1} \vec{f}(\vec{u}_{(n)})\;\;.\label{6-3}
\end{equation}
The map converges if we are close enough with the initial guess 
$\vec{u}_{(0)}$
and if the Newton matrix $M$ is invertible.
To obtain a good initial guess it is useful to consider some limiting
case of the system - e.g. the antiintegrability limit or something else.
Once a solution is found, we can change the parameters of the 
Hamiltonian and/or the period of the breather by small amounts and
(thus assuring that we have a good guess of a new solution, namely the
old one) repeat the iteration. This yields a systematic method
of tracing a solution.

Typically one needs about 10-20 iterations for one solution.
One disadvantage of this method appears 
whenever the Newton matrix is not (or nearly not) invertible. Still
the breather solution can continue to exist! 
A second problem can be computation time.
For each step in the iteration we have to perform $N$ integrations
of the equations of motion of the system - over the given period $T$.
This will give us the Newton matrix.
Still that is not the main source of computing time consumption
- its computing time grows as $N$. 
The calculation of the inverse of the Newton matrix can become
the main source of time problems, since the computing time grows
as $N^2$ at least.

\subsubsection{Fourier coefficients}
\label{s612}

The choice of the Fourier coefficients as the variables of the
Newton method does not change much. We have to increase the number
of variables per site - typically about 10-30 per site. Still
the $A_{kl}$ can be a good choice - especially if we can explicitly
write down the equations (i.e. that we can explicitly encode them
in a program). That implies that the nonlinear terms of the equations
of motion are finite order polynomials in the displacements.
This allows for an explicit dependence of the elements of the Newton
matrix on the $A_{kl}$. We will need some effort to write the program
code, but the final iteration will proceed very fast! 

If the nonlinear terms of the equations of motion are not finite
order polynomials, then one has to numerically perform
a Fourier integration over the nonlinear terms, which are
functions of the $A_{kl}$ \cite{sf95-pre-2}. This implies that we have
to numerically obtain the Newton matrix.

The advantage of this method occurs whenever the equations
of motion are low order polynomials in the phase space variables.
Still it requires 
at every iteration step 
the computation of the inverse of the Newton matrix,
which can cause computing time problems for large systems. The
method will again fail whenever the Newton matrix is not invertible.

\subsection{Steepest descent}
\label{s62}

Recall the steepest descent method for solving $f(x)=0$.
We define $g(x)=f^2(x) \geq 0$. If $x=s$ is a root which solves
$f(x)=0$ then $g(x)$ has a minimum at $x=s$ with $g(s)=0$.
To find the minimum we calculate the gradient (here simply the derivative)
of $g(x)$ and perform small steps in $x$ in the direction of
the negative gradient. The numerical implementation needs a variable
step size, which has to change whenever a subsequent step does not
lower the value of the function $g(x)$. Eventually we will end
in a minimum of $g(x)$ (this is signaled by the step size becoming
smaller than say the precision of the machine). If our initial
starting point $x$ was sufficiently close to $s$ the function $g(x)$
will be zero at the determined minimum - a signature that we
have solved the problem $f(x)=0$. 

\subsubsection{Phase space variables}
\label{s621}

Using the equations (\ref{6-1}),(\ref{6-2}) and the definitions from above
we can generalize the function 
\begin{equation}
g(\vec{u}) = \vec{f} \cdot \vec{f} \;\;. \label{6-4}
\end{equation}
The gradient is given by
\begin{equation}
\nabla g = 2 \vec{f} \cdot (\nabla \cdot \vec{f} ) \label{6-5}
\end{equation}
with the elements of the gradient given by
\begin{equation}
(\nabla g)_l = 2 \vec{f} \cdot \frac{\partial \vec{f} }{\partial u_l}
\;\;\;. \label{6-6}
\end{equation}
To compute the gradient we have to perform $N$ integrations of
the equations of motion (as with the Newton method). Once we have
a gradient we use the negative of its direction in the variable's space
to go as far as possible - under the condition that $g$ is minimized.
That defines the next 'step' at which we have to compute the gradient
again. 
To obtain a good initial guess it is again useful to consider some limiting
case of the system - e.g. the antiintegrability limit or something else.
Once a solution is found, we can change the parameters of the
Hamiltonian and/or the period of the breather by small amounts and
thus assuring that we have a good guess of a new solution, namely the
old one) repeat the iteration. This yields a systematic method
of tracing a solution.

The implementation of the method itself is slightly more complicated
(but still a basic task for any experienced programmer)
than the Newton routine - which is a simple map. However this method
does not compute and inverts matrices! All we have to do is to compute
gradients (vectors). This method was used for the calculation of
discrete breathers in three-dimensional systems \cite{fkm96}.
The structure of the program is in fact trivial if one splits the
formal search for the minimum of $g$ from the system-specific calculation
of the gradient. Another advantage of this method is that it does 
not break down when the Newton matrix ceases to be invertible. Consequently
this method traces discrete breathers through dynamical instabilities
without any problem. If at an instability (bifurcation) other breather
families appear, the method will simply contract onto one of them.
Combined with the Newton method this can be efficient in studying 
bifurcations, because the Newton method is sensitive to bifurcations, 
where $M$ is no longer invertible. Since the Newton matrix
yields the relevant eigenvectors of perturbations, one can then use
the structure of the eigenvectors and with the help of the steepest descent
follow any of the new periodic orbit branches. 
The disadvantage of the steepest descent is that there is no guarantee
that the minimum of $g$ which is always found corresponds to a zero in $g$
or better in $\vec{f}$. However we did not encounter any practical
problems of that kind. 

\subsubsection{Fourier coefficients}
\label{s622}

Again nothing will change if we use Fourier coefficients instead
of phase space variables. As for the Newton method this change
is only reasonable if the nonlinear terms of the equations of motion
are finite order polynomials in the phase space variables. 
Otherwise the advantages/disadvantages are not affected.

\subsection{Other methods}
\label{s63}

Campbell and Peyrard \cite{cp90} have also used a Newton method
to generate breathers. However the choice of the variables there
was quite different from the above. Namely the discretization
of time which is needed in any numerical intergration, has been
used in \cite{cp90} in order to define a coupled map lattice
in two dimensions - one discrete spatial and one discrete time
dimensions. Defining appropriate boundary conditions one can
indeed search for discrete breather solutions. The disadvantage
of this method is that in order to keep computational time 
within reasonable limits one has to choose quite large  time
steps - in \cite{cp90} 1/64 of the breather's period. Compare
this with typical time steps of 1/1000 of the breather's period
when numerically integrating the equations of motion. 

There are specific cases when one can use symmetries of a system
to design special methods. Take the case of homogeneous potentials
as an example. The equations for the time-independent amplitudes
are given in (\ref{5-7}). These equations can be generated by
the extrema of the function ($m \geq 2$) 
\begin{equation}
S= \sum_l\left[ \frac{1}{2} A_l^2 - \frac{v_{2m}}{2m}A_l^{2m}
- (A_l - A_{l-1})^{2m} \right] \;\;. \label{6-7}
\end{equation}
Because $S$ is a positive quadratic form of the variables $A_l$ 
close to zero and a negative $(2m)$ form far from zero the
only nontrivial extrema will be saddles - minima on a rim
which surrounds the valley with the local minimum $A_l=0$.
A very efficient method of computing these saddles is to
choose a certain direction (which essentially encodes the
breather solution one is searching for) and to find the maximum
of $S$ when departing from the origin in this direction. Once we
have done that, we are on the rim! Now we have to use steepest descent
to minimise $S$ on the rim - by adjusting after each step the point
to the rim again (in leading order we will always go down the rim, but
there will be always a small deviation from the rim itself).
This method will yield breather solutions extremely fast - also
for higher lattice dimensions, because the method is local (if the
initial direction was local of course)\footnote{
See also \cite{ff93-ap}.}.

Yet another method was reported in \cite{sf95-pre-2} when working with
Fourier coefficients $A_{kl}$. It consists of writing down each
of the equations $f_n=0$ (see above) in the way of a map.
Since there are different ways of doing that, one can choose
between maps which are unstable at small amplitudes (repelling)
or stable (attracting). Depending on the solution to be obtained
the global map, which is a set of all the local maps, is constructed
and it contains the code of the discrete breather to be searched for 
by encoding the map choice at each lattice site. Then one chooses
a small initial condition (e.g. $A_{11}=0.1$ and all other coefficients
at zero) and iterates. There is no guarantee that the iteration converges
to something meaningful. Yet if it does, it happens very quickly.
In \cite{sf95-pre-2},\cite{bp95} 
different solutions were obtained and analyzed. 
The major flaw of this method is its unpredictable convergence
criteria.

Neuper et al use a quasicontinuum method combined with
an iterative procedure to obtain DB solutions \cite{nmf94}.
Laedke et al use an iteration of a Fredholm type equation to
both prove and numerically obtain discrete breathers in the
DNLS \cite{lks96}.

\section{Structural stability}
\label{s7}

In the introduction we discussed the breather solution of
the sG PDE. Especially we mentioned that it is structurally
unstable. What about the structural stability of discrete breathers?

Structural stability of a solution implies that if the equations
generating the solution are slightly perturbed (changed) then
the new equations posess a solution close to the original one -
in the limit of the perturbations going to zero both solutions
should coincide. This is a tricky thing - one has to be careful
in defining the space of allowed perturbations (note that e.g. if
we would add a small friction, then all dynamical solutions would
disappear!).

That discrete breathers are structurally stable, follows essentially
from the existence proof of MacKay and Aubry \cite{ma94}. 
They are structurally stable
with respect to changes of the parameters of the Hamiltonian, unless
a resonance with the linear spectrum is encountered. Another way
of looking at the problem is to consider (\ref{4-3}) and to analyze
the properties of the map of the Fourier coefficients. Since the
fixed point $A=0$ is hyperbolic if the frequency of the supposed
existing discrete breather fullfills the nonresonance condition
(all its multiples are outside the linear spectrum) then the 
dimension of the stable and unstable manifolds is exactly one half of
the phase space dimension of the map. There is a problem in that
the phase space dimension is in fact infinite. Still we can 
find the following. If the DB solution exists, the two invariant
manifolds have intersections in common points. For any even space
dimension it then follows that such an intersection is either structurally
stable against small perturbations of the manifolds, or there
exists at least one perturbation such that the intersection becomes
structurally stable afterwards \cite{sf94}. This argument does not
depend on the space dimension of the map, 
so we can consider the limit of infinite
dimension afterwards.  
But changes in the Hamiltonian parameters will indeed only perturb
the invariant manifolds.Then we find that discrete breathers are 
structurally stable provided the nonresonance condition holds \cite{sf94}.

\section{Dynamical stability}
\label{s8}

In this section we will discuss the stability of discrete
breather solutions with respect to small perturbations
in some initial conditions. In contrast to the structural
stability we are not changing the Hamiltonian, but slighlty
departing in phase space from the DB periodic orbit.
We can not in general expect then to stay on a periodic orbit
again. The question is then, which properties does the new
trajectory have? In particular we will be interested in
the question: what are the topological properties of the 
new trajectory relative to
the original DB trajectory? These questions have not been
extensively studied. There exist analytical and numerical results
of 
Campbell and Peyrard \cite{cp90},
Marin and Aubry \cite{ma96}, analytical results by Bambusi \cite{db96}
and numerical results by Flach et al
\cite{fwo94},\cite{fkw94}. In the first part of this section we
will discuss the linear stability analysis, which considers
the Floquet eigenvalue problem assotiated with the linearized
phases space flow around the DB orbit. 
In this part we will follow closely a paper by Aubry \cite{sa96}.
In the second part we
will discuss results on the DB stability in the original system
(i.e. without linearizing the phase space flow).

\subsection{Linear stability analysis}
\label{s81}

Assume a discrete breather solution to be given (either analytically
or numerically): $u_l(t)=u_l(t+T_b)$.Add a small perturbation $\delta_l$ to
it (i.e. change say the initial conditions at $t=0$ slightly)
and obtain the equations of motion for $\delta_l$ using its smallness
(i.e. take into account only terms linear in $\delta_l$):
\begin{equation}
\ddot{\delta}_l = -\frac{\partial ^2 H}{\partial u_l \partial u_m}
\delta_m \;\;. \label{8-1}
\end{equation}
This equation describes the linearized phase space flow around the
discrete breather solution. 
Note that it has to contain the subset of perturbations which simply
continue the discrete breather solution!
Because of the nonlinearity of the problem
the right-hand side of (\ref{8-1}) contains parameters which depend on
time. Moreover these parameters are periodic in time with period $T_b$
and are spatially localized, so that far away from the breather
(\ref{8-1}) will essentially correspond to the equations of motion
of the original system when linearized around the groundstate. This type
of equations is a generalization of Hill's equations. It can be associated
with keywords like Bloch band theory, 
Floquet theory,
parametric resonance and others,
indicating its general applicability to many situations of interest.

The Bloch theorem states that all bounded solutions of (\ref{8-1})
can be represented in the form
\begin{equation}
\delta_l(t)={\rm e}^{i\omega_{\nu} t} \Delta_l
^{(\nu)}(t)\;\;, \label{8-2}
\end{equation}
where $\Delta_l$ is periodic with the period of the  parameters in
(\ref{8-1}) i.e.
\begin{equation}
\Delta_l(t)= \Delta_l(t+T_b)\;\;. \label{8-3}
\end{equation}
The task is then to find the eigenfunctions $\Delta(t)=\{ \Delta_l(t) \}$
which are periodic in time and fullfill (\ref{8-1}) after inserting
(\ref{8-2}). This can be simplified by recognizing that
the phase space flow defined by (\ref{8-1}) carries an initial 
condition $\delta_l$ into ${\rm e}^{i\theta_{\nu}} \delta_l$ if
the $\nu$-th eigenfunction has been excited. Here the Floquet
multiplier ${\rm e}^{i\theta_{\nu}} = {\rm e}^{i\omega_{\nu} T_b}$. 
Then we can consider the Floquet map $F(T_b)$ which maps the phase space
$\{ \delta_l , \dot{\delta}_l \}$ onto itself:
\begin{equation}
\{ \delta_l(t+T_b) , \dot{\delta}_l (t+T_b) \} =
F(T_b) \{ \delta_l (t) , \dot{\delta}_l (t) \} \;\;. \label{8-4}
\end{equation}
This map is symplectic which implies that for any
two initial conditions
the symplectic product
\[
\sum_l \left[ \delta^{(1)}_l(t) \dot{\delta}^{(2)}_l(t)-
\delta^{(2)}_l(t) \dot{\delta}^{(1)}_l(t) \right]
\]
is constant in time \cite{sa96}. The Floquet map is then a matrix
of the rank of the phase space of the problem. The Floquet multipliers
are the eigenvalues of this matrix. Those of them which are 
located on the unit circle
($\theta_{\nu}$ is real) correspond to bounded solutions of the
original problem (\ref{8-1}). In general if $\lambda$ is an eigenvalue
of $F(T_b)$ so are $1/\lambda$, $\lambda^*$ and $1/\lambda^*$.

The Floquet matrix can be computed numerically. It is in fact
nothing but the Newton matrix (which has to operate however on the
whole phase space, and not as discussed above on a subset $\{
\dot{u}_l=0\}$). One has to perform $2N$ integrations of the equations
of motion (\ref{8-1}), where the input is the time-periodic breather
solution itself (it defines the time-dependent parameters). One can use
all available symmetries (irreducible representations) like some
spatial symmetries assotiated with the DB solution in order to
reduce the numerical effort of diagonalization.

Numerically it is quite hard to find that a Floquet multiplier
is exactly on the unit circle. Aubry has suggested that the extension
of the diagonalization procedure may help both with better
interpreting the eigenvalue properties as well as with
the analysis of bifurcations - cases when the original DB periodic orbit
turns unstable. For that one adds to say the right hand side
of (\ref{8-1}) a linear term $E \delta_l$ \cite{sa96}. Now we can repeat the
whole construction and obtain a Floquet matrix parametrically depending
on $E$. Solving for each value of $E$ (and of course working in
the irreducible representations, i.e. taking into account all
symmetries of the matrix) we find a set of bands $E_{\nu} (\theta)$.
These bands are of course $2\pi$-periodic in $\theta$.
The original Floquet multipliers are obtained by putting $E=0$.
First of all this approach allows for a safer elimination of errors
in determining the Floquet multipliers located on the unit circle,
because we have now a 'correlation' given by $E(\theta)$ (at the
expense of more computations). 
Another important feature of these bands is that they immediately
predict the stability/instability of the periodic orbit under investigation.
Indeed, the periodic orbit is stable only if all Floquet multipliers
are located on the unit circle. Suppose we start with such a case
and vary the parameters of the periodic orbit (by e.g. changing
the frequency of the DB). Then multipliers can 'collide' and
leave the unit circle - either in pairs if the collision happened at
$\pm1$ or in two conjugated pairs if the collision happened elsewhere
on the unit circle. Once such a collision happens, the periodic orbit
becomes linearly unstable - because the corresponding perturbations
$\delta_l$ will exponentially grow in time without limitation.
Usually (but not always) new families of periodic orbits (usually stable)
bifurcate from the unstable periodic orbit. 

However often Floquet multipliers can cross each other on the unit
circle without any collision (i.e. they stay after crossing). 
Clearly from the above a loss of some Floquet multipliers
happens whenever a band $E(\theta)$ happens to loose two zeros
$E=0$ at $\theta=0$ or $\theta=\pi$ (or four zeroes - two at 
$\theta_+ \neq 0$ and two at $\theta_- = -\theta_+$). Thus Aubry has
refined the Krein signature which is a necessary condition for
collision to a sufficient condition - namely the behaviour of
zeroes of a band $E(\theta)$ \cite{sa96}. Only Floquet multipliers yielding
zeroes of one and the same band can collide and leave the unit circle.
From this it also follows, that whenever a multiplier
comes close to $\pm 1$ on the unit cicle, a collision is 
inavoidable.

There is one consequence of this analysis. Namely it follows
immediately that  one-site breathers continued from the noninteracting
case (antiintegrability limit) are linearly stable - at least up
to some finite interaction strength \cite{sa96}.

Given the above it is a matter of numerical calculation
to analyze the linear stability of discrete breathers. 
Marin and Aubry \cite{ma96} show discrete breathers
can be linearly stable or unstable and all of the above discussed
collisions can be observed.

Because we usually do not know the analytical form of the 
DB solution and because we study a lattice, it is hard to
predict analytically the stability behaviour of DBs. Indeed
the Floquet matrix diagonalization yields both extended and
localized eigenvectors. The existence of localized eigenvectors
has been long suggested in the literature (stability/instability
of even/odd parity modes \cite{sps92},\cite{cku93}, 
XC/XN modes \cite{dpw92},\cite{kc93}).
In fact in many cases the spatial symmetries of DB solutions have been
correlated with the dynamical linear stability. This aspect deserves
more clarification. Especially it seems plausible that the 
XC discrete breather - which is just an one-site breather
in the antiintegrable limit - has to be stable at least for weak
interactions. This result has been also found in the numerical
analysis of Campbell and Peyrard \cite{cp90}. There for some parameter
ranges the XN solution (a two-site breather) was found to
be linearly unstable, whereas the XC solution was stable. We will
discuss these aspects in section~\ref{s10} when discussing band edge
plane wave bifurcations.

Suppose we solve the linear stability problem for an infinite
system. All extended Floquet eigenvectors will then have 
the form of linear plane waves far from the breather center. Consequently
we know that all extended eigenvectors have $\theta=\omega_q T_b=
\omega_q/\omega_b 2\pi$. Because the DB is exponentially localized,
it can cause only a finite number of Floquet eigenvectors to be 
localized. Consequently there will be an infinite number of extended
eigenvectors, whose eigenvalues will densely fill the linear spectrum.
We can immediately derive some dangerous DB frequencies causing
instabilities (this result was derived by Flach and Willis 
in another way \cite{fw93}):
\begin{equation}
\omega_q/\omega_b = k/2\;\;,\;\; k=1,2,3,...\;\;\;. \label{8-5}
\end{equation}
Even $k$ correspond to collisions of Floquet multipliers at
+1, and odd $k$ at -1.
The cases $k$ even correspond to the resonance conditions 
when discrete breathers stop existing (section~\ref{s4}). 
In fact we can expect
that for $k=2$ the breather disappears, but for larger even values
of $k$ the breather becomes a nanopteron \cite{jpb90} 
- i.e. a localized object
with nondecaying tails. Consequently nanopterons are dynamically
unstable (if they exist at all)! Indeed Marin and Aubry report
about these nanopterons in numerical calculations \cite{ma96}.
The cases $k$ odd correspond to period doubling bifurcations.
The breather can still exist but will be dynamically unstable. The
relevant perturbations however will be extended, so that the new
periodic orbits bifurcating from the old one must again have
the spatial shape of nanopterons!

\subsection{Going beyond linearization}
\label{s82}

What happens to the solutions $\delta_l$ if we do not impose
the linearization (\ref{8-1}) but instead solve the original
equations of motion? One answer has been recently found by
Bambusi \cite{db96}. He showed that periodic breather orbits
are exponentially stable in the antiintegrable limit. That means
that after a suitable chosen metric has been introduced, the
perturbation of a periodic DB orbit stays close to the
orbit up to exponentially large times. Let us be more precise:
if the distance $d$ of the perturbed orbit from the periodic one
is smaller than a value proportional to the square root
of the interaction strength, then the distance stays close
to the periodic orbit for times proportional to the exponential
of some inverse power of the interaction (times a constant).
The frequency of the 
DB in the antiintegrable limit (zero interaction) has 
to be not only in nonresonance with the degenerate linear
spectrum, but it has to be diophantine with the linear spectrum.
Details can be found in \cite{db96}.  
Further Bambusi shows that a 'quasiperiodic' breather object although
not existing as a precise solution can be a good approximation
to a real trajectory for times again exponentially large (in
the same meaning as above). The small errors eventually acumulating
with time are referred to as radiation.

This radiation has been studied numerically. Indeed the
numerical experiments for long-lived localized excitations
discussed in section~\ref{s3} in fact study perturbed discrete
breather solutions. What can we learn from those results?
First, as long as the object is not a periodic DB, it radiates
energy in the form of small amplitude plane waves - exactly
as predicted in section~\ref{s4} and obtained by Bambusi. This decay
can be very weak - in fact so weak that it is hard to measure!
Estimates of the decay come up with a loss of 
'breather' energy per 'period' of oscillation (note that we
are not dealing with a strict time-periodic solution) of the order
of  0.01\% - 0.001\% in units of the 'breather' energy \cite{fwo94}!
From this circumstance it follows that even though discrete
breathers form a subset of measure zero in the phase space of
nonlinear lattice, their phase space support can be finite and
in fact the lifetimes of objects associated with DB solutions
can be tremendously large, making the breather concept highly 
reasonable.

Another interesting question is - what happens to a perturbed
breather if we wait long enough? Will the object radiate its
whole energy away and vanish or not?
In Fig.9 
%\begin{figure}
%\caption{}
%\end{figure}
we show the dependence of the object's energy on time
for different initial conditions (for details see \cite{fwo94}).
The internal timescale is of the order of 10. 
As long as the radiation is weak, we can view the object as
a quasiperiodic breather whose parameters (energy, frequencies)
are slowly
time-dependent. Then we can characterize the object by those 
parameters which are slowly varying functions. Several results
have been observed. For some initial perturbations
the object radiates away energy and consequently slowly approaches
the family of periodic breather orbits. Clearly the radiation
becomes weaker and weaker as the periodic orbit family is approached.
This behavior reminds us of an attractor in a dissipative system.
In fact because the radiation becomes exponentially weak, on finite
computing time scales we even observed a seemingly forever living
quasiperiodic breather. To see that we plot in Fig.10
%\begin{figure}
%\caption{}
%\end{figure}
a Poincare map of a perturbed DB of a two-dimensional lattice,
where consecutive points are connected with straight lines. 
The attractor-like behaviour is seen, together with a seemingly
remaining nonzero distance from the fixed point (periodic
orbit) in the middle of the figure (for details see \cite{fkw94}).

For some trajectories we find that internal resonances between
the internal frequencies describing the quasiperiodic evolution
on short time scales cause an 'explosion' of the breather-like object
(see Fig.9 and \cite{fwo94}), which in a local analysis was
due to a chaotic dynamics inside the breather - remarkably that
does not cause the whole structure to disappear! Moreover, after
some radiation which is still weak, of the order of 1\% of the
breather energy per breather period, the radiation suddenly stopped
at energies 0.35. A local phase space analysis yielded that 
the strong chaotic trajectories disappear around these energies.
Consequently we find that perturbed discrete breathers can slowly
radiate energy away and thus are either attracted to the periodic orbit
family (consequently stabilized) or repelled. In the second case
internal resonances can trigger chaotic local dynamics which causes
an increase of the radiation rate by two orders of magnitude. Still
even those cases eventually end up with a breather-like object 
which is again weakly decaying, most probably into an exact time-periodic
breather.

This indicates that the linearized stability analysis can be of help but
does not tell important things - because it is likely that all
the evolutions of breathers discussed here were initiated by perturbing
linearly stable discrete breathers!

\section{Movability}
\label{s9}

So far we have been discussing discrete breather solutions which
are time periodic. By definition these solutions are stationary,
i.e. they do not 'move' through the lattice. We have been also
discussing small perturbations of DB solutions, which were
supposed to stay small (at least for several DB periods). Again
such trajectories do not correspond to a 'moving' entity.
However we know that e.g. for the sG PDE the breather solution
can be Lorentz-boosted, due to the invariance of the sG PDE
under Lorentz transformations. Such a boosted breather would
actually move, i.e. its center is propagating with
constant velocity. That is certainly not a time-periodic
solution anymore. 
If we consider a lattice we loose the continuous symmetry in space
and replace it with a discrete symmetry.
What can be said about the existence of
objects similar to boosted sG breathers in lattices?

\subsection{Numerical findings}
\label{s91}

A good way to start with are numerical simulations of lattices.
Takeno et al  (see e.g. \cite{ht92-jpsj-2},\cite{ht92-jpsj-1},%
\cite{thor91},\cite{thom91},\cite{thor90}) indeed have reported
on moving breathers in one-dimensional FPU lattices. Apparently
the breathers are very localized in space and move over
long distances without significantly radiating energy or changing 
their shape
(which would be a sign that the object does not correspond to a
moving breather). Note that the internal frequencies are above the
linear spectrum, and the observed velocities are below the maximum
group velocity of linear plane waves. A careful numerical study relating
dynamically unstable DBs with moving breather-like objects has been
performed by Sandusky, Page and Schmidt \cite{sps92}. Sievers and
Page summarized numerical results on moving breathers in \cite{sp95}.

A different  picture emerges for one-dimensional 
Klein-Gordon lattices. There
moving breather-like objects are typically only found if the
amplitude of the object is small enough - which in turn implies
that these objects are only weakly localized in space. If the amplitude
becomes larger (and the objects stronger localized) then typically
one observes radiation and a subsequent stopping of the object. Once it stops
the object is similar to a   perturbed DB (see Bang and Peyrard \cite{bp95}). 

Yet another result comes if we consider two-dimensional lattices.
While nothing drastically changes for Klein-Gordon lattices
\cite{trp95},
the two-dimensional FPU lattice resists in producing moving
breather-like objects which are strongly localized - as opposed
to the corresponding one-dimensional case \cite{bkp90-prb}.

So the emerging picture is quite inhomogeneous. Moving
breather-like objects are detected  whenever the amplitudes
are fairly small  and the objects are weakly localized, and
for some systems (1d FPU lattices) even for large amplitudes
(objects are strongly localized). Moving objects are not
detected whenever the amplitudes are large and the localization
is strong (for Klein-Gordon lattices), and for lattice
dimensions larger than one (for FPU lattices). Note that
even when we say that moving breathers are detected that does not
imply that there exist lossless exact solutions - it just means
that during the time of the numerical experiment the object
did not radiate an appreciable amount of energy away ( typically
the threshold is 1\%-5\%). 

One emerging pattern seems to be that small-amplitude breathers
can move - at least over much larger times (and distances) than
large-amplitude breathers. There is a natural explanation of this
observation. If the breather-like object is weakly localized, its
envelope is slowly varying in space. As discussed in section~\ref{s2},
Kosevich and Kovalev used this circumstance to derive in some
lowest order a partial differential equation! Even though this
is not exact, it certainly can describe a real trajectory over
an increasingly larger time window with decreasing amplitude of
the object. But since the corresponding PDE possesses a continuous
symmetry (in space) its stationary solutions can be boosted, such
that they move! So moving breather-like objects appear to be
quite reasonable for weakly localized small-amplitude objects.
Most probably they will not move forever because we are actually
solving a lattice problem (unless there is some other symmetry
involved). But in the limit of small amplitudes the time scale
over which the objects appreciably change (decay) might diverge,
thus escaping from any finite-time window in a numerical experiment.

Of course there is still the possibility that exact moving
breather solutions exist! 
And they do exist - for the Ablowitz-Ladik lattice, which is integrable
\cite{al76}
\footnote{Fischer \cite{ff93-pla} considered a three-particle
FPU system which is integrable. Solutions similar to stationary
discrete breathers and moving breathers do exist there. However nothing
similar is known for the case of an infinite lattice.}.
However neither do there exist existence proofs for other lattices,
nor are the numerical results conclusive enough to support the
existence of moving breather in nonintegrable lattices. 
The situation is clearly different from the circumstances
of finding stationary discrete breathers, where numerical investigations
strongly suggested that time-periodic DBs exist independent
on the integrability properties of the lattice.

\subsection{What can we learn from moving lattice kinks?}
\label{s92}

Before asking whether moving breather solutions exist 
it is useful to see what is known about moving kinks in lattices.
Recall that a kink solution is defined by some nontrivial
boundary condition $(u_{l\rightarrow + \infty}
- u_{l\rightarrow - \infty}) = \eta \neq 0$. 
If a Klein-Gordon PDE supports static kink solutions, it again
supports boosted (moving) kink solutions due to the continuum symmetry.
Things are different for lattices. First of all there exist
exact moving kinks in the Toda lattice - which is however integrable.
McLeod \cite{bml95} has given an analytical proof of continuation
of moving Toda kinks into other one-dimensional FPU systems. 
Duncan et al \cite{defw93} have verified this result numerically
up to machine precision. At the same time MacLoyd was unable to continue
moving kinks into Klein-Gordon one-dimensional lattices, 
and the numerical results of
Eilbeck  et al suggest that moving kinks do not exist in a sG lattice -
instead an object similar to the nanopteron was found - namely a kink-like
structure which becomes a plane wave in the tails.

For Klein-Gordon chains where exact moving kinks
seem not to exist, kinks carry a topological charge \cite{degm82}.
This implies esentially that in order to remove  one
kink, one needs to overcome an infinitely high energy barrier
(for an infinite system). So even though a moving kink does not
exist, the kink can not disappear (note that there is no topological
charge associated with moving FPU kinks). Thus a Klein-Gordon kink
when boosted can only radiate some energy away and finally stop to
become a static kink solution! In order to account for these effects
a collective coordinate approach was developed 
\cite{wes86},\cite{sweb86},\cite{bsw88},\cite{webs89},%
\cite{bw89},\cite{wb90},\cite{dbcw92},\cite{bw92}. Within this
approach one performs a canonical transformation to new coordinates
some of whom are collective (nonlocal in the old coordinates).
These collective coordinates describe the kink.  As a result
one can obtain equations of motion of the kink, which are of course
coupled to the infinite number of the remaining degrees of freedom.
Using some constraints imposed on the kink's position one can
evaluate the energy of a static kink at different positions. This
energy will be periodic with the period of the lattice, and in general
be not constant. Consequently it is labeled Peierls-Nabarro potential,
because in the limit of a nearly static kink it appears as a potential
in the nearly Newtonian equations of motion for the kink. The potential
will have at least one minimum and maximum between two lattice sites.
These extrema define the true static kink solutions on the lattice,
and the difference of the heights of the extrema is called Peierls-Nabarro
barrier $E_{PN}$. 
The kink corresponding to the maximum of the Peierls-Nabarro 
potential is unstable with respect
to small perturbations (it corresponds to a saddle point in the
potential energy relief of the original Klein-Gordon system), 
and the kink corresponding to the minimum of Peierls-Nabarro potential
is stable with respect to small perturbations (it corresponds
to a local minimum in the potential energy relief 
of the original Klein-Gordon system). 
It is argued that exactly the energy $E_{PN}$ is needed by the
stable kink solution in order to overcome the lattice pinning
and to move. In turn if a moving kink is losing energy by radiating
plane waves, it should be trapped by the lattice at some moment
in time.

This concept has been proven to be very successful. Qualitatively
its predictions indeed take place. Quantitatively we have to take
into account the kink interaction with the plane waves, in order
to obtain radiation. This can be still done up to some degree 
analytically, and we think it is possible to say that the
Peierls-Nabarro concept is an effective way to reduce the 
problem of handling the infinite-dimensional phase space flow
to a finite-dimensional one. Note that the concept works 
however only in systems which allow for static kinks - nothing similar
has been done e.g. for FPU chains lacking static kink solutions but
allowing for exact moving kinks!

The success of this approach is based on the fact that
we can reduce the problem of the Peierls-Nabarro barrier
to the energy difference of certain extrema of a potential
function. Indeed, if we consider a local minimum of such
a function, then we need to find the lowest-lying saddle in order
to escape. No matter what we do, if the additional energy supply to
the local minimum state is lower than the difference to the lowest lying
saddle, we can not escape! So the $E_{PN}$-value becomes a meaningful
depinning energy which can be used e.g. in a statistical analysis.

\subsection{The movability separatrix}
\label{s93}

It is tempting to use the approach for lattice kinks as a
description of moving breathers. The ultimative goal would
be the calculation of a depinning energy. Indeed some numerical
indications for discrete breathers are quite reasonable in that
context - there exist seemingly always stable and unstable 
stationary DB solutions, and moving breather-like objects
radiate energy and eventually stop (are pinned by the lattice)
etc \cite{dpw93},\cite{bp95},\cite{bp96}. 

Such a task has been performed 
in \cite{ckks93},\cite{kc93}. Discrete breather solutions were parametrized
(internal frequency and position on the lattice). A projection
of the phase space flow onto a low-dimensional subspace 
yielded indeed energy values which have been coined Peierls-Nabarro
energies, in analogy to moving kinks. In the following we will
have to extend
the notion of a Peierls-Nabarro energy to a movability separatrix,
which does not allow the definition of a depinning energy in general.
There is no depinning energy for discrete breathers.

First of all we mention the numerical analysis of Bang and Peyrard
\cite{bp95}. They have analysed the movability properties of 
discrete breathers. By considering a multiple-scale expansion
for the equations of motion up to third order (which is only
useful for small amplitudes) moving breather solutions were obtained.
Of course these solutions are approximative, so Bang and Peyrard
used these approximative solutions as initial conditions for
the true equations of motion. Their very extensive analysis showed
that the projections of the phase space flow done in \cite{ckks93}
yielded quantitative discrepances of orders of magnitude.
 
Let us now immediately answer the question: why is it impossible
to introduce a depinning energy? The answer is: simply because
the stationary discrete breather solutions we want to depin are
not isolated  -  i.e. they come in one-parameter families.
Suppose that we have a certain solution of this family
of solutions
which at some time $t=0$ is given by a certain point $P_1$
in the phase space of the system. The energy of this point
is $E(P_1)$.
Now assume we can define a depinning energy which
has to be added to this solution in order to depin it from the lattice
(to let it move). That means that we can add a certain perturbation
of the initial conditions of the discrete breather such that the
new point $P_2$ in phase space will have a higher energy $
E(P_2) > E(P_1)$ and that the trajectory which is defined
by $P_2$ corresponds to some moving breather-like object.
Then we could always choose another stationary discrete breather
solution on the original one-parameter family with a different
energy. Since there exist no upper bounds on energies of
discrete breathers in general, this energy can be larger than
$E(P_2)$. Adding to this second discrete breather a perturbation
which brings it right back to $P_2$ we thus can depin a discrete
breather by adding negative energy! In fact it is clear that we
can find any situation, in other words we are able to add or 
substract energy from a discrete breather and by choosing the right
perturbation depin it! Certainly this makes clear that the concept
of depinning energies does not make the least sense when applied
to discrete breathers. 

In order to understand more we can follow 
a phenomenological approach which has been 
successfully tested in numerical experiments \cite{fw94}
To describe a periodic elliptic (linearly stable) DB we need to introduce one
degree of freedom, which describes the one-parameter
set of DB solutions. We will work in the action-angle phase $(J,\theta)$
space and name this degree of freedom $J_1$. Its corresponding
frequency will be $\omega _1=\dot{\theta _1}=\partial H/\partial J_1$. Here $H$
denotes the full Hamiltonian of the lattice. We assume that
there exists a certain transformation between the original
variables (positions, momenta) and the actions and angles. This
does not imply integrability of the system as well as it does
not imply the inverse.
Since the DB solutions
are regular solutions (at least on moderate time scales)
there is no need in introducing stochasticity
(cf.
\cite{fwo94} for details). We assume that the DB solutions
have a spatial symmetry.
To excite
a moving NLE we have to excite an additional degree of freedom $J_3$.
Exciting $J_3$ we destroy the spatial symmetry of the linearly stable DB.
But since
it is always possible to perturb the DB conserving the symmetry,
we have to include an additional symmetryconserving degree of freedom $J_2$
into the
consideration. Thus we end up with the simplest generic case
of a Hamiltonian problem with three degrees of freedom:
\begin{equation}
H=H(J_1;J_2;J_3) \;\;\;, \;\;\; \omega_i=\dot{\theta_i}=
\frac{\partial H}{\partial J_i}, \;\;i=1,2,3 \;\;. \label{9-1}
\end{equation}
According to our notation $i=3$ labels the symmetrybreaking
degree of freedom. If it is excited strongly enough we expect
to hit a separatrix which separates stationary DBs from
moving ones.
We will name this separatrix movability separatrix.
All three degrees of freedom are assumed to be of local character,
especially they can be well defined in the reduced problem for the DB.

Let us state the general condition for the
movability separatrix we are looking for. Since on the
movability separatrix
a trajectory will for infinite times asymptotically reach a hyperbolic state
(which is a linearly unstable DB manifold
 and its space-symmetric
perturbations) the corresponding frequency of the 3d degree
of freedom
\begin{equation}
\omega_3=\frac{\partial H}{\partial J_3}=f(J_1;J_2;J_3) \label{f}
\end{equation}
has to vanish on the movability separatrix i.e.,
\begin{equation}
f(J_1;J_2;J_3) = 0 \;\;\;, \label{9-2}
\end{equation}
which implies an equation for a surface in the three-dimensional
subspace of the actions $(J_1;J_2;J_3)$. We can always eliminate $J_2$
using the expression for the energy $E=H(J_1;J_2;J_3)$, so that
(\ref{9-2}) yields:
\begin{equation}
f(J_1;J_2;J_3)=\tilde{f}(E;J_1;J_3) = 0 \;\;\;.\label{9-3}
\end{equation}
From (\ref{9-3}) it follows that there exists a critical value
for $J_3$ on the movability separatrix:
\begin{equation}
J_3^s=g(J_1;J_2)=\tilde{g}(E;J_1) \;\;\;. \label{9-4}
\end{equation}
The critical value $J_3^s$ depends both on $E$ and $J_1$.
Only for nongeneric cases when $\tilde{g}$ is constant would
we be able to introduce an extended notion of a depinning energy
by considering a subspace of relevant perturbations. However
as shown in \cite{fw94} this is not possible in general, and thus
we have to accept that stationary linearly stable discrete breathers,
which come in one-parameter families and thus form two-dimensional
manifolds in phase space, are separated by a movability separatrix
from possible moving solutions. 
The concept of a depinning energy therefore has no meaning.

The existence of a movability separatrix implies that if we
consider a trajectory which corresponds to breather-like moving object
(at least over some period of time) then the presence of the separatrix
will show up in a modulation of the time-dependence of  the object
if parametrized in some meaningful way. This modulation is due
to the presence of the frequency $\omega_3$ from above. It follows
that this frequency must be related to the periodicity of the lattice
through the motion of the object.
However we measure the averaged propagation of the object $v$
where distance is measured in units of lattice spacing - it will be
thus related to $\omega_3$ through the relation $\omega_3 = 2 \pi v$.
Sandusky, Page and Schmidt have observed precisely this relation by
independently measuring $v$ and $\omega_3$ in careful numerical
experiments (Figs. 7 and 8 in \cite{sps92}).

\subsection{Breather-breather collisions}
\label{s94}

In a series of numerical experiments Peyrard et al \cite{dpw93},%
\cite{bp96},\cite{fpm94}
have analyzed the collisions between discrete breathers. Typically
moving breather-like objects were excited and the effects of many
collisions studied. Because of our lack of understanding of
moving solutions the approach to describe the observed phenomena
in collisions is heuristic. It has been observed that the breathers
survive collisions, but exchange energy. The energy exchange seems
to depend on some phase differences (note that 
we have to introduce a parametrization of moving breathers, a rather
delicate task). Eventually breathers which collect enough energy
become too localized in space and trap on the lattice, stopping
their energy collection. This problem needs further clarification.

\subsection{Do moving breathers exist?}
\label{s95}

We have still no answer to the question whether moving breathers
exist. By requiring that after some traveling time a breather
is exactly recovered at some other location in the lattice we
would have to solve for differential equations with advanced
and retarded terms. Moving breathers would correspond to certain 
homoclinic orbits of this equation. We have no knowledge about
solutions which correspond to moving discrete breathers.

\section{Plane wave bifurcations and discrete breathers}
\label{s10}

So far we did not pose the question: how are discrete breathers
connected to the plane wave solutions of the linearized equations
of motion? In fact there exists a well-known approach in the literature
which is coined modulational instability. Originally this appoach
was designed for the study of waves in continuous media - some
older papers use also the term Benjamin-Feir instability instead
\cite{gbw74}. Within this approach a plane wave solution
of the linearized equations of motion is continued into the
weakly nonlinear regime. Small plane wave perturbations (of different
wave length) are then added and the stability of the perturbed wave
is analyzed. This approach has been very helpful in connecting the
instability of certain modulated plane waves with spatially
localized solutions, which exist because the nonlinearity of the system
effectively prevents a dispersing of the object.
Thus the stability study of plane waves can become crucial when
predicting the existence of localized solutions without actually
calculating the latter.

Modulational instability has been analyzed for lattices
with respect to discrete breathers in a number of 
publications by Kivshar and Peyrard \cite{kp92},
Flytzanis, Pnevmatikos and Remoissenet \cite{fpr85},
Tsurui \cite{at72} and Sanduski and Page \cite{sp94}.

Here we will follow the approach described in \cite{sf96},
where the stability analysis was performed for finite systems.
Strictly speaking we have to analyze
the stability properties of a periodic orbit (plane wave).
For that we have to linearize the phase space flow around the
periodic orbit - just as we did for the discrete breather.
Then we have to find the eigenmodes and eigenfrequencies.
Thus it is reasonable to perform a 
stability analysis of plane waves for a large but
finite system. This will then lead to results which depend
on the size of the system. The stability dependence on the 
size of the system is crucial when understanding properties
of discrete breathers in different lattice dimensions (cf.  
section~\ref{s11})!
Moreover in this case we can connect the instability of a plane
wave with a bifurcation of new periodic orbits - a connection
hard to make for the infinite system approach. Further we will
be able to prove that the new bifurcating periodic orbits
can not be invariant under discrete translations along the lattice -
as expected for discrete breather solutions!

First we will argue why we have to consider
band edge plane waves when we want to understand the 
connection to discrete breathers.
Secondly we will review
the plane wave stability analysis for finite systems and discuss
its consequences for discrete breathers. Finally we will make 
some predictions.

\subsection{Why the band edge plane wave?}
\label{s10.1}

If discrete breathers are connected to plane waves, we have to
look for breather solutions which are weakly localized on the lattice.
This seems to happen only when the 
frequency of the discrete breather is
 close to a band edge of the linear spectrum.
That in turn implies that the amplitude of the breather center
is small.
Consequently it appears to be logical to expect that discrete breather
periodic orbits which belong to a family of solutions which contains
weakly localized small-amplitude DBs appear through bifurcations of
band edge plane waves, i.e. plane wave periodic orbits which in
the limit of small amplitude correspond to normal modes with frequencies
at the edge of the linear spectrum. Recall that there can be several
band edges of the linear spectrum in principle. This follows from the
fact that regardless from what periodic orbits DBs bifurcate, once
the frequency of the DB is close to the band edge of the linear spectrum,
the DB becomes by definition a weakly modulated band edge plane wave, 
in other words its internal spatial symmetry will be very close to the
spatial symmetry (defined by the corresponding wave vector) of the
band edge plane wave. Consequently we can rule out, that a DB family
might bifurcate from any other plane wave but the band edge plane wave.
But then it follows that the frequency (period) of the DB close to the
bifurcation is very close to the frequency (period) of the band edge plane
wave, i.e. the bifurcation is expected to be a tangent one! By that we
mean that the Floquet multipliers of the linearized stability analysis
of the band edge plane wave have to collide at $+1$ on the unit circle!
With these statements in mind we will then perform the Floquet analysis
of the band edge plane wave. 

Before we proceed we mention another possibility
of DB occurence which has not
yet been studied. The concept of the nanopteron (a breather-like
object with some resonances of multiples of its frequency with
the linear spectrum) allows for cases when DBs occur through analytical
continuation of nanopterons. Consider a linear spectrum which has
two gaps at least. Number the gaps $1,2$ with increasing frequency
values. Then a breather can exist with $\Omega_b$ in gap 1, a second
harmonic in gap 2, and the rest of the harmonics above the linear
spectrum. It is also possible to choose the linear spectrum such,
that either lowering or increasing $\Omega_b$ will lead
to a resonance of $2\Omega_b$ with the linear spectrum first. When lowering
$\Omega_b$ the second harmonic
$2\Omega_b$ touches the upper band edge of the linear spectrum
located between gaps 1 and 2,
and increasing $\Omega_b$ yields a first touch of
$2\Omega_b$ with the lower band edge of the linear spectrum band located
above gap 2. Then in both cases we can expect the DB to become
a nanopteron. Consequently this DB family will never make contact
with a plane wave. Thus for complicated linear spectra the existence
of some families of DB solutions might be not connected to the stability
of band edge plane waves at all!  

\subsection{Tangent bifurcations of band edge plane waves}
\label{s10.2}

Consider system (\ref{3-1a}) for a finite size $N$ of the lattice.
We assume periodic boundary conditions
\[
u_{l}=u_{l+N}\;\;,\;\;\dot{u}_l=\dot{u}_{l+N}\;\;.
\]
Then system (\ref{3-1a}) exhibits permutational symmetry. The
permutational operator $\hat{P}$ is defined by
\begin{equation}
\hat{P}g(u_1,u_2,...,u_N,\dot{u}_1,\dot{u}_2,...,\dot{u}_N) =
g(u_2,u_3,...,u_N,u_1,\dot{u}_2,\dot{u}_3,...,\dot{u}_N,\dot{u}_1)\;\;.
\label{10-1}
\end{equation}
Clearly $\hat{P}^N=\hat{1}$ and $\hat{P}H=H$.

Let us introduce normal coordinates
\begin{equation}
Q_q=\frac{1}{N}\sum_{l=1}^{N} {\rm e}^{iql}u_l \;\;. \label{10-2}
\end{equation}
The wave number $q$ can take any of the values
\[
q=\frac{2\pi}{N}n\;\;,\;\;n=0,1,2,...,(N-1)\;\;.
\]
The inverse transform of (\ref{10-2}) is given by
\begin{equation}
u_l = \sum_q {\rm e}^{-iql}Q_q \;\;. \label{10-3}
\end{equation}
The equations of motion for the normal coordinates $Q_q$ read
\begin{equation}
\ddot{Q}_q = \frac{1}{N}\sum_{l=1}^N {\rm e}^{iql}\ddot{u}_l =
-\frac{1}{N}\sum_{l=1}^N {\rm e}^{iql}\frac{\partial H}{\partial u_l}
\;\;.\label{10-4}
\end{equation}
Using (\ref{3-1a}) and (\ref{3-1b}),(\ref{3-1c}) we obtain the following
lengthy expression
\begin{eqnarray}
\ddot{Q}_q=-\Omega_q^2Q_q - \sum_{\mu=3}^{\infty}v_{\mu}
\sum_{q_1,q_2,...,q_{\mu-2}}\left[ \prod_{\nu=1}^{\mu-2}Q_{q_{\nu}}
\right] Q_{q-\sum_{\nu=1}^{\mu-2}q_{\nu}} -\;\;\;\;\;\;\;\;\;\;
\;\;\;\;\;\;\;\;\;\;\hspace*{2cm} \nonumber \\
-\sum_{l=1}^N {\rm e}^{iql} \sum_{\mu=3}^{\infty}\phi_{\mu}
\left[ \left\{ \sum_{q'}(1-{\rm e}^{iq'}){\rm e}^{-iq'l}Q_{q'}\right\}^{\mu-1}
- \left\{ \sum_{q"}({\rm e}^{-iq"}-1){\rm e}^{-iq"l}Q_{q"}\right\}^{\mu-1}
\right]\;\;.\;\; \label{10-5}
\end{eqnarray}
Here $\Omega_q$ abbrevates the eigenfrequencies of the linearized (in $Q_q$)
equations of motion and is given by the dispersion relation
\begin{equation}
\Omega_q^2 = v_2 +4\phi_2 {\rm sin}^2\left( \frac{q}{2}\right)\;\;.
\label{10-6}
\end{equation}
Let us give the solutions for two periodic orbits of the considered
lattice, which correspond to the band edge plane waves ($q=0$ and
$q_{N/2}=\pi$) in the limit of small energies:
\begin{eqnarray}
{\rm I:} \;\;Q_{q\neq 0}=0\;\;,\;\;\ddot{Q}_{q=0}=-\sum_{\mu=2}^{\infty}
v_{\mu}Q_{q=0}^{\mu-1}\;\;, \label{10-7a} \\
{\rm II:}\;\;Q_{q \neq \pi}=0\;\;,\;\;\ddot{Q}_{q_{N/2}}=
-\sum_{\mu=2,4,6,...}^{\infty}\bar{v}_{\mu}Q_{q_{N/2}}^{\mu-1}\;\;.
\label{10-7b}
\end{eqnarray}
The paramter $\bar{v}_{\mu}$ is given by
\[
\bar{v}_{\mu} = v_{\mu} + 2^{\mu}\phi_{\mu} \;\;\;.
\]
In case II we have to demand
\[
{\rm II:}\;\; v_{2m+1}=0
\]
in order to be able to continue the upper band edge plane wave
to finite energies in the given form.
The terms $\phi_{2m+1}$ can be in general nonzero,
but they simply do not contribute to (\ref{10-7b}) because of the
odd symmetry of the upper band edge plane wave \cite{sp94}.

\subsubsection{Tangent bifurcation of orbit I}
\label{s10.21}

Let us consider a small perturbation $\left\{ \delta_{q}\right\}$
of the periodic orbit I
\[
Q_q \rightarrow Q_q + \delta_{q}\;\;.
\]
Linearizing the equations of motion for the perturbation we obtain
\begin{equation}
\ddot{\delta}_q = -\omega_q^2\delta_q - \sum_{\mu=3}^{\infty}
(\mu-1)v_{\mu}Q_{q=0}^{\mu-2}\delta_q \;\;.\label{10-8}
\end{equation}
For $q=0$ equation (\ref{10-8})
describes the continuation of the periodic orbit itself. All other
perturbations do not couple with each other, so that we can consider
(\ref{10-8}) for each value of $q$ separately. If we increase the
energy of the periodic orbit
\begin{equation}
E_I = \frac{1}{2}\dot{Q}_{q=0}^2 + V(Q_{q=0}) \label{10-9}
\end{equation}
then the first tangent bifurcation will occur if $q=q_1=2\pi /N$
is choosen in equation (\ref{10-8}). 
This is so because the associated linear mode frequency is the closest
to the band edge frequency.
Here the tangent bifurcation implies a collision of two Floquet
multipliers at $+1$. The accounting of the bifurcation
is nothing but a calculation of Arnold tongues in the theory
of parametric resonance \cite{via92}. The explicit calculation
can be done in perturbation theory. Perturbation theory can be applied
only if the resulting bifurcation amplitudes or related variables ($E_I$)
are small. The reader will find all details in \cite{sf96}. 
For large values of $N$ we
have
\[
\Omega_{q_1}^2=v_2 +4\phi_2\frac{\pi^2}{N^2}
\]
and consequently obtain (see \cite{sf96} for a detailed
evaluation)
the following bifurcation energy $E_I^c$:
\begin{eqnarray}
{\rm i):} \;\; \phi_2=0\;\;,\;\;E_I^c\;\; {\rm arbitrary}\;\;, \label{10-10} \\
{\rm ii):}\;\; E_I^c = \frac{1}{N^2}\frac{12\pi^2 v_2^2 \phi_2}
{10v_3^2 - 9v_2v_4}\;\;. \label{10-11}
\end{eqnarray}
Solution ii) is correct for large system sizes, because in the limit
of large $N$ the critical amplitude of the band edge plane wave 
is inverse proportional to $N$, and thus the application of perturbation
theory is justified.
Since we have to require positive values for $E_I^c$ and $\phi_2$
tangent bifurcation can take place only if
\begin{equation}
\frac{v_4}{v_2} \leq \frac{10}{9}\frac{v_3^2}{v_2^2} \;\;\;. \label{10-12}
\end{equation}
Condition (\ref{10-12}) is equivalent to the condition, that the frequency
of the lower band edge plane wave decreases with increasing energy.
In other words, the sufficient condition for a tangent bifurcation
of the lower band edge plane wave is the repelling of its frequency
from the linear spectrum (\ref{10-6}) with increasing energy.

The energy $E_I$ is on the scale of total energy per particle (cf.
(\ref{3-1a}),(\ref{10-2}),(\ref{10-9})). Consequently the
amplitude threshold of
the tangent bifurcation decreases as $N^{-1}$ for
the individual $u_l$ amplitudes. 
In the limit $N \rightarrow \infty$ the threshold
goes to zero.

\subsubsection{Tangent bifurcation for orbit II}
\label{s10.22}

We again derive the linearized equations for the perturbation $\delta_q$:
\begin{eqnarray}
\ddot{\delta}_{q_{a,b}}=-\omega_{q_{a,b}}^2 \delta_{q_{a,b}}
\sum_{\mu=4,6,...}^{\infty}(\mu-1)\bar{\bar{v}}_{\mu,q_{a,b}}Q_{q_{N/2}}
^{\mu-2} \delta_{q_{a,b}} \nonumber \\
-i\sum_{\mu=3,5,...}^{\infty}(\mu-1)\bar{\bar{\phi}}_{\mu,q_{b,a}}
Q_{q_{N/2}}^{\mu-2}\delta_{q_{b,a}}\;\;. \label{10-13}
\end{eqnarray}
Here we have used the following notations:
\begin{eqnarray}
\bar{\bar{v}}_{\mu,q_{a,b}} = v_{\mu} + {\rm sin}^2(\frac{q_{a,b}}{2})
2^{\mu}\phi_{\mu} \;\;, \label{10-14} \\
\bar{\bar{\phi}}_{\mu,q_{a,b}}=-{\rm sin}(q_{a,b})2^{\mu}\phi_{\mu}
\;\;. \label{10-15}
\end{eqnarray}
The two wave numbers $q_{a,b}$ are related to each other by
\begin{equation}
q_b=q_a\pm \pi \;\; {\rm mod}2\pi\;\;.\label{10-16}
\end{equation}
In contrast to the previous case we have a coupling between a pair
of normal coordinates (\ref{10-16}) in (\ref{10-13}). Notice
that the coupling term is given by the second sum on the right hand
side of (\ref{10-13}) and is zero if $\phi_{2m+1}=0$. Also this coupling
term is proportional to $i$, which causes a mixing of real and imaginary
parts of the perturbations.
Interestingly for the pair $q_a=\pi,q_b=0$ we obtain again no coupling,
because (\ref{10-15}) vanishes for both wavenumbers. Consequently
 for $q=\pi$
(\ref{10-13}) describes  the continuation of the periodic orbit II.

\subsubsection{The case $\phi_{2m+1}=0$}
\label{s10.23}

If we assume $\phi_{2m+1}=0$ then the equations for the perturbations
$\delta_{q_{a,b}}$ decouple. In analogy to the case of orbit I
we have the first tangent bifurcation of orbit II if we consider
the perturbation $q=\pi(1-2/N)$. Consequently we obtain for the
bifurcation energy $E_{II}^c$
\begin{eqnarray}
{\rm i):} \;\;\phi_2=0\;\;,\;\;E_{II}^c\;\; {\rm arbitrary}\;\;,
\label{10-17} \\
{\rm ii):}\;\; E_{II}^c=\frac{1}{N^2}\frac{v_2+4\phi_2}{3(v_4+16\phi_4)}
4\pi^2\phi_2\;\;.\label{10-18}
\end{eqnarray}
Again we observe that a bifurcation will take place only if $\bar{v}_{\mu}$
is positive, i.e. only if the frequency of the periodic orbit II is
repelled from the linear spectrum with increasing energy.

\subsubsection{The case $\phi_{2m+1} \neq 0$}
\label{s10.24}

Now the coupling between two perturbations has to be taken into account.
One could expect that in the limit of large $N$ the coupling between
$q_a$ and $q_b$ vanishes. That is indeed so if we require $v_2 \neq 0$
(linear spectrum is optical-like) but turns out to be wrong for the
case $v_2=0$ (linear spectrum is acoustic-like). The details of
these subtleties are given in \cite{sf96}. Here we proceed to
the final result for the tangent bifurcation energy $E_{II}^c$:
\begin{eqnarray}
{\rm i):}\;\;E_{II}^c=\frac{1}{2\phi_3^2}\left( v_2+4\phi_2\right)
\left(\phi_2+\frac{3}{16}v_2\right) \phi_2\;\;, \label{10-19} \\
{\rm ii):} E_{II}^c=\left\{
\begin{array}{lr}
\frac{4\pi^2}{N^2}\frac{v_2+4\phi_2}{3(v_4+16\phi_4)}\phi_2
& v_2\neq 0 \\
\frac{16\pi^2}{N^2}\frac{\phi_2^3}{3\phi_2(v_4+16\phi_4)-64\phi_3^2}
& v_2=0
\end{array}
\right.
\;\;. \label{10-20}
\end{eqnarray}
As we can see solution i) (\ref{10-19}) is always positive (since
$\phi_2 > 0$ and $v_2 \geq 0$) but is not dependent on $N$.
Since we applied perturbation theory, (\ref{10-19}) is correct only
in the limit of small energies. For larger energies corrections apply.
Solution ii) (\ref{10-20}) is the one which gives
arbitrarily small bifurcation energies for sufficiently large $N$.
Since $\phi_3$ does not enter the energy dependence of periodic orbit II,
we then again obtain as the necessary condition for the
existence of the bifurcation,
that the frequency of the upper band edge plane wave
has to be repelled from the linear spectrum with increasing
energy. For the optical-like spectrum case $v_2 \neq 0$ this
is again a sufficient condition. However for the case
$v_2=0$ a more restrictive condition is obtained by demanding
\begin{equation}
3\phi_2(v_4+16\phi_4 ) \geq 64 \phi_3^2 \;\;. \label{10-21}
\end{equation}
Consequently for an acoustic spectrum case the condition - 
that the band edge plane wave frequency is repelled from the linear
spectrum with increasing amplitude - is only necessary but not
sufficient for a tangent bifurcation to occur.
It is interesting to note that condition (\ref{10-21}) has been obtained
with the help of multiple scale expansions already in 1972 by Tsurui
\cite{at72} 
and more recently by Flytzanis, Pnevmatikos and Remoissenet
\cite{fpr85} for systems with $v_{\mu}=0$.

\subsubsection{Symmetry breaking}
\label{s10.25}

At the bifurcation of the plane wave new periodic orbits occur. Because
the bifurcation is tangent, the new orbits have the same period
as the plane wave orbit (at the bifurcation point).
Any periodic orbit is a closed loop in the phase space of the system.
Consequently the new bifurcating orbits can be obtained by deformations
of the loop corresponding to the plane wave orbit at the bifurcation.
Flach has proven \cite{sf96} that there is no possibility to simultanously
deform the plane wave loop and to keep its invariance with respect
to the permutations (\ref{10-1}). Consequently there are at least
$N$ families of periodic orbits bifurcating from the plane wave orbits
at a tangent bifurcation. The spatial structure of these orbits
corresponds to exactly the spatial structure of a
discrete breather \cite{sf96}.

\subsection{Let us predict!}
\label{s10.3}

What follows from the stability analysis of band edge plane waves?
First, if a band edge plane wave undergoes a tangent bifurcation,
new periodic orbits bifurcate which i) are not invariant under
discrete translations, ii) have a spatial shape of one discrete breather.
Second, we know that for systems given by (\ref{3-1a}) 
and an optical-like spectrum ($v_2 \neq 0$) the repelling of the
band edge plane wave frequency with increasing amplitude is a sufficient
condition for the tangent bifurcation to occur! For acoustic-like spectra
$v_2=0$ the repelling condition is necessary but not sufficient.

That enables us to predict the existence of discrete breathers -
without actually finding the solutions. A well-studied example is
the FPU lattice in one dimension (acoustic linear spectrum). 
If we choose $\phi_4 > 0$ and increase $|\phi_3|$ starting
from zero, we will loose breathers which bifurcate from the upper
band edge plane wave - exactly when (\ref{10-21}) is violated.
That has been demonstrated by Sanduski and Page by performing
careful numerical experiments \cite{sp94}. Another prediction:
there exist no discrete breathers in a Toda chain (this is an
integrable system
with nearest neighbour interaction $\Phi(z)=({\rm e}^{-z}+z-1)$ and without
on-site potentials $V(z)=0$ \cite{mt89}, 
thus the linear spectrum is acoustic).
Indeed, expanding the Toda interaction $\Phi(z)$ we obtain
$v_{\mu}=0$, $\phi_2=1$, $\phi_3=-1/2$, $\phi_4=1/6$. Consequently
(\ref{10-21}) is violated. Indeed numerical efforts to obtain
discrete breathers in a Toda lattice have been fruitless 
\cite{kbs93},\cite{at95}.

If we consider a system with an optical-like spectrum, we see
that the repelling of the plane wave frequency from the linear spectrum
with increasing amplitude is sufficient. Consider now a narrow linear
spectrum. Then in general (note that exceptions apply) either
both band edge plane wave frequencies are increasing with increasing
amplitude, or they are decreasing. Thus usually one of the band edge
plane waves will fullfill the repelling criterion and one will not.
The plane wave which fulfills the repelling criterion thus gives
rise to discrete breathers! We obtain another way of understanding
the existence proof of MacKay and Aubry \cite{ma94} where the antiintegrability
limit was nothing else but the limit of a narrow optical-like 
linear spectrum! 
Further we can predict that when the Toda lattice is modified
by e.g. alternating heavy and light masses, one changes the linear
spectrum and obtains in addition to the acoustic band an optical band.
Then discrete breathers should exist which bifurcate from
one of the edges of the optical band. Indeed numerical simulations
have demonstrated the existence of discrete breathers in the Toda
lattice with alternating masses!

\section{Lattice dimension effects}
\label{s11}

In the first part of this section we will give some general
remarks on dimension effects on discrete breathers.
In a second part we will discuss
recent results by Flach, Kladko and MacKay \cite{fkm96}, who show
a profound effect of the lattice dimension on energy properties
of DBs. Namely energies of DBs in two- and three-dimensional
lattices have nonzero lower bounds. In contrast one-dimensional
lattices can allow for DBs with arbitrary small energy values.

\subsection{General remarks}
\label{s11.1}

As it follows from the preceeding analysis the existence
of discrete breathers is not crucially related to the dimension
of the lattice. Thus we are dealing with localized excitations
which can exist e.g. in one-, two- and three-dimensional lattices.
Consequently the dimension of the lattice can have only an effect
on the properties of discrete breathers, but apparently not on their 
existence.

As we discussed in section~\ref{s4} the dimension can affect the spatial
decay of breathers which bifurcate from acoustic band edge plane
waves. This happens if the system does not possess a symmetry $H(x)
=H(-x)$, because in that case even Fourier number components
are excited in the breather solution. The dc component will 
then resonate with the lower band edge of the acoustic band
of the linear spectrum. As we conjectured, the spatial decay
of the dc component will be then alebraic instead of exponential.
The power of the algebraic decay will depend on the lattice
dimension.

\subsection{Energy thresholds for discrete breathers}
\label{s11.2}

The tangent bifurcation analysis of band edge plane waves in 
section~\ref{s10} has been done for finite one-dimensional lattices.
The analysis can be carried out in higher dimensional lattices
\cite{sf96}. For convenience assume that we are discussing
a $d$-dimensional hypercubic lattice with periodic boundary
conditions and a total number of sites $N$. Then the critical amplitude
of the band edge plane wave at the bifurcation is given
by 
\begin{equation}
Q_q \sim N^{-1/d}\;\;\;, \label{11-1}
\end{equation}
i.e. the critical amplitude is inverse proportional to the
linear size of the system. This happens because the distance
between the eigenvalues of the linear spectrum at a band edge
are inverse proportional to the squared linear size of the system
\cite{sf96}. Let us assume that the energy of a solution is 
essentially quadratic in the amplitudes for small amplitudes.
It follows
for the critical energy of the plane wave at the bifurcation
point \cite{sf96}
\begin{equation}
E \sim N^{1-2/d} \;\;. \label{11-2}
\end{equation}
Note that this energy is not to be confused with the one-particle
energies $E^c$ (or energy densities) from section~\ref{s10}.
The energy of the bifurcating new periodic orbits have to
be the same as (\ref{11-2}) at the bifurcation.
Result (\ref{11-2}) is surprising, since it predicts that 
the energy of a discrete breather for small amplitudes
should diverge for an infinite lattice for $d=3$ and stay finite (nonzero)
for $d=2$, whereas for $d=1$ the discrete breather energy will tend
to zero (as initially expected) in the limit of small amplitudes
and large system size. 
Because the energy of a discrete breather can not be zero
for nonzero amplitudes of the breather center, it follows that
DB energies have nonzero lower bounds for two- and three-dimensional
lattices. That finding can be extremely important for any experimental
application.

Let us estimate
the discrete breather energy in the limit of small amplitudes and
compare the result with (\ref{11-2}).
Denote the largest amplitude of a given DB by $A_0$ where we define
that the site $l=0$ is the one with the largest amplitude. Then the
amplitudes away from the breather center will decay in space according
to an exponential law $A_l \sim A_0 {\rm e}^{-\delta |l|}$.
To estimate the dependence of the spatial decay exponent $\delta$ on
the frequency of the time-periodic motion $\Omega_b$ (which is
close to the edge of the linear spectrum) it is enough to
consider the dependence of a frequency of the linear spectrum $\Omega_q$
on the wave vector $q$ when being close to the edge. Generically this
dependence is quadratic $\Omega_E-\Omega_q \sim |q-q_E|^2$ where $\Omega_E
\neq 0$ marks the frequency of the edge of the linear spectrum
and $q_E$ is the corresponding edge wave vector.
Then analytical continuation of $(q-q_E)$ to ${\rm i} (q-q_E)$ yields a
quadratic dependence $|\Omega_b - \Omega_E| \sim \delta^2$.
Finally we need to know how the detuning of the breather frequency from
the edge of the linear spectrum $|\Omega_b - \Omega_E|$ depends on
the small breather amplitude. This is done using perturbation
theory for weakly nonlinear oscillators \cite{ahn93}. 
For that we need to know the exponent $(\mu-1)$ of the first
nonlinear term in the equation of motion for the band edge plane 
wave for small amplitudes (see (\ref{10-7a}) and (\ref{10-7b}) as
an example). Note that we slightly changed the definition of $\mu$
as compared to \cite{fkm96}.
Depending
on $\mu$ we obtain $|\Omega_b - \Omega_E| \sim A_0^z$
where $z=\mu-2$ for even $\mu$ and $z=2\mu -4$ for odd $\mu$.

Now we are able to calculate the energy of the discrete breather
replacing the sum over the lattice sites by an integral
\begin{equation}
E_b \sim A_0^2 \int r^{d-1} {\rm e}^{-\delta r} {\rm d}r \sim A_0
^{(4-zd)/2}\;\;. \label{11-3}
\end{equation}
We find that if $d > d_c=4/z$ the breather energy diverges
for small amplitudes, if $d=d_c$ the DB energy stays finite and nonzero
for small breather amplitudes, and for $d < d_c$ the DB energy tends
to zero for small amplitudes. 
The tangent bifurcation analysis was performed using $\mu=3,4$.
Inserting $\mu=3,4$ we obtain $d_c=2$
which is in accord with the exact results on the plane wave stability
\cite{sf96}. 

An immediate consequence is that if $d \geq d_c$ the energy of a breather
can not be zero, i.e. there is a nonzero lower bound for DB energies.
This happens because for any finite amplitude the breather energy
can not be zero. We  obtain an energy threshold for
the creation of discrete breathers for $d \geq d_c $. This new
energy scale is set by combinations of the expansion coefficients
in (\ref{3-1a}).

Another consequence is that for $d \geq 2$ the energy threshold
of DBs is always nonzero, independent of the value of $\mu$
(note that by definition $\mu \geq 3$). A one-dimensional system
can have nonzero lower energy bounds for DBs only if $\mu \geq 5$.

Numerical calculations of discrete breathers have been performed
in \cite{fkm96} in order to test these results. Discrete breathers
were numerically continued starting from large energy breathers which
were strongly localized. The method chosen was the steepest descent method,
using phase space variables. The dimension of the lattice was
$d=1,2,3$. A typical size for a three-dimensional lattice was
$30 \times 30 \times 30$. Details can be found in \cite{fkm96}.
In Fig.11
%\begin{figure}
%\caption{}
%\end{figure}
we show the dependence of the DB energy on the DB amplitude 
for a DNLS system with $\mu=4$. Clearly the energy thresholds are
observed. Other Hamiltonians have been also successfully tested.
The analysis of the DB solution which corresponds to a minimum
of energy shows that these breathers are still strongly localized
on a few lattice sites \cite{fkm96}.

The existence of a minimum of the energy on the one-parameter
family of DB solutions has further consequences. It immediately
follows that at the minimum energy solution a saddle-node bifurcation
appears, which is responsible for the fact that no DB solutions
exist below the threshold energy. 
Consequently either DB solutions with smaller or larger amplitudes have
to be dynamically unstable! 
Since in the case of a nonzero energy threshold the DB solutions 
increase in energy as the amplitude is decreased, a similar analysis
of the tangent bifurcation of the band edge plane wave leads to the
conclusion that the DB branch with smaller amplitudes is dynamically
unstable, whereas the one with larger amplitudes (with respect to the
amplitude which extremizes the DB energy) will be dynamically stable.
These conclusions support also the fact that the steepest descent method
is insensitive against instabilities of DB solutions.

\section{ Discrete symmetries}
\label{s12}

In this section we will discuss the relation between discrete
breathers and discrete symmetries of the equations of motion.
We will not discuss continuous symmetries because the only
continuous symmetry in the case of (\ref{2-1}) is the symmetry
with respect to shifts in time, i.e. any solution $X_l(t)$ will
generate another solution $X_l(t+a)$. For convenience and practical
purpose we consider Hamiltonian problems, which usually imply
that the trajectories in phase space defined by $X_l(t)$ and $X_l(t+a)$
are identical, i.e. $t \rightarrow (t+a)$ corresponds to a shift
along the trajectory. Note that this is not the case for e.g.
damped equations of motion (then the equations still have the continuous
symmetry, but the trajectories in phase space are not invariant).

Then the only symmetries we are left with are discrete symmetry operations.
E.g. (\ref{3-1a}) is invariant under time reversal $t \rightarrow
-t$. Another important discrete symmetry is the discrete translational
symmetry. In addition we can have optional symmetries due to
the lattice properties and a parity symmetry (displacement
reversal) $u_n \rightarrow -u_n$
for all $n$ (the last one happens in (\ref{3-1a}) if
$\phi_{\mu} = v_{\mu} = 0$ for all odd $\mu$). All these symmetries
imply that
when 
the corresponding symmetry operation 
is applied to a given solution, a new solution of the equations of
motion is obtained.
Each symmetry can
be represented as a function acting on the phase space of the system. These
functions map the phase space into itself. The relevant question is then,
whether a given trajectory in phase space is invariant if the
symmetry function is applied, or not. For a trajectory to be invariant
means that all points of the trajectory when transformed according to
the symmetry function are again points on the same trajectory.  

Note that since we are solving nonlinear equations in the phase space
variables, the invariance of a trajectory under a symmetry operation
is not granted at all. This is different from linear equations which
yield eigenvalue problems. Then the corresponding eigenvectors
will always reflect the symmetry of the equations (of the matrix).
Exceptions can occur when there are degeneracies (e.g. 
traveling wave solutions ${\rm cos}(i(kn - \omega t))$ of
a linear lattice with
periodic boundary conditions 
are not invariant under time reversal: 
this is due to degeneracies with respect to $\pm k$, which allows
for symmetric and antisymmetric standing waves (invariant under
time reversal) and travelling waves with $\pm k$ (not invariant
under time reversal).

Discrete breathers correspond to trajectories in phase space which
are not invariant under discrete translations along the lattice.
Consequently we can generate new discrete breather solutions 
by applying discrete translations. The Fourier coefficients $A_{kl}$
describing the solution are required to fullfill 
\begin{equation}
A_{kl} = A_{-k,l}^* \;\;, \label{12-1}
\end{equation}
were $A^*$ denotes the complex conjugate of $A$. In general the solutions
do not have to be invariant under time reversal. 
MacKay and Aubry apply the existence proof of DBs however to 
DB solutions which are invariant under time reversal (in the case
when the equations are invariant). Also all practical calculations
performed so far have actually been dealing with DB solutions
invariant under time reversal. This implies in addition to
(\ref{12-1}), that for solutions invariant under time reversal
there exists an origin of time such that
\begin{equation}
A_{kl} = A_{-k,l}\;\;\;, \label{12-2}
\end{equation}
i.e. that all Fourier coefficients are real.

Another restriction on the Fourier coefficients is obtained, if
the equations are invariant under displacement reversal, and
if we look for solutions which are also invariant under
displacement reversal. Then it follows that
\begin{equation}
A_{kl} = 0 \;\;,\;\; k=0,2,4,6,... \;\;\;. \label{12-3}
\end{equation}
This can help to reduce computation time for finding numerical
DB solutions. The same applies to lattice reflection symmetries.
However one should bear in mind that in principle DB solutions
can lack these discrete symmetries, just as they lack the
discrete translational symmetry!

Let us mention some related results. First MacKay and Aubry
\cite{ma94} explain that DB solutions in the antiintegrable
limit can be obtained for i) systems without discrete translational
symmetry (disorder); ii) Hamiltonian systems without time
reversibility. 
%MacKay and ... \cite{???} have shown the existence
%of DB solutions in coupled ...
Further the existence of phase-locked DB solutions
systems with damping and time-periodic driving will be
briefly discussed  in section~\ref{s19}. These DB solutions are obtained
for systems which are not invariant under time reversal and
even under continuous time translations (the continuous 
time translation is replaced by a discrete one)!

If one is looking for moving breather solutions, then 
the solutions can not be invariant under time reversal.
It could thus be that exact moving breather solutions
can occur through bifurcations from plane waves which
are not band edge plane waves (this is so because typically
band edge plane waves are not degenerate in the linearized
equations, and thus band edge plane waves keep the time
reversibility of the system).

As a result we can conclude that discrete breather solutions
are not only quite robust under system perturbations
which leave the systems symmetries unchanged, but discrete
breathers apparently are quite common among systems with
different symmetry properties.

\section{A conceptual approach}
\label{s13}

Very often one is dealing with a rather complicated system
(as a crystal with several atoms per unit cell), so that
it appears to be rather complicated to check numerically
whether discrete breathers exist. It amounts to taking into
account e.g. about ten degrees of freedom per unit cell
(this number can easily increase to twenty!). Solving then
for spatially localized oscillations can become a very hard
numerical task. It is then important to have a conceptual
approach of analyzing the system without actually solving
for DBs. As a result we want to be able to predict the existence
of discrete breathers.

This can be accomplished using the fact that DB solutions
bifurcate from band edge plane waves. In the following we will
explain this approach.

First one has to solve the linearized equations of motion - 
still a complicated task. This has been done using
either numerical diagonalizations, but one can also use
results of scattering experiments (usually neutron scattering,
but that is up to the experimental setup). We then know
the linear spectrum of the system. 

The linear spectrum has to be analyzed in order to find all
allowed frequency regions for potential discrete breather
solutions by taking into account the nonresonance condition
of all harmonics of a DB with respect to the linear spectrum.

After accomplishing that task, we should try to find 
narrow optical bands in the linear spectrum, whose surrounding
forbidden frequency regions belong to the allowed frequency
range of a discrete breather. Call these bands selected
optical bands. Here narrow implies that the width of those
bands is smaller than their absolute position on the frequency
axis.

Now we have to make some estimate on the nonlinearities in
the system. Especially we need nonlinearities in the 
normal mode variables of the linearized system which correspond
to the band edges of the selected optical bands.
We need only a perturbative treatment of the nonlinear part
which is valid for small amplitudes.

Next we have to calculate the change of the frequency of the
band edge normal modes of the selected optical bands in
perturbation theory with increasing amplitude. We have to select
the band edges whose plane wave frequencies are repelled from
the selected optical band with increasing amplitude.

Now we can expect that discrete breathers exist with frequencies
in the gaps of the linear spectrum which touch the selected band
edges of the selected optical bands. The spatial symmetries
of these breathers should correspond to the eigenvectors of the
normal modes of the selected band edges.

We should take into account that for two-dimensional systems
(e.g. surfaces) or three-dimensional systems these discrete
breathers will have nonzero lower energy thresholds.

\section{Plane wave scattering by discrete breathers}
\label{s14}

In this section we will discuss results on the scattering
of plane waves by discrete breathers. There are only few results
known, and certainly there remains much to be done in this field.
First we will discuss small amplitude lattice plane wave scattering.
This problem is closely tied to the dynamical stability analysis
of DBs. 
Some numerical results can be found in \cite{fw95}. Recent
results have been provided by Cretegny \cite{tc96}.
In a second approach we will discuss the scattering
of electrons by discrete breathers (here the discrete breather
is a localized vibration of a crystal lattice which is interacting
with electrons). 
Details can be found in \cite{fk96}.

Conceptually these two cases can be separated
with respect to their plane wave frequencies. In the first case
(phonons scattered by discrete breathers) the time scale defined
by the plane wave excitations is of the same order as the time scale
of the DB dynamics. In the second case (electron scattering) 
the time scale of the electronic plane wave is orders of magnitude
shorter than the time scale of the breather dynamics. These differences
are crucial for the concepts used.

\subsection{Phonon scattering}
\label{s14.1}

We use the term phonon here only to stress that the plane
waves considered here are small perturbations of the discrete
breather solution itself. These plane waves are connected
to the eigenmodes of the linear stability analysis of a DB.

Not much is known about phonon scattering, neither numerically nor
analytically. Up to now the only published results on this topic
are for one-dimensional lattices of the type (\ref{3-1a}), which
occured in Ref. \cite{fw95} where
a numerical analysis of the transmission of small-amplitude plane
waves was performed. The full equations of motion were used
in this analysis. The discrete breather solution was obtained
with precision $10^{-4}$ in the amplitudes, which limited the
observation to transmitted wave amplitudes of the same amplitude.
On the other hand the incident plane wave had to have a small enough
amplitude in order to avoid nonlinear effects in its propagation
besides the interaction with the discrete breather. In particular 
the superposition should still hold approximately. Thus the incident
wave amplitudes were $10^{-2}$. That defined a lower observation
threshold on the transmission coefficient $T \geq 0.01$.

The study was performed for different wave numbers of the incident
wave. Reflection and transmission were measured. Reflection was 
observed
due to an increase in the energy density on the incident
side of the DB, whereas
the transmission coefficient was  smaller than 1.
The scanned wavenumber interval was $0.1\pi < q < 0.9 \pi$. 
The results showed that the transmitted wave intensity was
modulated in time (see Fig.12). 
However this modulation could not be easily related
to the internal dynamics of the breather. The amplitude of these
modulations was of the order of the averaged transmission itself.
The time period of these modulations is of the order of $20$
breather periods, more precisely about $280$ time units of (\ref{3-1a})!

The most important result of this study was the exponential
decay of the squared transmission coefficient with increasing
wavenumber as shown in Fig.13
%\begin{figure}
%\caption{}
%\end{figure}
(triangles). 
Note that the filled circles represent the time-averaged
transmitted intensity, and the exponential decay in $q$ is
observed only for the $q$-region studied.
Thus discrete breathers appear to strongly reflect
phonon plane waves, which can be of importance whenever we
consider heat flow. The reflectivity properties are strongly
wavenumber dependent.

Further numerical  and analytical results have to be obtained.
There are several important aspects to be clarified - the influence
of lattice dimensions, discrete breather properties  and
of the plane wave wave
vector  on the scattering. In the following we will present
some recent results by Aubry and Cretegny which appear to be
very stimulating for further studies.

The idea is to derive the scattering result from the linear
stability analysis of a discrete breather. 
Indeed if we solve the Floquet problem of the linear stability
analysis of a discrete breather, we know the eigenvectors of
the spatially extended states. We could then combine these eigenvectors
in order to obtain a scattering setup - an incident and reflected
waves on one side of the obstacle (here the DB) and a pure transmitted
wave on the other side. However in order to do so it requires that
the eigenvalues of the two eigenvectors are exactly equal. This poses
a problem for finite systems, as the discussion in the following shows.

Recall that the linear stability analysis is defined by equation (\ref{8-1}),
which is a set of coupled linear differential equations for the 
small perturbations $\delta_l$, where the coefficients of the equations
are functions of the breather solution, and thus time-periodic.
The corresponding Floquet map (\ref{8-4}) defines the eigenvalue problem.
Since the Floquet map operates on the phase space, the space dimension
is $2N$ where N is the total number of degrees of freedom. Because
of the symplectic structure of the map there are at most $N$ independent
Floquet multipliers. Consider now the case that no breather is excited.
Still we can solve the Floquet eigenvalue problem. Assume periodic boundary
conditions (PBC). Then the Floquet multipliers are double degenerate. This happens
because of the familiar degeneracy of eigenvalues for running waves with
opposite wave numbers. The degeneracies are a consequence of the nonabelian
group formed by the discrete translational symmetry (a result of periodic
boundary conditions) and the reflection symmetry around any lattice site
\cite{LLIII}. Note that we can avoid these degeneracies by imposing
fixed or free boundary conditions.
The PBC are attractive because we want to study travelling waves.
However for any finite system the scattering problem is not defined
regardless of the boundary conditions.

If we have a breather solution, equation (\ref{8-1}) is not translationally
invariant anymore, even for PBC. As a consequence we are left with an
abelian group of reflection symmetry around the center of the DB, which
causes all degeneracies in the Floquet spectrum to be removed \cite{LLIII}. 
All Floquet eigenstates are now strictly symmetric or anti-symmetric
with respect to the reflection operation. There exists no pair
of symmetric and anti-symmetric eigenstates with the same eigenvalues.
This holds for any finite system regardless of the boundary conditions.

Far away from the breather location any extended Floquet eigenstate will
have the structure of the extended eigenstates of the Floquet map
without a DB. Thus for PBC the true eigenstates far away from the
breather can be represented schematically as $(a|q_+> + b|q_->)$, where
$|q_{\pm}>$ denotes a running wave with wave vector $\pm q$.
Note that these running wave vectors are not eigenvectors of the
Floquet map without a DB solution, because the eigenvalues will
not match. The wave number $q$ is determined from the dispersion
relation for the infinite system without a DB solution. The corresponding
frequency $\omega(q)$ is determined from the numerically obtained
Floquet phase $\theta$:
\begin{equation}
\theta = \omega T_b + 2 \pi m\;\;\;, \label{14-1}
\end{equation}
where $T_b$ is the DB period and $m$ is any integer. The resulting
frequencies
\begin{equation}
\omega = \frac{\theta}{T_b} - m \frac{2\pi}{T_b} \;\; \label{14-2}
\end{equation}
have to belong to the linear spectrum $\omega(q)$. In general
this allows for several values $\omega_q$. However we know that
the DB frequency and all its multiples have to be outside the
linear spectrum. This limits the number of allowed $q$-values.

If we can find another eigenvector of the Floquet map with
the same eigenvalue, we can combine both in an appropriate way
to obtain the scattering setup as described above. That
is impossible for any finite system because there are no degeneracies
left once the DB solution is assumed to exist. What happens in the
limit a large system size $N \rightarrow \infty$? Suppose the DB solution
is itself symmetric ( or anti-symmetric). Then the anti-symmetric
(symmetric) Floquet eigenstates will be unchanged as compared to the
Floquet map without a breather! 
Consequently the eigenvalues of the anti-symmetric (symmetric) eigenstates
did not change either.
Thus we have to study the $N$-dependence of the symmetric (anti-symmetric)
eigenstates in the presence of the breather. The changes of these
eigenstates due to the breather presence give the values for the
eigenvalue splittings due to the lifting of the degeneracies.

We can obtain an upper estimate for the splittings. The splittings
can not become larger than the averaged spacing, since the spectrum
of each subgroup (symmetric and anti-symmetric) is confined by
the band edges of the linear spectrum and crossings within each
subgroup are not allowed in general. The averaged spacing is of
the order $1/N$ in the middle of the linear spectrum and decreases
to $1/N^2$ and the band edges. Consequently the splittings decrease
with increasing system size. In the limit $N \rightarrow \infty$
we can find degeneracies and thus construct a scattering case.
If (\ref{14-2}) is fulfilled for more than one frequency $\omega(q)$
then scattering involves several wave numbers. 

Cretegny has calculated the  transmission coefficient of the  
case in Fig.13 (solid line). Qualitative and partially quantitative
agreement with the
numerical simulation of the full system is found. Moreover
a decrease of the transmission coefficient at small wave numbers
is predicted (this is in analogy to the scattering by a single
defect in a linear system, for details see Economou \cite{ene83}).
However this linear approach is not capable of explaining the observed
modulations (see above). These modulations appear for (\ref{14-2})
having exactly one solution $\omega_q$. Also finite-size effects
can be ruled out. Splittings in the Floquet spectrum will cause
some modulations, but the amplitudes should be small, and the estimated
modulation period is three  orders of magnitude larger than the
observed one! Apparently there is still a lot to understand in
the scattering problem. Also the solid line in Fig.13 indicates
that total reflection is possible for wave numbers $q \approx 2.1$.
However the numerical experiment (filled circles in Fig.13) do not
indicate any total reflection. 

\subsection{Electron scattering}
\label{s14.2}

If a discrete breather is excited in a crystal lattice, we obtain
a localized crystal lattice vibration. Often lattice vibrations
couple to electronic degrees of freedom. From a mathematical point
of view the adding of electronic degrees of freedom corresponds
to additional degrees of freedom with eigenfrequencies orders of magnitude
larger than the eigenfrequencies of lattice vibrations. 
This is different from the interaction of a breather with a lattice
plane wave (see preceeding subsection).

Let us consider a classical nonlinear lattice which allows for
breather solutions.
Generally the excitation of the relevant breather degrees of freedom
leads to a localized polarization of the lattice.
In the classical ground state of this
system
the polarization vanishes.
If we excite a discrete breather, then it will induce
a (time-periodic) multipole field at distances large compared
to the breather size. Generally
the first nonvanishing moment will be a
dipole moment. The induced polarization will be spatially
localized, in accord with the strong localization properties of
the breather solution.

Let us consider the interaction of a single electron with a discrete
breather in the case when the distance between the electron and the breather
is much larger than the breather size.
Since we are describing the lattice degrees of freedom classically,
we can use the adiabatic approximation \cite{asd73}. This means, that
the motion of the electron is described by using the positions
of the lattice degrees of freedom as parameters. Thus the electron feels
a multipole field which originates from the breather. This kind of treatment
of the electron is similar to the consideration of electrons in a lattice
with defects \cite{jm73}.
The difference is, that i) the breather (dynamical defect)
does not posess an uncompensated charge and ii) the breather is
slowly (as compared to the inverse electron frequency) changing its multipole
field.

Since the multipole field of the breather contains in general a dipole
component, we can study the scattering of an electron in a
dipole field. We consider the case when  the
electron can follow a path which does not
come close to the breather location. If this assumption is not
true anymore, the electron can be trapped by the breather, as will be
shown in the next section.

The potential of a dipole in a three-dimensional lattice is given by
\begin{equation}
V_d(\vec{r}) = \frac{1}{\epsilon}\frac{\vec{d}\vec{r}}{r^3}\;\;.\label{14-3}
\end{equation}
Here $\vec{d}$ is the dipole moment induced by the breather (which
is actually slowly periodically oscillating with time). The dielectric
permeability $\epsilon$ describes the reduction of the dipole
field due to polarization effects.

The motion of an electron with isotropic effective mass
$m^*$ and charge $e$ will be then
described by the Hamiltonian $H$ and the wave function $\Psi(\vec{r},t)$
\cite{jm73}
\begin{equation}
H = -\frac{\hbar ^2}{2m^*} {\rm \Delta} + eV_d(\vec{r}) \;\;,
\;\; i\hbar \frac{\partial \Psi}{\partial t} = H \Psi\;\;.
\label{14-4}
\end{equation}
The dipole potential (\ref{14-3}) does not posess localized states.
This can be easily determined by considering the corresponding classical
motion in the potential (\ref{14-3}). Clearly there exist no periodic
orbits having some finite distance from the potential center $\vec{r}=0$.
Thus there appear no localized
electron states which are weakly localized as compared to the breather
size.This situation is opposite to the Coulomb field problem where periodic
orbits exist and lead to the existence of hydrogen like localized states,
as used in the description of Wannier excitons. To find trapped electronic
states induced by a discrete breather we have to take into account
the internal breather structure, which will be studied in the next section.

To account for the elastic electron reflection in the
dipole potential (\ref{14-3})
we can use Born's approximation \cite{asd73},
which holds if the interaction energy
between the electron and the breather will be small compared to the
kinetic energy of the electron. Denoting by $|\vec{k}>$ the plane wave
states of the electron in the absence of a breather, we have to
calculate the matrix elements
\begin{equation}
<\vec{k}|V_d|\vec{k}'> = \int {\rm e}^{i(\vec{k}-\vec{k}')\vec{r}}
V_d(\vec{r}){\rm d}r^3\;\;.\label{14-5}
\end{equation}
Straight forward integration gives
\begin{equation}
<\vec{k}|V_d|\vec{k}'> = -i \frac{4e \pi}{\epsilon}
\frac{\vec{d}(\vec{k}-\vec{k}')}
{|\vec{k}-\vec{k}'|^2} \delta(E_{\vec{k}} - E_{\vec{k}'})\;\;. \label{14-6}
\end{equation}
The electronic energies $E_{\vec{k}}$ measure the energy of the incoming
and outgoing plane waves.
All other quantities related to the electron scattering in the used
approximation can be obtained from these matrix elements.

In the nongeneric symmetric case that the breather
does not posess a dipole moment,
the quadrupole field tensor $D_{\alpha \beta}$ has to be considered
(note that we use the definition $D_{\alpha \beta}=\sum_i e_i x^{(i)}_{\alpha}
x^{(i)}_{\beta}$).
The corresponding potential is given by
\begin{equation}
V_q(\vec{r}) = \frac{1}{2\epsilon} D_{\alpha \beta} \frac{\partial}
{\partial x_{\alpha}}\frac{\partial}{\partial x_{\beta}} \frac{1}{r}
\;\;. \label{14-7}
\end{equation}
Again there are no bound states in potential (\ref{14-7}) as in the dipole
case. The matrix elements can be obtained by integrating:
\begin{equation}
<\vec{k}|V_q|\vec{k}'> = -i \frac{2e \pi}{\epsilon}
 \frac{D_{\alpha \beta}k_{\alpha}k_{\beta}}{|\vec{k}-\vec{k}'|^2}
\delta(E_{\vec{k}} - E_{\vec{k}'})\;\;.
\end{equation}

\section{Trapping of plane waves by discrete breathers}
\label{s15}

In the preceeding section we discussed the scattering of
plane waves by discrete breathers. A natural next question
would be, whether these plane waves can be trapped by the
discrete breather. In other words, we look for bound states
of the medium which provided the plane waves to be scattered
by a discrete breather. This is not to be confused with the 
discrete breather solution itself - which is a bound state too,
but not due to an obstacle which is breaking the translational
symmetry.

We will first discuss trapping of phonons. These bound states
will be closely connected to the discrete breather solution itself,
as well as to its dynamical stability properties.
Then we will discuss the trapping of electrons by breathers.
We will review results of Flach and Kladko \cite{fk96}, 
Aubry \cite{sa96} and Vekhter and Ratner \cite{vr94},\cite{vr95}, 
and add some thoughts.

\subsection{Phonon trapping}
\label{s15.1}

We consider a discrete breather solution. Then we add small
perturbations to the initial conditions and ask whether
we can find bound states for the perturbations. The task
is thus again to analyze the linearized stability problem
(\ref{8-1}). First we know that the DB solutions come in one-parameter
families. Thus equation (\ref{8-1}) will always have two
eigenvectors with Floquet multipliers both equal to $+1$.
This is so because the one-parameter family of DB solutions
(which are time-periodic and thus one-dimensional manifolds
in phase space) is of dimension two. Given any point $P$ on this
manifold, we can always find two orthogonal directions (no matter
what scalar product we use) which are tangent to the manifold
in $P$. For instance one direction is simply pointing
in the direction of the periodic orbit which is defined by $P$.
The second one will point to neighboring periodic orbits.

Both eigenvectors will be exponentially localized on the
lattice, because they correspond to differences of two
neighbouring DB solutions on the manifold. Thus formally
we can call them bound phonon states. This is another interpretation
of the same fact that discrete breathers come in one-parameter
families. The term 'bound state' implies here that we can add
energy to the DB solution such that the additional energy is not dispersed
over the entire lattice.

Another possibility is that the eigenvalue problem (\ref{8-1})
allows for additional spatially localized eigenstates. If the
Floquet multipliers of these eigenstates are located on the unit circle,
i.e. if these perturbations are linearly stable, then we obtain
bound states which are not merely a consequence of the existence
of the DB solution itself. Such eigenstates were observed in many
numerical studies \cite{sps92},\cite{cku93},\cite{dpw93},\cite{fw93},%
\cite{fwo94},\cite{fw94},\cite{fkw94}, and especially 
studied systematically by Marin and Aubry in \cite{ma96}.

What happens if we consider the full nonlinear equations instead
of the linearized phase space flow around a breather solution?
The two-dimensional manifold of one-parameter DB families is still
present, so that the first type of bound states survives - but of
course these bound states are somehow trivial. The second nontrivial
type of bound states would if it survives correspond to quasiperiodic
breathers. Generally we do not expect them to exist (see section~\ref{s4}). 
Strictly speaking we are then left with the first type of trivial
bound states. However as also discussed in sections~\ref{s3} and \ref{s8},
the life-time of the second type of bound states can be very
large compared to the time periods of the DB solution.

\subsection{Electron trapping}
\label{s15.2}

As already indicated in the previous section, the treatment
of electrons interacting with discrete breathers is quite
complicated. We will have to make a lot of additional assumptions
in order to come to published results. 
First we have to consider just one electron moving on a lattice -
so we do not know the effect of electron-electron interaction
\cite{pf95} on what
is to come. Next we are not aware of any rigorous treatment of both
lattice and electronic degrees in a quantum mechanical way. So what
we will consider is one electron whose motion is described within
a tight binding model. The electron will be coupled to a 
lattice of classical interacting degrees of freedom. 

\subsubsection{Model structure}
\label{s15.21}

The electron description in a tight-binding representation
implies that at every lattice site we can find a complex
probability amplitude $\Psi_l(t)$ such that $|\Psi_l(t)|^2$
gives the probability distribution for the electron.
The tight-binding Hamiltonian for the electron is then
\begin{equation}
H_e = \sum_l \left[ \epsilon _l \Psi_l \Psi_l^*
+ T_l(\Psi_l \Psi_{l+1}^* + cc) \right] \;\;\;. \label{15-1}
\end{equation}
Here $T_l$ is some hopping matrix element, and $\epsilon_l$ is
a one-site energy. 
The equations of motion are given by
\begin{equation}
\gamma \dot{\Psi}_l = i \frac{\partial H_e}{\partial \Psi_l^*}
\;\;\;. \label{15-2}
\end{equation}
Here the parameter $\gamma$ regulates the frequency scale of the
electronic system. 
Thus the consideration of just one electron reduces the
whole quantum mechanical task for the electron
to a classical Hamiltonian
problem (\ref{15-1}),(\ref{15-2}) where we have besides the
energy another integral of motion - the number of electrons $N_e$
\begin{equation}
N_e = \sum_l |\Psi_l|^2\;\;\;. \label{15-3}
\end{equation}

The coupling to the classical variables
$u_l(t)$ of (\ref{3-1a}) is then obtained by assuming a certain
functional form of the parameters $\epsilon_l(\{u_{m} \})$ and
$T_l(\{u_m \})$. Usually the coupling has some locality properties.
Consequently we arrive at some extended lattice problem - instead
of only (\ref{3-1a}) we have in addition (\ref{15-1}), which increases
the number of degrees of freedom per lattice site. As we already
pointed out, the number of degrees of freedom is not crucial for
the existence of discrete breathers. Thus we can search for
discrete breathers in the extended system with an electron, and if
we find solutions, they will correspond to both a vibrational localization
of the lattice as well as a localization of the electron.

\subsubsection{Holstein model and nonlinear generalizations}
\label{s15.22}

The original Holstein model is obtained by setting
$v_2=1$, $v_{\mu \neq 2} = \phi_{\mu}=0$ in (\ref{3-1b}),(\ref{3-1c}),
and $T_l=T$, $\epsilon_l = u_l$. Aubry considered a nonlinear
generalization of this model by allowing $V(z)$ of (\ref{3-1b})
to be any anharmonic potential (Aubry still requires $\Phi(z)=0$)
\cite{sa96}. 

The Holstein model has been used to discuss polaron formation.
The polaron is a groundstate solution of the electron-lattice
system which is not invariant under discrete lattice translations.
Moreover the polaron solution corresponds to a localized electron
combined with a localized static displacement field. The polaron
is indeed often a consequence of an interaction $\epsilon_l=u_l$
because this interaction term shifts the equilibrium position
of the $u_l$ variables in the presence of some nonzero electronic
density. 
Let us consider the polaron solution for the original
Holstein model. There $V(z)=u_l^2/2$. The condition $\dot{u}_l=0$
leads to 
\begin{equation}
u_l = -|\Psi_l|^2\;\;\;. \label{15-3a}
\end{equation}
Then we obtain a DNLS for the electronic amplitudes
\begin{equation}
\gamma \dot{\Psi}_l = i\left[ -|\Psi_l|^2\Psi_l
+ T(\Psi_{l-1} + \Psi_{l+1}) \right] \;\;\;. \label{15-3b}
\end{equation}
We have already discussed this equation. It allows for breather
solutions. Consequently the polaron groundstate is 
exponentially localized.
Once the groundstate is not invariant under translations,
so will be other trajectories which can be regarded as excitations
above the given groundstate. Because of the localization properties
of the groundstate some excitations above the groundstate may stay
localized too. This is in complete analogy to the excitations above
a kink groundstate in a kink-bearing one-dimensional lattice 
\cite{bw89}.
All we have to check is whether the spectrum of the localized excitations
is not in resonance with the linear spectrum of the extended excitations
above the polaron groundstate. This is in complete analogy to
the discrete breather analysis.

Aubry indeed proves the existence of localized excitations above the
polaron groundstate by analytical continuation from the antiintegrability
limit $T=0$ of the nonlinear generalization of the Holstein
model \cite{sa96}. For $T=0$ the lattice sites are not interacting,
and the solution at one lattice site with $|\Psi_l|^2 = \rho$ reads
\begin{eqnarray}
u_l(t)=G(t;\omega_b,\alpha)=G((t+2\pi/\omega_b);\omega_b,\alpha)
\;\;,\;\;\ddot{G} = -V'(G) - \rho \;\;\;, \label{15-4} \\
\Psi_l(t)=\rho^{1/2} {\rm e}^{i(\beta + \gamma^{-1} f_l(t))}\;\;,\;\;
f_l(t)=\int_0^t u_l(t') {\rm d}t'\;\;\;. \label{15-5}
\end{eqnarray}
Here $\omega_b$ is the frequency of the $l$-th oscillator and
$\alpha$ is a phase.
Since for $\rho \neq 0$ the $u_l(t)$-dependence will show up
with a nonvanishing dc term $\overline{u_l(t)}$, the solution $\Psi_l(t)$ is 
given by
\begin{equation}
\Psi_l = \xi_l(t) {\rm e}^{i (\overline{u_l(t)}/\gamma)t}\;\;,\;\;
\xi_l(t)=\xi_l(t+2\pi/\omega_b)\;\;\;. \label{15-6}
\end{equation}
Consequently the lattice variable $u_l$ is strictly periodic
and its Fourier series representation contains all multiples
of the fundamental frequency $\omega_b$. The electronic probability
amplitude is quasiperiodic, but its Fourier series representation
is still an equidistant spectrum with frequencies $(\overline{u_l(t)}/\gamma
+k\omega_b)$. This should make it easy to escape resonances
with the linear spectrum of our problem. The linear spectrum
for $T \neq 0$ is given by a degenerate value $\sqrt{v_2}$ -
the lattice contribution, and a band of width $T/\gamma$ located
at the origin - the electronic tight binding band. For small values
of $T$ (independent of the value of $\gamma$) we can indeed escape
resonances of both solution parts with the linear spectrum
(note that we have to use polarization vectors here to describe
the linear spectrum - consequently the dc part of the lattice 
contribution does resonate with the electronic band, but due to
the orthogonality of the polarization vectors these resonances
do not prevent the argument for localization from being valid). 

The exponent of the spatial localization of the electron depends
essentially on the distance of the electron's frequency
from the electronic band. This distance can be chosen to be large
for large $u_l$ amplitudes, which is then in principle even
possible for finite values of $T$. Thus dynamical excitations
above a polaron groundstate are possible where the whole
excitation energy is concentrated on a few lattice sites.

\subsubsection{Related models}
\label{s15.23}

Flach and Kladko have studied a model different from
the Holstein model \cite{fk96}. There the one-site
energies are dropped completely $\epsilon_l=0$ and
the coupling between the electron and the lattice
is given by $T_l=1+\beta_1 (u_l - u_{l+1})$. Also the
interaction potential on the lattice $\Phi(z) \neq 0$.
Adiabatic approximation (which is here essentially
a separation of time scales) was used. The lattice itself was assumed
to posess discrete breather solutions in the absence of
any coupling $\beta_1 = 0$. Then there exist bound electron-breather
states for $\beta_1 \neq 0$ where the electron can be strongly
localized even for small values of $\beta_1$. Again that
is accomplished in the limit of large amplitude DB solutions,
which are also strongly localized. 

Vekhter and Ratner \cite{vr94},\cite{vr95} have studied numerically
a similar model. There a bound electron-breather state was
found. These numerical studies showed several interesting
features (e.g. a weak radiation of the electron out of
the localized object) which deserve further investigation.

\subsubsection{Bound states without polarons}
\label{s15.24}

All above discussed models tend to the formation of polarons
(we know that for the Holstein model, and suspect it for the
other models). This is because the interaction term
is proportional to $u_l$. What happens if we choose e.g.
a Holstein model with $\epsilon_l = u_l^2$? There will be
no polaron solution, so the groundstate of the system with
$T\neq 0$ is invariant under translations. However we can easily
use Aubry's antiintegrability scheme \cite{sa96}
and
find localized solutions in the coupled system for $T$ smaller
than some finite values. 
Indeed for T=0 there is only one change in (\ref{15-5}) namely
$f_l(t)=\int_o^t u_l^2(t')dt'$. Again we find that both parts
$u_l(t)$ and $\Psi_l(t)$ have Fourier time series corresponding
to an equidistant spectrum. Again we can easily avoid resonances
with the linear spectrum. 
These solutions will be very similar
to the discrete breathers discussed before, in fact these
solutions are discrete breathers. Thus the trapping of an electron
by a discrete breather is also a slightly irritating term,
which is helpful in understanding the physics, but irritating
in making mathematical connections. If the presence of an electron
creates a polaronic groundstate, then it should be of no
surprise that there exist localized excitations above the 
groundstate (one can still have a hard time in showing that).
If the presence of the electron does not create a polaronic
groundstate, localized excitations can still exist, because
the coupling between the electron and the lattice is already
a nonlinear term in the equations of motion.

\section{Disorder and Nonlinearity}
\label{s16}

So far we discussed the existence of discrete breathers
in systems which are invariant under discrete translations.
In this section we will briefly discuss the effect of disorder
on discrete breathers. The reader will find additional 
information in a recent review of Sievers and Page \cite{sp95}.

Let us first consider the effects one isolated impurity will have.
Using the antiintegrability scheme it follows that 
discrete breathers of networks of coupled oscillators allow
for breather solutions also in the presence of a defect \cite{ma94}.
In other words an isolated impurity will have no serious impact
on the existence of DB solutions. Of course the DB solutions will
smoothly change, where the changes will occur locally at the defect 
site. There will be also smooth changes in all properties of
DB solutions as linear stability, scattering, bound states etc.
These statements are of course true only if the system without
a defect is nonlinear.

It is well-known that a linear system with a defect allows
for localized solutions too. Because of the linearity the
defect problem is reduced to the diagonalization of a matrix.
If the defect strength is large enough some eigenvectors
(at least one) will be exponentially localized on the lattice.
What happens to these localized solutions if nonlinear terms
are present? It was long believed that in general nonlinearities
destroy the local defect modes. That belief stems from the
observation that any nonlinearity is equivalent to an interaction
between the normal modes of the linear system. Consequently
it was believed that this interaction will cause the energy of the
local defect mode to be dispersed among all other extended normal
modes. This heuristic argument is indeed true for most trajectories
in the phase space of the system. That must be so because typically
nonlinearities destroy integrability and cause the phase space
to be densely filled with an Arnold web of chaotic trajectories
\cite{suz88},\cite{bvc79}. 
The trajectories in this web will come close to any
point in phase space in the course of evolution. That circumstance
is nothing but the celebrated ergodic property of a nonintegrable
system. Trajectories from this web can not be localized in some
subpart of the phase space then.
However the same argument can be applied to the cases we discussed
before - most of the systems allowing for DB solutions are nonintegrable.
The reason why DB solutions can exist (apart from the nonresonance
condition they have to fullfill) is that typically any nonintegable
Hamiltonian system will have a phase space also densly filled with
periodic trajectories. These trajectories are not ergodic of course.
Thus the above given heuristic expectation of the loss of
localized defect modes due to nonlinearities has to be corrected.
If a localized mode exists in the linear system its frequency
will be located outside the linear spectrum of the extended
eigenmodes. The frequency of the mode does not change as we
change the amplitude of the mode - as a result of the linearity
of the system. However the energy of an eigenmode surely changes
with amplitude. Bambusi \cite{db96} has shown that the continuation
of periodic orbits can be done by considering orbits with fixed
energy (as opposed to fixed frequency). 
Adding nonlinearities will then cause the periodic orbit to be 
slightly deformed, but still to exist! In other words, a local defect
mode will surely survive the presence of nonlinearities. 
We can be sure about that because the continuation of the 
linear defect mode into the nonlinear regime does not meet the
problem of resonances. Of course we have to test first whether
all higher multiples of the local mode frequency do not resonate
with the linear spectrum of extended states. It would be of interest
to test whether the continuation of such a mode into the nonlinear
regime and a subsequent removing of the defect strength continuously
transforms the local defect mode into a discrete breather. We suspect
that cases exist where it indeed will happen.

There exist studies of the interaction of a moving breather-like
entity with defects. Since we do not know precisely how to 
treat the moving object, it is again hard to judge the results
of the numerical studies \cite{wfb94},\cite{ysk93-prb},\cite{dpw93},%
\cite{thom91},\cite{th93},\cite{ht92-pla},\cite{tip96},\cite{hw95},%
\cite{fpm94}. 
Observations range
from trapping of moving DBs by a defect to reflection of
DBs by a defect\footnote{See also \cite{ej88}.}. 

An even more complicated problem arises if the system is complelety
disordered, e.g. that there is a finite density of defects
(see also \cite{gk92}).
The linear problem is nothing but the Anderson localization
\cite{LHLXI}. Although this problem has been widely studied, the
methods to prove nontrivial statements are quite elaborate
\cite{LHLXI}, essentially because it is hard to treat the disorder.
Adding nonlinearities makes the problem even more complicated.
Albanese et al \cite{af88},\cite{afs88},\cite{af91} and
Frohlich et al \cite{fsw86} studied
the nonlinear problem with disorder. 
Albanese et al prove the existence of periodic orbits
in the disordered DNLS where the orbits are exponentially localized
on the lattice. MacKay and Aubry \cite{ma94} can use their
continuation scheme for any system of coupled oscillators (with
disorder incorporated). The continuation of periodic orbits
which correspond to one-site breathers in the ordered lattice
is straightforward. But to obtain the exponential localization
on the lattice, a notion of distance in the space of periodic
orbits is needed and a restriction of the coupling
to be of exponential decay itself on the lattice (see also 
\cite{hrtg95}).

\section{Quantum breathers}
\label{s17}

So far we were discussing properties of coupled nonlinear
differential equations. In terms of physics we were discussing
classical many-particle systems (lattices). Often the classical
description of a physical problem can be an approximation to
the quantum mechanical (QM) properties of the system. Then it
is legitimate to ask what is the trace of the classical discrete
breather in the quantum mechanical description. For continuum
problems these questions have been studied e.g. in 
\cite{dhn74},\cite{gj75}.

First we have to specify the correct correspondence relation
between a classical model and its quantum mechanical counterpart.
This will be the conventional first or second quantization
procedures. There exists some arbitrariness, because the classical
variables commute. Consequently there exist families of QM models
which yield the same classical model in the classical limit.
In the cases studied this degeneracy was not important.

What is a quantum breather? We are interested in eigenstates
to the Hamiltonian. Since the QM systems we are looking for
are translationally invariant, and the QM eigenvalue problem is
by definition a linear problem, we can not expect that spatially
localized eigenstates exist. On the contrary all eigenstates
will be delocalized in space. 
How can that be connected to the existence of classical DB solutions?
A possible way is 
that
quantized DB solutions can tunnel from site to site.
Then in the classical limit the tunneling rate will vanish,
and we obtain classical DB solutions. That also implies that 
in the classical limit we expect to find certain groups of eigenstates
(with exactly $N$ members where $N$ is the number of lattice sites)
where the eigenvalues in each group are degenerate. Yet another
way of saying the same is that in the classical limit we expect
to find zero-width bands of $N$ eigenstates.
Away from the classical limit we expect these bands to broaden,
but not to change in some qualitative way to be specified later.

Because of the bosonic commutation relations we can thus
define quantum breathers as bound boson states. A quantum breather
state belongs to a band of $N$ states. Each bound state of such
a quantum breather band is characterized by a quantum number - 
a wave number. The particle property of such a bound state
can be probed with the help of correlation functions. These correlation
functions should show (exponential) decay
in the distance between parts of splitted bounded bosons.

There exist several studies where classical DB solutions are
quantized using various approximations \cite{sh91},\cite{st90-jpsj-2}%
,\cite{kit92}
to obtain
estimations on the gap and width of breather bands.

The existence of quantum breathers raises two 
questions. The first one is: what QM models allow for
quantum breathers, and is there any correspondence to
classical models allowing for DB solutions? The second one is:
how is the DB solution obtained in a classical limit of
a QM model allowing for quantum breathers?
We will address both questions in the following.

\subsection{Weakly interacting oscillators}
\label{s17.1}

Let us first discuss a system of weakly interacting oscillators
given e.g. by (\ref{3-1a}). We can always choose a local basis
given by the noninteracting system $\Phi(z)=0$. In this case
we can formally solve the one-site problem of a particle
moving in the potential $V(z)$. Denote the eigenvalues of
the corresponding one-site Hamiltonian with $\epsilon_n$
with $n$ being the quantum number. 

If the potential $V(z)$ is harmonic we find an equidistant
eigenvalue spectrum e.g. $\epsilon_n = n$. Then still having
$\Phi(z)=0$ the eigenvalues $E_n$ of the system of $N$
oscillators are given by $E_n=n$ where $n$ is defining
the number of bosons excited. The linearity of $\epsilon_n$
in $n$ implies that these bosons are noninteracting (do not
mix this interaction with the lattice interaction $\Phi(z)$).
Consequently the degeneracies of $E_n$ are larger than $N$ except
for $n=0,1$: $E_0$ is nondegenerate, $E_1$ is $N$-fold degenerate,
$E_2$ is $N(N+1)/2$-fold degenerate, $E_3$ is $N(N^2+2)/3$-fold
degenerate etc. Switching the interaction $\Phi(z)$ on we
can expect all these degeneracies to be lifted. However except
for the one boson band $n=1$ all other bands will consist
of many more than $N$ states. Indeed we do not expect quantum 
breather bands for a harmonic function $\Phi(z)$.

Assume now that the potential $V(z)$ is anharmonic. Then the
one-site eigenvalue spectrum $\epsilon_n$ is not equidistant
in $n$ anymore. This change has consequencies for the eigenvalue
spectrum of the system of $N$ noninteracting oscillators.
Apart from changes in the eigenvalues the degeneracies are drastically
reduced. The old eigenvalue $E_2$ splits now into two
eigenvalues - $\epsilon_2$ with two bosons
\footnote{Note that here we are using the notion of a boson loosely
to paraphrase the number state - a product state with each site
excited to the n-th level (n bosons at that site).}
on one lattice site (exactly
$N$ states) and $2\epsilon_1$ with two bosons on different lattice sites
and $N(N-1)/2$-fold degeneracy. 
The relative position of the two eigenvalues depends on the
type of anharmonicity in $V(z)$.
The old eigenvalue $E_3$ splits
into three eigenvalues - $\epsilon_3$ with three bosons on one
lattice site (exactly $N$ states), $(\epsilon_2 + \epsilon_1)$ with
two bosons on one lattice site and one boson on a separate lattice
site ($N(N-1)$-fold degenerate) and $3\epsilon_1$ with all three
bosons on different lattice sites and $N(N-1)(N-2)/3$-fold
degeneracy. Thus the anharmonicity in $V(z)$ creates an infinity
of states with $m$ bosons on one lattice site,
which are exactly $N$-fold degenerate. We expect these states to
be quantum breather states. The subsequent increase of $\Phi(z)$
will then lead to a finite band width of these breather bands
with preserving the particle-like nature of the bound states.

These expectations (for nonvanishing interactions) have not been
proven yet for a general case. One of the reasons is that
the noninteracting case conserves the number of bosons, but
the interacting case in general does not. 

From the above it follows that quantum breather bands are likely
to exist for systems with an antiintegrability limit $\Phi(z)=0$.
The necessary condition is the anharmonicity of $V(z)$, which is
also the necessary condition of finding classical DB solutions.
Thus in this particular limit we expect a strong correspondence
between the existence of classical DB solutions and the 
existence of quantum breather bands. Increasing the interaction $\Phi(z)$
the quantum bands broaden, and eventually overlap (hybridize) with
other bands. Further increase of the interaction may or may not 
destroy the particle-like properties of the bound states. 
The correspondence between loosing/keeping quantum breather bands
and classical DB solutions is a field yet to be explored.

A numerical analysis of quantum breather bands has been recently
performed by Bishop et al \cite{wgbs96}. The systems were one-dimensional,
with a total number of lattice sites $N=4,8$. 
For $N=4$ seventeen states per site were taken into account, and 
six states per site for the case $N=8$. Low lying quantum breather
bands were found (note that these bands consist
of 4 or 8 states only). The particle-like properties of the
breather states were successfully 
probed with the help of correlation functions.
The smallness of the considered systems is due to computational
limitations.
Another recent computation of quantum breather states has been 
performed by Schofield eta al \cite{sww96}. These authors aim
at the understanding of local bond excitations in molecules and
consider a system of six Morse oscillators coupled through nearest
neighbour couplings on a ring (i.e. with periodic boundary conditions).
Instead of calculating correlation functions Schofield et al
calculate survival probabilities of local excitations and observe
for weak coupling a bound state with approximately five bosons
excited initially on a single oscillator.

\subsection{Systems with boson number conservation}
\label{s17.2}

Much more is known for finite systems with an additional integral of motion
(besides the energy) - the total number of bosons \cite{hs94},%
\cite{esse92},\cite{seg94},\cite{bes90}.
This is so because the infinite dimensional Hilbert space separates
then into an infinite set of finite dimensional noninteracting
subspaces. Each subspace contains only states with the same total
number of bosons. Consequently one deals now with the problem of 
diagonalizing each subspace separately, which requires diagonalizing
only
a finite dimensional matrix as opposed to an infinite dimensional one.
As a result
physically we have to solve only the problem of a finite number of
interacting particles. 

Let us consider the DNLS as a reference system. The classical
equations are given in (\ref{2-5}). The quantum problem
is defined by replacing the complex variables $\psi_l$ 
and $\psi^*_l$ by
an annihilation operator $a_l$ and a creation operator
$a^+_l$ with boson commutation relation $[a_l,a^+_l]=1$
 and choosing the Hamiltonian
\begin{equation}
H= \kappa \sum_l (a_la^+_l)^2 +  \sum_l (a_la^+_{l+1}+
a^+_la_{l+1} )\;\;\;. \label{17-1}
\end{equation}
Note that the second sum in (\ref{17-1}) plays the role of a kinetic
energy.

Bound states exist because of the boson number conservation.
Several bosons initially close to each other can not
separate because that would violate energy conservation. 
The existence of bound states for both attracting
($\kappa < 0$) or repulsive ($\kappa > 0$)
boson-boson interaction is due to the discreteness of the system.
The discreteness limits the kinetic energy from above, and thus
even repelling bosons can form bound states. 

\subsubsection{Two bosons}
\label{s17.21}

Eilbeck et al \cite{egs93} (see also \cite{seg94}) have extensively
considered the case of two bosons in one-dimensional lattices
including the quantum DNLS (\ref{17-1}), in which they found
bound states.
A typical result is shown in Fig. 14
%\begin{figure}
%\caption{from \cite{seg94}}
%end{figure}
where the  band of bound states is well separated from the
continuum of unbounded states. 
In the case of the quantized DNLS we can immediately
find one bound state for $d$-dimensional lattices without frustration.
This bound state is constructed out of a linear combination of
local states with two bosons on one site. The prefactors in the
linear combination have all the same absolute value and opposite
signs for interacting sites. Note that this particular bound state
does not contain any states with the two bosons separated! 
This state corresponds to the $k=\pm 50$ bound states in 
Fig. 14.
Since the two boson problem is essentially a two-body problem,
it is particularly evident that bound states can exist.

\subsubsection{Many bosons}
\label{s17.22}

Less is known for the case of
more than two bosons. The only system studied
extensively is the DNLS with two sites (dimer). 
Note that the discrete translational symmetry is here reduced
to a mere permutational symmetry, i.e. the Hamiltonian is
invariant under permutations of the two different site labels.
This system
is integrable due to the existence of two integrals of motion
(energy and boson number). The classical version can
be completely solved. Bernstein et al \cite{bes90},\cite{ljb93}
and Aubry et al \cite{afko96} have studied the expected splitting
of degenerate pairs of eigenvalues in the quantum system.
The results demonstrate that there is a one-to-one correspondence
between the existence of classical orbits which are not invariant
under permutational symmetry, and the existence of exponentially
small splittings of eigenvalues of corresponding eigenstates in
the quantum problem. Aubry et al \cite{afko96} have moreover
demonstrated that the existence of a bifurcation of periodic orbits
and the existence of a separatrix in the classical model is clearly
reflected in the eigenvalue spectrum  of the quantum model.
Perturbation theories for the splittings of the corner states
in \cite{bes90} and for all states in \cite{afko96} have been
successfully applied.

Practically nothing is known for systems with many lattice
sites and many bosons. Usage of the perturbation theory initially
applied to the dimer suggests exponentially small bandwidths
for quantum breather bands with large boson
numbers \cite{bes90},\cite{sa96}. 

\subsection{Open questions}
\label{s17.3}

So far we lack a proof for the existence of bound states.
An exception are systems with boson number conservation
and two bosons. We think that it should be possible to construct
a rigorous proof for a general case of interacting oscillators.
The analytical continuation of eigenstates and eigenenergies
is easy, because in a linear eigenvalue problem there is no room
for bifurcations (i.e. the number of eigenstates is fixed).
The nontrivial part of a proof must show that certain
properties of eigenstates can be continued, such as the
property of an eigenstate to be a bound state of many
bosons. Starting with zero interaction (antiintegrability
limit) the continuation of low-lying bound states (i.e. where only
a few bosons are involved) should be possible, because there
is a finite distance to eigenvalues of other states.

A much harder problem is the description of many-boson bound states,
because even in the limit of zero interaction of coupled
oscillators there are plenty of other states with nearly
the same eigenvalues. So one has to focus on why the 
overlap with those states is negligible small.

Further numerical diagonalizations following the
promising results of Bishop et al \cite{wgbs96} will
be of help in studying the properties of bound states.
Especially it is interesting to investigate the fate
of states which  in the limit of zero interaction between 
oscillators correspond to several groups of bosons
(e.g. three bosons on one site and two bosons on another site).
These states could correspond to classical multi-site
breather solutions (see \cite{ma94}).

\section{Possible experimental realizations}
\label{s18}

In this section we will briefly discuss the experimental 
situations where discrete breathers can be expected to exist.
The shortness of this
section is not due to expected limitations of applicability,
but rather due to the fact that only few things are
known yet. 

One possible situation to observe discrete breathers seems
to be lattice dynamics of crystals. Anharmonicity is well known
to exist and to contribute to several phenomena, e.g.
heat expansion, structural phase transitions etc. We think
that molecular crystals are particularly promising
candidates. Especially molecular crystals and long molecules 
with intramolecular
interactions being large compared to intermolecular interactions
appear to be similar to the discussed systems of weakly 
interacting oscillators. Fillaux et al \cite{fc90},\cite{fck91},%
\cite{fck92},\cite{fcp94},\cite{fci95} are e.g.
studying the inelastic neutron scattering by the molecular
crystal 4-methyl-pyridine and observe quantum bound states
of phonons. Breathers in polyacetylene were studied
in \cite{bcl84},\cite{bs96} and breathers in conjugated polymers
in \cite{pbh89}. The problem of single bond excitation in molecules
is a field of activity for itself. Many of the concepts recently emerging
there can be unified with the concept of bound states in anharmonic
lattices \cite{tu91},\cite{lsp94}. 

Sievers and Page have reviewed the interrelation between discrete
breathers and impurity modes in alkali halides \cite{sp95}.

Similar effects can be expected for ferromagnetic and
antiferromagnetic systems \cite{bmrr85},\cite{tk92}. 
We expect bound states of spin
excitations to exist. These solutions will be quantum 
counterparts of classical spin breathers which are
similar to the breather solutions of the DNLS.

The dynamics of coupled arrays of Josephson junctions 
is described with the help of a discretized sine Gordon
equation, which is system (\ref{3-1a}) with $\Phi(z)\sim z^2$
and $V(z) \sim {\rm cos} z$. In addition there are 
damping terms. To excite any stationary solution 
an external driving is required.  This is realized by
applying an external current. Orlando et al \cite{wszo95} have
measured the resulting voltage. By applying an ac driving
(ac current)
we expect discrete breathers to be excited. However their
impact will be the creation of
local fluctuations of the voltage, in contrast
to a dc voltage which is typically measured.

Recent results on photonic band gap materials are
of interest with respect to breathers. Photonic
band gap materials are systems with a spatially periodic
modulation of dielectric properties. Typically the
wavelength is of the order of micrometers (visible light).
Theoretical calculations suggest that Maxwell's equations
if solved in such a medium give rise to allowed frequency
bands of travelling light as well as to forbidden frequency
gaps. Although there are some profound differences to
the problem of electrons in a lattice (Schroedinger's
equation) due to some symmetries, the idea is essentially
that what works for electrons should also work for photons.
The interested reader will find details in the reviews
of Pendry \cite{jbpe96}, John \cite{sj93} and Joannopoulos
\cite{jmw95}. 
All calculations that have been done are using
linear optics. Consequently the solutions are always
travelling light waves. The optical periodicity
of the medium causes gaps to occur in the dispersion
relation of light. There has been much discussion that
the addition of defects opens the possibility of localizing
light, and even guiding light along defined channels.
Our point is that if media with optical
nonlinearities \cite{smk95} are used, then localization of light
can appear even without adding defects. This can be
of interest because no permanent changes to the material
have to be done (as in the case of defects).
The reason is that the dispersion relation of light in
a homogeneous medium is $\omega \sim q$. Spatially periodic
modulations yield gaps which are equidistant in $q$ and thus
also equidistant in $\omega$. Thus we can fulfill the nonresonance
condition for localized time-periodic solutions! For one-dimensional
systems Sanchez et al \cite{ssbv92} have demonstrated the existence
of breathers.

Breathers have been observed in electrical lattices
\cite{mbr95}.
We mention the demonstration of discrete breathers
in macroscopic models of pendula with magnetic dipole-dipole
interaction \cite{rzed96}. Discrete breathers can be easily
observed in these systems provided the friction is not too strong.
These systems are extremely easy to make, and can be a nice
tool for demonstration. 

Finally we discuss some properties of discrete breathers
that might be important in designing experiments. The energy
thresholds of discrete breathers in two- and three-dimensional 
lattices can be used - if the supplied energy is below the
threshold, no breathers can be excited, and thus the energy
is dissipated on a certain time scale due to phonons. If the
supplied energy is above the threshold, discrete breathers
are formed, and the decay is now defined by the lifetime of
the discrete breather (which can be orders of magnitude larger
than the typical phonon mediated decay time). 

Another important
feature is the scattering of plane waves by discrete breathers,
which leads to strong
reflection in the presence of a breather. Especially in
one-dimensional systems one can probe the existence or nonexistence
of a discrete breather by testing the transmittion properties
of the system with respect to plane waves.
A similarly interesting event is the possibility of trapped
electron-breather states.

Moving discrete breathers can be of practical interest too.
However we need to know more about movability in first
place.

\section{Conclusions}
\label{s19}

\subsection{Summary}
\label{s19.1}

In this report we discussed existence and properties
of discrete breathers. These solutions  of Hamiltonian
lattices are time-periodic and exponentially localized.

Discrete breathers are generic and structurally stable solutions
because the neccessary nonresonance condition is easily
fulfilled for a lattice. This condition requires the DB frequency
as well as all of its multiples to not resonate with the linear
spectrum of the system. Nonresonance is easy to achieve
because the linear spectrum of a Hamiltonain lattice is bounded.

The nonresonance condition explains why breather solutions
are nongeneric and structurally unstable in the opposing
case of Hamiltonian fields.

Numerical studies show that long-lived localized
excitations typically created in numerical experiments
show up with periodic or quasiperiodic temporal behaviour.
The periodic objects are discrete breathers. The quasiperiodic
objects are phase space perturbed discrete breathers. Numerical
studies suggest that quasiperiodic objects radiate
energy in the form of small-amplitude waves and thus are
not real solutions. Indeed the nonresonance condition is shown
to be never fulfilled for the generic case of a quasiperiodic
solution, although nongeneric exceptions can apply.

Discrete breathers are periodic orbits. For generic Hamiltonian
systems periodic orbits occur in one-parameter families, and
so do discrete breathers. The parameter describing the family
can be the amplitude of a breather, or its frequency, or any other
meaningful observable.

The spatial decay of a discrete breather is typically
exponential. A Fourier series representation of the time-periodic
discrete breather leads to Fourier number dependent exponents
of the spatial decay. A linear theory can account for these
exponents by using the frequency of the breather as an input
parameter. In some cases nonlinear corrections apply, still allowing
for a calculation of the exponents. Numerical calculations
underpin the findings.

The decay of the Fourier components with increasing Fourier
number is obtained. The essential finding are power laws with
exponents depending on the lattice site.

Existence proofs of discrete breathers are reviewed for systems
of weakly interacting oscillators and Fermi-Pasta-Ulam-like
systems. We discussed the antiintegrability approach of Aubry.

Several numerical methods to obtain discrete breather solutions
are reviewed, and their advantages and weak points are discussed.

We discuss structural and dynamical stability of discrete
breathers. Our essential finding is that discrete breathers
are structurally stable, and typically can be dynamically
stable (note that discrete breathers can as easily be also
dynamically unstable).

The movability  of discrete breathers is discussed at length,
illuminating the conceptual problems. The main result is
that a Peierls-Nabarro barrier can not be introduced in the
way it had been done for moving lattice kinks.

The appearance of discrete breathers through bifurcations
of band edge plane waves is discussed. The bifucation
analysis of band edge plane waves for finite systems is
demonstrated, and the results are used to predict the
existence or nonexistence of discrete breathers for
systems without actually solving for the discrete breather.
Instead the band edge plane wave stability is analyzed.

The effect of lattice dimension on discrete breathers is
discussed. Discrete breathers are not limited to certain
dimensions of the lattice. However the lattice dimension
has strong impact on the energy properties of discrete breathers.
For lattice dimensions $d\geq 2$ the energy of discrete breathers
has nonzero lower bounds, leading to an activation energy
of discrete breathers.

We discuss the effect of discrete symmetries on discrete breathers.
Discrete breather solutions are not invariant under discrete
translations. However they can be invariant under other
discrete symmetries (provided the Hamiltonain is).  This has
practical consequencies, since the number of independent Fourier
coefficients is reduced with every additional discrete symmetry.

We refer the reader to section~\ref{s13} for a complete conceptual
summary of the first twelve sections.

We review phonon and electron scattering by discrete breathers.
Discrete breathers turn out to be strong scatterers of
phonons and electrons
and will thus be of interest whenever heat or charge transport
is considered.   

Plane wave trapping is especially important in the case
of electrons. Bound electron-discrete breather states
exist, and can be of conceptual interest in many areas
of solid state theory.

The connection between disorder and nonlinearity with respect
to discrete breathers is briefly discussed. The main point
is that discrete breathers could be directly (or through bifurcations)
connected to the well-known impurity states of linear systems.

The correspondence between classical discrete breathers 
and quantum bound states (quantum breathers) is
discussed and some questions in this new area are raised.

Finally we briefly mention possible experimental realizations
of discrete breathers.

\subsection{Unsolved problems}
\label{s19.2}

We are unable to list all unsolved problems with respect to discrete
breathers.
However in the following we list problems that arose in
this report.

In connection with the discussion of field equations in
the introduction we expect Boyd's nanopteron \cite{jpb90} to
be of further interest. Discrete breathers will most probably
become partially delocalized nanopterons once a higher harmonics
of the breather frequency hits the linear spectrum. In this
sense there can exist a natural connection between discrete breathers
and nanopterons or breathers in some field equations. 
Discrete breathers as well as breathers of field equations
are solutions of finite energy. Nanopterons are solutions of infinite
energy. Staying within the class of finite energy solutions
there is most probably no link between discrete breathers
and breathers of field equations. However including solutions
with infinite energy the missing link might be recovered.

The spatial decay of discrete breathers in systems with 
acoustic spectra has to be studied. The lack of displacement
reversal symmetry leads to the excitation of even Fourier components.
The dc component corresponds to zero frequency and will thus
resonate with the lower bound of the acoustic band of the linear
spectrum. It is not clear (although we presented some arguments)
how this resonance is reflected in the localization properties
of discrete breathers, especially for two- and three-dimensional
lattices. 

A problem related to the above one is the importance of 
polarization vectors for the spatial decay. Suppose a
lattice has more than one degrees of freedom per unit cell.
Then scenarios are possible where optical and acoustic
spectra coexist. What happens if a breather is assumed to
bifurcate off a band edge plane wave of the optical spectrum,
provided it contains a dc component due to lack of symmetry?
Will this dc component be affected by the resonance with the
acoustic spectrum? What is again the influence of lattice
dimension on the findings?

Concerning the dynamical stability the radiation of perturbed
breather solutions (quasiperiodic objects) has to be
quantified. Also the apparent differences in the nonlinear
stability analysis depending on the choosen perturbed solutions
need further investigation.

Do moving breather solutions exist or not? Thus far
we are lacking proofs and complete numerical investigations.

What about the possibility of thermally exciting discrete breathers?
Dauxois et al \cite{dpb93}, Burlakov et al \cite{bkp90-prb}
and Flach and Siewert \cite{fs93} report on numerical evidence
that for some models finite temperature simulations
show the existence of breather-like objects with finite lifetime
(the lifetime in the cited cases was considerably larger than
the breathers inverse frequency, thus the identification is meaningful).
Anomalous slow relaxations in the time-dependent correlator
of energy density fluctuations has been connected with the existence
of breathers \cite{fm94}, \cite{ta96}.
How is the observed lifetime connected to our analysis of
dynamical stability of discrete breathers? 

Further studies are required to clarify the domain of
existence of discrete breathers. Adding a friction will
certainly make discrete breathers disappear. However addition
of ac driving can actually phaselock discrete breathers 
\cite{bam94},\cite{rp96},\cite{cgjr97}.
For some results on phase-locked sG field breathers check
\cite{gk93},\cite{kn78}. 

Much more is needed to be known about the scattering properties,
bound states, quantum breathers. Recent results by
MacKay et al \cite{ms94} and Pikovsky et al \cite{pa96} use
the breather concept for finding corresponding solutions for
networks of weakly coupled bistable units and weakly coupled
symplectic maps respectively. 

And finally we are waiting for serious experimental 
evidence that discrete breathers are detected. This
hope is based on the fact that discrete breathers are
generic solutions of nonlinear Hamiltonian lattices
in one, two, three etc dimensions. In some sense quantum bound
states of bosons and fermions can be regarded as quantum
breathers.

\subsection{Some closing words}
\label{s19.3}

We reviewed a new area of research which is studying
discrete breathers. The aim of the review was to give
the reader both an overview of the achieved results, and to
provide linking thoughts in order to embed the results
into the general field of physics. Most certainly we
forgot to appropriately mention several results, and for this
we sincerely apologize\footnote{An update of DB related publications
is located at the web page: \\
http://www.mpipks-dresden.mpg.de/$^{\sim}$flach/breather.DIR/db.bib}. 
At this place we want to thank many
colleagues for sharing their views and discussing
related problems with us. Especially important were discussions
with S. Aubry, U. Bahr, L. Bernstein, O. Bang, D. Campbell, 
Th. Cretegny, C. Eilbeck, V. Fleurov, P. Fulde, K. Kladko,  
R. Livi, R. S. MacKay,
E. Olbrich, T. Orlando, M. Peyrard, M. Wagner, and many others.
This work was funded in parts by the Deutsche Forschungsgemeinschaft. 
Finally S.F. wants to thank Elja and Nora for the indescribable that
helped so much.

\section{What's new?}

This section is added in print. It refers to the previous
sections in this work and briefly mentions new results which appeared
in the past months after the original work was submitted. 
\\
\\
\\
NEWS IN \ref{s44}
\\
\\
The spatial decay considered in chapter \ref{s44} was considered
for lattices with nearest neighbour interaction. What matters however
is the dispersion of the phonon band at the band edges. Since the
spatial decay of the breather solution far from its center is given
by the Green's function
\begin{equation}
G_{\lambda}(n)=\int \frac{{\rm cos}(kn)}{\lambda^2+\Omega_k^2-1}{\rm d}k
\end{equation}
for large values of $n$, the fast oscillation of the numerator
allows for an expansion of the denominator around the band edges.
If the dispersion is quadratic at the band edge, the poles of 
the Green's function are located on the imaginary axis, and exponential
decay in space follows. If however the interaction is long-range, the
dispersion relation near the band edge will change, e.g. for an
interaction $1/r^s$ in one-dimensional lattices Gaididei et al \cite{gmcr97}
find 
\begin{equation}
\Omega_k^2=1+C\zeta(s-2)k^2 \;\;(s > 3)\;\;,\;\;
\Omega_k^2=1+2Ca(s)k^{s-1} \;\;(1<s<3)\;\;, 
\end{equation}
with $\zeta(s)$ being the Riemann Zeta function. Consequently in this 
example 
\begin{equation}
G_{\lambda}(n) \sim {\rm e}^{-\lambda n} \;\;(s>3)\;\;,\;\;
G_{\lambda}(n) \sim n^{-s} \;\; (1<s<3)\;\;. 
\end{equation}
The algebraic decay in space found for $s < 3$ follows from the fact
that the poles of the Green's function leave the imaginary axis starting
with $s=3$, caused in turn by the change in the dispersion relation
at the band edge.
\\
\\
\\
NEWS IN \ref{s47}
\\
\\
The heuristically predicted algebraic decay of the dc-component
has been recently confirmed for $d=1$ by an analytical proof of
the existence of acoustic breathers \cite{slm97}. The existence
of the static kink structure 
predicted in \ref{s47} and found numerically is proven.
\\
\\
\\
NEWS IN \ref{s51}
\\
\\
As already mentioned,  Livi, MacKay and Spicci \cite{slm97} proved
the existence of acoustic breathers, i.e. breathers in the absence
of on-site potentials. A diatomic chain was considered. The limit
of infinite heavy masses yields a starting point for the continuation
of spatially localized periodic orbits. The general problem is that
the Newton operator is not invertible if considered in the original
variables. This happens because the phonon spectrum degenerates into
two values in the considered limit - one optical value (nonzero frequency)
and one value zero. This zero is removed in \cite{slm97} by considering
new variables. The continuation is only possible if a corresponding
strain field of dc distortions is imposed. This distortion turns out
to be of a simple kink structure. 
\\
\\
\\
NEWS IN \ref{s9}
\\
\\
A recent work by Chen et al \cite{cat96} attempts to create moving
breathers by considering stable stationary breathers with a localized
eigenstate of the corresponding Floquet matrix (orthogonal to the
breather family). Perturbations in the direction of the localized
eigenstate will cross the movability separatrix as the amplitude
of the perturbation exceeds some critical value (this was already
discussed in \cite{fw94}). Chen et al find trajectories similar to
moving breathers over long periods of time.
\\
\\
\\
NEWS IN \ref{s11}
\\
\\
As already mentioned, the spatial decay of the breather is essentially
governed by the phonon dispersion at the band edges. In order to
obtain energy thresholds for breathers, exponential decay in space
is needed, with exponents depending on the breather parameter. Thus
the results from \ref{s11} can be extended to systems with nonlocal
interactions, provided the quadratic dispersion at the band edges 
holds.
\\
\\
\\
NEWS IN \ref{s17.22}
\\
\\
A recent study of a trimer \cite{ff97} considered the fate
of tunneling pairs of the quantum dimer when the phase space of the
classical problem becomes mixed. Tunneling pairs survive avoided
crossings with single states and other tunneling pairs, and get finally
destroyed when the classical phase space which supports these states
turns chaotic. An application of the results to the concept of
doorway states for local bond excitations in molecules confirms
that simplistic arguments like the lifetime of a state being proportional
to the density of nearby lying eigenstates can be very misleading.
\\
\\
\\
NEWS IN \ref{s18}
\\
\\
Bound states of phonons have been measured in the following systems:
Hydrogen on Si(111) surface \cite{pgs91}, crystal CO$_2$ \cite{bssj93},
CO on Ru(001) surface \cite{pj96}. Corresponding measurements for
small molecules (benzene, naphtalene, anthracene) are known for a long
time \cite{sla76}.

A recent work by Floria et al \cite{fmmfa96} demonstrates theoretically
that in an anisotropic ladder realization of Josephson junctions
rotobreathers can be excited, which would lead to the generation
of a dc voltage and thus measurable.

Kisilev et al have calculated breathers in ionic crystals \cite{ks97},
and R\"ossler et al proposed optically controlled generation of
breathers \cite{rp97}.

\newpage

FIGURE CAPTIONS
\\
\\
\\
Fig.1
\\
\\
$\Psi_b(x,t)$ from (\ref{1-2}) versus $x$ for 26 different
times equally spaced and covering one breather period and $m=0.5$.
\\
\\
\\
Fig.2
\\
\\
$e_{(5)}$ versus time (dashed line). Total energy of the chain, solid 
line. Inset: energy distribution $e_l$ versus particle number for the
same solution as in Fig.2 measured for $1000 < t < 1150$.
\\
\\
\\
Fig.3
\\
\\
Fourier transformed FT$\left[u_l(t \geq 1000)\right](\omega)$
with initial condition as in Fig.2 for $l=0$. Inset: for
$l=\pm 1$.
\\
\\
\\
Fig.4
\\
\\
Poincare intersection between the trajectory and the 
subspace $\left[ \dot{u}_1,u_1,u_0=0, \dot{u}_0 > 0 \right]$ for the 
symmetric reduced three-particle problem and energy $E=0.58$.
\\
\\
\\
Fig.5
\\
\\
$e_{(5)}(t)$ dependence. Upper short dashed line - total energy
of all simulations; solid lines (4) - initial conditions of
fixed points in islands 1,2 from Fig.4 and larger torus in
island 1 and torus in island 2 from Fig.4; long dashed line -
initial condition of torus in island 3 in Fig.4; dashed-dotted
line - initial condition of chaotic trajectory in Fig.4.
\\
\\
\\
Fig.6
\\
\\
Energy distribution for the breather solution with initial energy
$E=0.3$ after waiting time $t=3000$. The filled circles
represent the energy values for each particle; the solid lines
are guides to the eye. Inset: Time dependence of the breather
energy $e_{(5)}$.
\\
\\
\\
Fig.7
\\
\\
Numerical solution for the Fourier components of a discrete 
breather with $v_2=1$, $v_4=-1$, $\phi_2=0.1$, $N=100$ and
frequency $\omega = 0.8$. The absolute values of the components
$A_{kl}$ are shown as functions of the lattice site $l$ in a 
window of 30 lattice sites around the breather center. The open
squares are the actual results. The lines are guides to the eye
and connect components with same Fourier number $k$. $k$ increases
from top to bottom as $k=1,3,5,7,...,23,25$. Fourier components with
even $k$ are zero because of the displacement reversal symmetry
of the potential.
\\
\\
\\
Fig.8
\\
\\
Slopes of the lines in Fig.7 as a function of $k$ (correspond
to the exponents of the decay of the corresponding Fourier
components, cf. text) are shown as open squares. The solid line
connects the points of the theoretical prediction using the
eigenvalues of the linearized map (cf. text).
\\
\\
\\
Fig.9
\\
\\
Same as in Fig.5 but longer time scale. Note the rapid decrease
of $e_{(5)}$ for the solid line around $t=6000$. This is due to
an internal resonance in the quasiperiodic breather, which induces
locally chaotic dynamics, changes the local time spectrum from
discrete to continuous and increases the radiation strength
by two orders of magnitude.
\\
\\
\\
Fig.10
\\
\\
Poincare intersection (as in Fig.4) where $u$ and $v$ are the
position and velocity of the nearest neighbour(s) of the central
particle for a two-dimensional lattice (for further details see
\cite{fkw94}). The filled circles are the actual mapping results.
The lines are guides to the eye and connect the circles in the order
of their appearance. The spiral-like form of the broken line
indicates the evolution of the contraction of the breather-like
object to a fixed point which corresponds to an exact time-periodic
discrete breather solution.
\\
\\
\\
Fig.11
\\
\\
Breather energy versus maximum amplitude for the discrete nonlinear
Schr\"odinger system in one, two and three lattice dimensions (cf. text).
Parameters $C=0.1$ and $\mu=4$ for all cases.
System sizes: $d$=1 - $N$=100; $d$=2 - $N$=$25^2$; $d$=3 - $N$=$31^3$.
\\
The estimated points ($A;E$) of bifurcation of the band edge plane wave are:
$d=1$ - (0.014;0.024); $d=2$ - (0.064;5.53); $d=3$ - (0.097; 237).
\\
\\
\\
Fig.12
\\
\\
Discrete energy density distribution versus lattice site $l$ of a
phonon scattering experiment after a waiting time of 12000. The
incident phonon wave (from left) has energy density $e_l=10^{-4}$
and wave number $q=0.2 \pi$. The breather is positioned at $l=1500$.
Since the transmitted part is weak, the reflected wave combines
with the incoming wave into a standing wave with wavelength 
$\lambda/2=5$. The transmitted wave (right part) shows periodic
intensity modulations. Since the group velocity is about 0.04 sites
per unit of time, the frequency of the modulation can be estimated
to be $\approx 0.023$.
\\
\\
\\
Fig.13
\\
\\
The squared absolute value of the transmission coefficient $|t|^2$
as a function of wave number $q$ of the infalling phonon wave.
The value $q=\pi$ corresponds to the Brillouin zone boundary.
The filled circles are the results of numerical experiments
(example in Fig.12). Since the transmitted intensity varies in time,
the filled circles represent the time-averaged intensity.
The solid line is the result of the scattering calculation of
the linearized phase space flow around the discrete breather
by Cretegne (see text). 
\\
\\
\\
Fig.14
\\
\\
Quantum energy levels for the DNLS system with 101 sites and
two bosons (from \cite{egs93}).


\newpage


\begin{thebibliography}{100}

\bibitem{al76}
M.~J. Ablowitz and J.~F. Ladik.
\newblock Nonlinear differential-difference equations and fourier analysis.
\newblock {\em J. Math. Phys.}, 17:1011, 1976.

\bibitem{LHLXI}
E.~Akkermans, G.~Montambaux, J.~L. Pichard, and J.~Zinn-Justin, editors.
\newblock {\em Mesoscopic Quantum Physics}. Elsevier Amsterdam, 1995.

\bibitem{af88}
C.~Albanese and J.~Frohlich.
\newblock Periodic solutions of some infinite-dimensional hamiltonians
  associated with nonlinear partial differential equations i.
\newblock {\em Comm. Math. Phys.}, 116:475, 1988.

\bibitem{af91}
C.~Albanese and J.~Frohlich.
\newblock Perturbation theory for periodic orbits in a class of infinite
  dimensional hamiltonian systems.
\newblock {\em Comm. Math. Phys.}, 138:193, 1991.

\bibitem{afs88}
C.~Albanese, J.~Frohlich, and T.~Spencer.
\newblock Periodic solutions of some infinite-dimensional hamiltonians
  associated with nonlinear partial differential equations ii.
\newblock {\em Comm. Math. Phys.}, 119:677, 1988.

\bibitem{arm91}
L.~M. Alonso, E.~M. Reus, and E.~O. Moreno.
\newblock Breathers in 2+1 dimensions.
\newblock {\em Phys. Lett. A}, 159:384, 1991.

\bibitem{ma92}
M.~Aoki.
\newblock Self-localized mode in a diatomic nonlinear lattice.
\newblock {\em J. Phys. Soc. Japan}, 61:3024, 1992.

\bibitem{at95}
M.~Aoki and S.~Takeno.
\newblock Stationary anharmonic gap modes in the diatomic toda chain.
\newblock {\em J. Phys. Soc. Japan}, 64:809, 1995.

\bibitem{ats93}
M.~Aoki, S.~Takeno, and A.~J. Sievers.
\newblock Stationary anharmonic gap modes in a one-dimensional diatomic lattice
  with quartic anharmonicity.
\newblock {\em J. Phys. Soc. Japan}, 62:4295, 1993.

\bibitem{via89}
V.~I. Arnold.
\newblock {\em Mathematical Methods of Classical Mechanics}.
\newblock Springer-Verlag New York, 1989.

\bibitem{via92}
V.~I. Arnold.
\newblock {\em Ordinary Differential Equations}.
\newblock Springer Berlin, 1992.

\bibitem{sa94}
S.~Aubry.
\newblock The concept of anti-integrability applied to dynamical systems and to
  structural and electronic models in condensed matter physics.
\newblock {\em Physica}, D71:196, 1994.

\bibitem{sa96}
S.~Aubry.
\newblock Breathers in nonlinear lattices: Existence, linear stability and
  quantization.
\newblock {\em Physica D}, in print, 1997.

\bibitem{aa90}
S.~Aubry and G.~Abramovici.
\newblock Chaotic trajectories in the standard map. the concept of
  anti-integrability.
\newblock {\em Physica D}, 43:199, 1990.

\bibitem{afko96}
S.~Aubry, S.~Flach, K.~Kladko, and E.~Olbrich.
\newblock Manifestation of classical bifurcation in the spectrum of the
  integrable quantum dimer.
\newblock {\em Phys. Rev. Lett.}, 76:1607, 1996.

\bibitem{db96}
D.~Bambusi.
\newblock Exponential stability of breathers in hamiltonian networks of weakly
  coupled oscillators.
\newblock {\em Nonlinearity}, 9:433, 1996.

\bibitem{bp95}
O.~Bang and M.~Peyrard.
\newblock Higher order breathers solutions to a discrete nonlinear klein-gordon
  model.
\newblock {\em Physica}, D81:9, 1995.

\bibitem{bp96}
O.~Bang and M.~Peyrard.
\newblock Generation of high-energy localized vibrational modes in nonlinear
  klein-gordon lattices.
\newblock {\em Phys. Rev.}, E53:4143, 1996.

\bibitem{bes90}
L.~Bernstein, J.~C. Eilbeck, and A.~C. Scott.
\newblock The quantum theory of local modes in a coupled system of nonlinear
  oscillators.
\newblock {\em Nonlinearity}, 3:293, 1990.

\bibitem{ljb93}
L.~J. Bernstein.
\newblock Quantizing a self-trapping equation.
\newblock {\em Physica D}, 68:174, 1993.

\bibitem{bks93}
S.~R. Bickham, S.~A. Kisilev, and A.~J. Sievers.
\newblock Stationary and moving intrinsic localized modes in one-dimensional
  monoatomic lattices with cubic and quartic anharmonicity.
\newblock {\em Phys. Rev.}, B47:14206, 1993.

\bibitem{bssj93}
R.~Bini, P.~R.~Salvi, V.~Schettino and H.~J.~Jodl.
\newblock The spectroscopy and relaxation dynamics of three-phonon
bound states in crystal CO$_2$.
\newblock{\em J. Chem. Phys.}, 98:164, 1993.


\bibitem{bb94-2}
B.~Birnir.
\newblock Nonexistence of periodic solutions to hyperbolic partial differential
  equations.
\newblock In M.~Gyllenberg and L.~E. Persson, editors, {\em Analysis, Algebra,
  and Computers in Mathematical Research}, page~43. Marcel Dekker, Inc. New
  York, 1994.

\bibitem{bb94}
B.~Birnir.
\newblock Qualitative analysis of radiating breathers.
\newblock {\em Comm. Pure Appl. Math.}, XLVII:103, 1994.

\bibitem{bcl84}
A.~R. Bishop, D.~C. Campbell, P.~S. Lomdahl, B.~Horowitz, and S.~R. Phillpot.
\newblock Breathers and photoinduced absorption in polyacetylene.
\newblock {\em Phys. Rev. Lett.}, 52:671, 1984.

\bibitem{bs96}
S.~Block and H.~W. Streitwolf.
\newblock Nucleation of optically excited solitrons and breathers in
  trans-polyacetylene.
\newblock {\em J. Phys.: Cond. Matter}, 8:889, 1996.

\bibitem{bp91}
R.~Boesch and M.~Peyrard.
\newblock Discreteness effects on a sine-gordon breather.
\newblock {\em Phys. Rev.}, B43:8491, 1991.

\bibitem{bsw88}
R.~Boesch, P.~Stancioff, and C.~R. Willis.
\newblock Hamiltonian equations for multiple-collective-variable theories of
  nonlinear klein-gordon equations: A projection-operator approach.
\newblock {\em Phys. Rev.}, B38:6713, 1988.

\bibitem{bw89}
R.~Boesch and C.~R. Willis.
\newblock Exact determination of the peierls-nabarro frequency.
\newblock {\em Phys. Rev.}, B39:361, 1989.

\bibitem{bw92}
R.~Boesch and C.~R. Willis.
\newblock Removal of singularities from collective-variable theory by
  incorporating relativistic invariance.
\newblock {\em Phys. Rev.}, A45:5422, 1992.

\bibitem{bms95-prb}
D.~Bonart, A.~P. Mayer, and U.~Schr\"oder.
\newblock Anharmonic localized surface vibrations in a scalar model.
\newblock {\em Phys. Rev.}, B51:13739, 1995.

\bibitem{bms95-prl}
D.~Bonart, A.~P. Mayer, and U.~Schr\"oder.
\newblock Intrinsic localized anharmonic modes at crystal edges.
\newblock {\em Phys. Rev. Lett.}, 75:870, 1995.

\bibitem{bmrr85}
J.~P. Boucher, F.~Mezei, L.~P. Regnault, and J.~P. Renard.
\newblock Diffusion of solitons in the antiferromagnetic chains of
  $(cd_3)_4nmncl_3$: A study by neutron spin echo.
\newblock {\em Phys. Rev. Lett.}, 55:1778, 1985.

\bibitem{jpb90}
J.~P. Boyd.
\newblock A numerical calculation of a weakly non-local solitary wave: the cap
  phi 4 breather.
\newblock {\em Nonlinearity}, 3:177, 1990.

\bibitem{bk91}
V.~M. Burlakov and S.~A. Kisilev.
\newblock Molecular-dynamics simulation of the decay kinetics of uniform
  excitation of an anharmonic 1d chain.
\newblock {\em JETP}, 72:854, 1991.

\bibitem{bkp90-prb}
V.~M. Burlakov, S.~A. Kisilev, and V.~N. Pyrkov.
\newblock Computer simulation of intrinsic localized modes in one-dimensional
  and two-dimensional anharmonic lattices.
\newblock {\em Phys. Rev. B}, 42:4921, 1990.

\bibitem{bkp90-ssc}
V.~M. Burlakov, S.~A. Kisilev, and V.~N. Pyrkov.
\newblock Computer simulations of intrinsic localized modes in 1-d anharmonic
  lattices.
\newblock {\em Solid State Comm.}, 74:327, 1990.

\bibitem{bkr90-jetpl}
V.~M. Burlakov, S.~A. Kisilev, and V.~I. Rupasov.
\newblock Localized excitations of uniform anharmonic lattices.
\newblock {\em JETP Lett.}, 51:544, 1990.

\bibitem{bkr90-pla}
V.~M. Burlakov, S.~A. Kisilev, and V.~I. Rupasov.
\newblock Localized vibrations of homogeneous anharmonic chains.
\newblock {\em Phys. Lett.}, A147:130, 1990.

\bibitem{cbg95}
D.~Cai, A.~R. Bishop, and N.~Gr$\o$nbech-Jensen.
\newblock Spatially localized, temporally quasiperiodic, discrete nonlinear
  excitations.
\newblock {\em Phys. Rev.}, E52:R5784, 1995.

\bibitem{cp90}
D.~K. Campbell and M.~Peyrard.
\newblock {\em in: CHAOS - Soviet American Perspectives on Nonlinear Science,
  ed. by D. K. Campbell}.
\newblock American Institute of Physics New York, 1990.

\bibitem{cat96}
D.~Chen, S.~Aubry and G.~P.~Tsironis.
\newblock Breather mobility in discrete $\phi^4$ nonlinear lattices.
\newblock{\em Phys. Rev. Lett.}, 77:4776, 1996.



\bibitem{bvc79}
B.~V. Chirikov.
\newblock A universal instability of many-dimensional oscillator systems.
\newblock {\em Physics Reports}, 52:263, 1979.

\bibitem{cgjr97}
P.~L.~Christiansen, Yu,~B.~Gaididei, M.~Johansson and K.~O.~Rasmussen.
\newblock Breatherlike excitations in discrete lattices with noise
and nonlinear damping.
\newblock{\em Phys. Rev. B}, 55:5759, 1997.


\bibitem{ck93}
O.~A. Chubykalo and Y.~S. Kivshar.
\newblock Kink-profile vibrational modes in one-dimensional nonlinear lattices.
\newblock {\em Phys. Lett.}, A178:123, 1993.

\bibitem{cku93}
O.~A. Chubykalo, A.~S. Kovalev, and O.~V. Usatenko.
\newblock Stability of intrinsic localied modes in anharmonic 1-d lattices.
\newblock {\em Phys. Lett. A}, 178:129, 1993.

\bibitem{ckks93}
C.~Claude, Yu.~S. Kivshar, O.~Kluth, and K.~H. Spatschek.
\newblock Moving localized modes in nonlinear lattices.
\newblock {\em Phys. Rev.}, B47:14228, 1993.

\bibitem{tc96}
T.~Cretegny, S. Aubry and S. Flach.
\newblock {\em in preparation}, 1997.

\bibitem{dhn74}
R.~F. Dashen, B.~Hasslacher, and A.~Neveu.
\newblock Nonperturbative methods and extended-hadron models in field theory
  ii.
\newblock {\em Phys. Rev. D}, 10:4130, 1974.

\bibitem{dpb93}
T.~Dauxois, M.~Peyrard, and A.~R. Bishop.
\newblock Dynamics and thermodynamics of a nonlinear model for dna
  denaturation.
\newblock {\em Phys. Rev. E}, 47:684, 1993.

\bibitem{dpw92}
T.~Dauxois, M.~Peyrard, and C.~R. Willis.
\newblock Localized breather-like solution in a discrete klein-gordon model and
  application to dna.
\newblock {\em Physica D}, 57:267, 1992.

\bibitem{dpw93}
T.~Dauxois, M.~Peyrard, and C.~R. Willis.
\newblock Discreteness effects on the formation and propagation of breathers in
  nonlinear klein-gordon equations.
\newblock {\em Phys. Rev.}, E48:4768, 1993.

\bibitem{asd73}
A.~S. Davydov.
\newblock {\em Quantum Mechanics}.
\newblock Pergamon Press, New York, 1976.

\bibitem{jd93}
J.~Denzler.
\newblock Nonpersistence of breather families for the perturbed sine gordon
  equation.
\newblock {\em Commun. Math. Phys.}, 158:397, 1993.

\bibitem{jd95}
J.~Denzler.
\newblock Second order nonpersistence of the sine-gordon breather under an
  exceptional perturbation.
\newblock {\em Annales de l'institut henri poincare}, 12:201, 1995.

\bibitem{jd69}
J.~Deudonne.
\newblock {\em Foundations of Modern Analysis}.
\newblock Academic New York, 1969.

\bibitem{dbcw92}
P.~Tchofo Dinda, R.~Boesch, E.~Coquet, and C.~R. Willis.
\newblock Discreteness effects on the double-quadratic kink.
\newblock {\em Phys. Rev.}, B46:3311, 1992.

\bibitem{degm82}
R.~K. Dodd, J.~C. Eilbeck, J.~D. Gibbon, and H.~C. Morris.
\newblock {\em Solitons and Nonlinear Wave Equations}.
\newblock Academic Press New York, 1982.

\bibitem{defw93}
D.~B. Duncan, J.~C. Eilbeck, H.~Feddersen, and J.~A.~D. Wattis.
\newblock Solitons on lattices.
\newblock {\em Physica D}, 68:1, 1993.

\bibitem{dvw96}
R.~Dusi, G.~Viliani and M.~Wagner.
\newblock Breathing self-localized solitons in the quartic
Fermi-Pasta-Ulam chain.
\newblock{\em Phys. Rev. B}, 54:9809, 1996.


\bibitem{dw96}
R.~Dusi and M.~Wagner.
\newblock A gauss procedure for the construction of localized solitons in
  anharmonic chains.
\newblock {\em Physica}, B219-220:393, 1996.

\bibitem{ej88}
W.~Ebeling and M.~Jenssen.
\newblock Trapping and fusion of solitons in a nonuniform toda lattice.
\newblock {\em Physica D}, 32:183, 1988.

\bibitem{ene83}
E.~N. Economou.
\newblock {\em Green's Functions in Quantum Physics}.
\newblock Springer-Verlag Heidelberg, 1983.

\bibitem{egs93}
J.~C. Eilbeck, H.~Gilh$\o$j, and A.~C. Scott.
\newblock Soliton bands in anharmonic quantum lattices.
\newblock {\em Phys. Lett. A}, 172:229, 1993.

\bibitem{els85}
J.~C. Eilbeck, P.~S. Lomdahl, and A.~C. Scott.
\newblock The discrete self-trapping equation.
\newblock {\em Physica D}, 16:318, 1985.

\bibitem{ekns84}
V.~M. Eleonskii, N.~E. Kulagin, N.~S. Novozhilova, and V.~P. Shilin.
\newblock Asymptotic expansions and qualitative analysis of finite-dimensional
  models in nonlinear field theory.
\newblock {\em Teor. Mat. Fiz.}, 60:395, 1984.

\bibitem{emm89}
S.~Elytin, A.Maimistov, and E.~Manykin.
\newblock Numerical study of fast oscillating breathers in the self-induced
  transparency phenomenon.
\newblock {\em Phys. Lett. A}, 142:493, 1989.

\bibitem{esse92}
V.~Z. Enolskii, M.~Salerno, A.~C. Scott, and J.~C. Eilbeck.
\newblock There's more than one way to skin schr\"odinger's cat.
\newblock {\em Physica D}, 59:1, 1992.

\bibitem{fc90}
F.~Fillaux and C.~J. Carlile.
\newblock Inelastic-neutron-scattering study of methyl tunneling and the
  quantum sine-gordon breather in isotopic mixtures of 4-methyl-pyridine at low
  temperature.
\newblock {\em Phys. Rev. B}, 42:5990, 1990.

\bibitem{fci95}
F.~Fillaux, C.~J. Carlile, and A.~Inaba.
\newblock Relaxation kinetics of the sine-gordon breather mode in
  4-methyl-pyridine crystal at low temperature.
\newblock {\em Physica B}, 213 - 214:646, 1995.

\bibitem{fck91}
F.~Fillaux, C.~J. Carlile, and G.~J. Kearley.
\newblock Inelastic-neutron-scattering study at low temperature of the quantum
  sine-gordon breather in 4-methyl-pyridine with partially deuterated methyl
  groups.
\newblock {\em Phys. Rev. B}, 44:12280, 1991.

\bibitem{fck92}
F.~Fillaux, C.~J. Carlile, and G.~J. Kearley.
\newblock Inelastic neutron scattering study at low temperature of the quantum
  sine-gordon breather in 4-methyl-pyridine with partially deuterated methyl
  groups.
\newblock {\em Physica B}, 180 - 181:642, 1992.

\bibitem{fcp94}
F.~Fillaux, C.~J. Carlile, and M.~Prager.
\newblock Inelastic neutron-scattering study of methyl tunneling and the
  quantum sine-gordon breather mode in isotopic mixtures of
  2,6-dimethyl-pyridine at low temperature.
\newblock {\em Physica B}, 202:302, 1994.

\bibitem{ff93-ap}
F.~Fischer.
\newblock Self-localized single-anharmonic vibrational modes in two-dimensional
  lattices.
\newblock {\em Ann. Physik}, 2:296, 1993.

\bibitem{ff93-pla}
F.~Fischer.
\newblock Three-particle quartic-power dynamics: an integrable model.
\newblock {\em Phys. Lett.}, A182:417, 1993.

\bibitem{sf94}
S.~Flach.
\newblock Conditions on the existence of localized excitations in nonlinear
  discrete systems.
\newblock {\em Phys. Rev. E}, 50:3134, 1994.

\bibitem{sf95-pre-1}
S.~Flach.
\newblock Existence of localized excitations in nonlinear hamiltonian lattices.
\newblock {\em Phys. Rev. E}, 51:1503, 1995.

\bibitem{sf95-pre-2}
S.~Flach.
\newblock Obtaining breathers in nonlinear hamiltonian lattices.
\newblock {\em Phys. Rev. E}, 51:3579, 1995.

\bibitem{sf96-2}
S.~Flach.
\newblock Existence and properties of discrete breathers.
\newblock In E.~Alfinito, M.~Boiti, L.~Martina, and F.~Pempinelli, editors,
  {\em Nonlinear Physics Theory and Experiment}, page 390. World Scientific
  Singapore, 1996.

\bibitem{sf96}
S.~Flach.
\newblock Tangent bifurcation of band edge plane waves, dynamical symmetry
  breaking and vibrational localization.
\newblock {\em Physica}, D91:223, 1996.

\bibitem{ff97}
S.~Flach and V.~Fleurov.
\newblock Tunneling in the nonintegrable trimer - a step towards
quantum breathers.
\newblock{\em J. Phys.: Cond. Mat.}, accepted, 1997.


\bibitem{fk96}
S.~Flach and K.~Kladko.
\newblock Interaction of discrete breathers with electrons in nonlinear
  lattices.
\newblock {\em Phys. Rev.}, B53:11531, 1996.

\bibitem{fkm96}
S.~Flach, K.~Kladko, and R.~S. MacKay.
\newblock Energy thresholds of discrete breathers in one-, two- and
  three-dimensional lattices.
\newblock {\em Phys. Rev. Lett.}, 78:1207, 1997.

\bibitem{fkw94}
S.~Flach, K.~Kladko, and C.~R. Willis.
\newblock Localized excitations in two-dimensional lattices.
\newblock {\em Phys. Rev. E}, 50:2293, 1994.

\bibitem{fm94}
S.~Flach and G.~Mutschke.
\newblock Slow relaxation and phase space properties of a conservative system
  with many degrees of freedom.
\newblock {\em Phys. Rev. E}, 49:5018, 1994.

\bibitem{fs93}
S.~Flach and J.~Siewert.
\newblock Fast and slow dynamics in the one-dimensional $\phi^4$ lattice model
  - a molecular dynamics study.
\newblock {\em Phys. Rev.}, B47:14910, 1993.

\bibitem{fw93}
S.~Flach and C.~R. Willis.
\newblock Nonlinear localized excitations in a discrete klein-gordon model.
\newblock {\em Phys. Lett. A}, 181:232, 1993.

\bibitem{fw94}
S.~Flach and C.~R. Willis.
\newblock Movability of localized excitations in nonlinear discrete systems.
\newblock {\em Phys. Rev. Lett.}, 72:1777, 1994.

\bibitem{fw95}
S.~Flach and C.~R. Willis.
\newblock {\em in: Nonlinear Excitations in Biomolecules, ed. by M. Peyrard}.
\newblock Editions de Physique, Springer-Verlag, 1995.

\bibitem{fwo94}
S.~Flach, C.~R. Willis, and E.~Olbrich.
\newblock Integrability and localized excitations in nonlinear discrete
  systems.
\newblock {\em Phys. Rev. E}, 49:836, 1994.

\bibitem{fmmfa96}
L.~M.~Floria, J.~L.~Marin, P.~J.~Martinez, F.~Falo and S.~Aubry.
\newblock Intrinsic localization in the dynamics of a Josephson-junction
ladder.
\newblock{\em Europhys. Lett.}, 36:539, 1996.


\bibitem{fpr85}
N.~Flytzanis, S.~Pnevmatikos, and M.~Remoissenet.
\newblock Kink, breather and assymmetric envelope or dark solitons in nonlinear
  chains: I. monoatomic chain.
\newblock {\em J. Phys. C: Solid State Phys.}, 18:4603, 1985.

\bibitem{fpm94}
K.~Forinash, M.~Peyrard, and B.~A. Malomed.
\newblock Interaction of discrete breathers with impurity modes.
\newblock {\em Phys. Rev. E}, 49:3400, 1994.

\bibitem{fsw86}
J.~Frohlich, T.~Spencer, and C.~E. Wayne.
\newblock Localization in disordered nonlinear dynamical systems.
\newblock {\em J. Stat. Phys.}, 42:247, 1986.

\bibitem{pf95}
P.~Fulde.
\newblock {\em Electron Correlations in Molecules and Solids}.
\newblock Springer Berlin, 1995.


\bibitem{gmcr97}
Yu.~B.~Gaididei, S.~F.~Mingaleev, P.~L.~Christiansen and K.~\O.~Rasmussen.
\newblock Effects of nonlocal dispersive interactions on self-trapping
excitations.
\newblock{\em preprint}, 1997.


\bibitem{jg94}
J.~Geicke.
\newblock Logarithmic decay of phi4 breathers of energy less than or equal to
  1.
\newblock {\em Phys. Rev.}, E49:3539, 1994.

\bibitem{gj75}
J.~Goldstone and R.~Jackiw.
\newblock Quantization of nonlinear waves.
\newblock {\em Phys. Rev. D}, 11:1486, 1975.

\bibitem{gk93}
R.~Grauer and Yu.~S. Kivshar.
\newblock Chaotic and phase-locked breather dynamics in the damped and
  parametrically driven sine-gordon equation.
\newblock {\em Phys. Rev. E}, 48:4791, 1993.

\bibitem{gk92}
S.~A. Gredeskul and Yu.~S. Kivshar.
\newblock Propagation and scattering of nonlinear waves in disordered systems.
\newblock {\em Physics Reports}, 216:1, 1992.

\bibitem{pgs91}
P.~Guyot-Sionnest.
\newblock Two-phonon bound state for the Hydrogen vibration on the H/Si(111)
surface.
\newblock{\em Phys. Rev. Lett.}, 67:2323, 1991.


\bibitem{hs94}
M.~H. Hays and A.~C. Scott.
\newblock Quantizing the discrete self-trapping equation.
\newblock {\em Phys. Lett. A}, 188:21, 1994.

\bibitem{dh96}
D.~Hennig, K. O. Rasmussen, H. Gabriel and A. B\"ulow.
\newblock Solitonlike solutions of the generalized discrete nonlinear
Schr\"odinger equation.
\newblock {\em Phys. Rev. E}, 54:5788, 1996.

\bibitem{hrtg95}
D.~Hennig, K.~$\O$. Rasmussen, G.~P. Tsironis, and H.~Gabriel.
\newblock Breatherlike impurity modes in discrete nonlinear lattices.
\newblock {\em Phys. Rev. B}, 52:R4628, 1995.

\bibitem{hw95}
M.~Hisakado and M.~Wadati.
\newblock Dynamics of breather modes in a nonlinear model of dna with
  discontinuity.
\newblock {\em J. Phys. Soc. Japan}, 64:1910, 1995.

\bibitem{kh93}
K.~Hori.
\newblock Wavelet analysis of anharmonic self-localized modes.
\newblock {\em J. Phys. Soc. Japan}, 62:1819, 1993.

\bibitem{ht92-jpsj-2}
K.~Hori and S.~Takeno.
\newblock Low-frequency and high-frequency moving anharmonic localized modes in
  a one-dimensional lattice with quartic anharmonicity.
\newblock {\em J. Phys. Soc. Japan}, 61:4263, 1992.

\bibitem{ht92-jpsj-1}
K.~Hori and S.~Takeno.
\newblock Moving self-localized modes for the displacement field in a
  one-dimensional lattice system with quartic anharmonicity.
\newblock {\em J. Phys. Soc. Japan}, 61:2186, 1992.

\bibitem{ht92-pla}
K.~Hori and S.~Takeno.
\newblock Soliton scattering by an impurity on a nonlinear lattice.
\newblock {\em Phys. Lett.}, A169:355, 1992.

\bibitem{hsx93}
G.~Huang, Z.~Shi, and Z.~Xu.
\newblock Asymmetric intrinsic localized modes in a homogeneous lattice with
  cubic and quartic anharmonicity.
\newblock {\em Phys. Rev.}, B47:14561, 1993.

\bibitem{pj96}
P.~Jakob.
\newblock Dynamics of the C-O stretch overtone vibration of
CO/Ru(001).
\newblock{\em Phys. Rev. Lett.}, 77:4229, 1996.


\bibitem{jmw95}
J.~D. Joannopoulos, R.~D. Meade, and J.~N. Winn.
\newblock {\em Photonic Crystals}.
\newblock Princeton University Press Princeton, 1995.

\bibitem{sj93}
S.~John.
\newblock The localization of light.
\newblock In C.~M. Soukoulis, editor, {\em Photonic Band Gaps and
  Localization}, page 1993. Plenum New York, 1993.

\bibitem{jm73}
W.~Jones and N.~H. March.
\newblock {\em Theoretical Solid State Physics}.
\newblock Dover Publications New York, 1973.

\bibitem{kn78}
J.~D. Kaup and A.~Newell.
\newblock Theory of nonlinear oscillating dipolar excitations in
  one-dimensional condensates.
\newblock {\em Phys. Rev. B}, 18:5162, 1978.

\bibitem{sk91}
S.~Kichenassamy.
\newblock Breather solutions of the nonlinear wave equation.
\newblock {\em CPAM}, 44:789, 1991.

\bibitem{sak90}
S.~A. Kisilev.
\newblock Stationary vibrational modes if a chain of particles interacting via
  an even order potential.
\newblock {\em Phys. Lett.}, A148:95, 1990.

\bibitem{kbs93}
S.~A. Kisilev, S.~R. Bickham, and A.~J. Sievers.
\newblock Anharmonic gap modes in a perfect one-dimensional diatomic lattice
  for standard two-body nearest-neighbor potentials.
\newblock {\em Phys. Rev. B}, 48:13508, 1993.

\bibitem{kbs94-prb}
S.~A. Kisilev, S.~R. Bickham, and A.~J. Sievers.
\newblock Anharmonic gap mode in a one-dimensional diatomic lattice with
  nearest-neighbor born-mayer-coulomb potentials and its interaction with a
  mass-defect impurity.
\newblock {\em Phys. Rev.}, B50:9135, 1994.

\bibitem{kbs94-pla}
S.~A. Kisilev, S.~R. Bickham, and A.~J. Sievers.
\newblock Anharmonic impurity modes in a 1-d lattice with two-body potentials.
\newblock {\em Phys. Lett.}, A184:255, 1994.

\bibitem{kbs94-jl}
S.~A. Kisilev, S.~R. Bickham, and A.~J. Sievers.
\newblock Localized anharmonic defect modes in a one-dimensional lattice with
  standard two-body nearest-neighbor potentials.
\newblock {\em Journal of Luminescence}, 58:23, 1994.

\bibitem{kr90}
S.~A. Kisilev and V.~I. Rupasov.
\newblock Stationary vibrational modes of a polyatomic chain of particles
  interacting via an even order potential.
\newblock {\em Phys. Lett.}, A148:355, 1990.

\bibitem{ks97}
S.~A.~Kisilev and A.~J.~Sievers.
\newblock Generation of intrinsic vibrational gap modes in three-dimensional
ionic crystals.
\newblock{\em Phys. Rev. B}, 55:5755, 1997.


\bibitem{kit92}
T.~Kitamura and S.~Takeno.
\newblock Solitons and bound states of the self-consistent potential by the
  boson transformation method.
\newblock {\em Phys. Lett.}, A172:184, 1992.

\bibitem{kp92}
Y.~S. Kivshar and M.~Peyrard.
\newblock Modulational instabilities in discrete lattices.
\newblock {\em Phys. Rev.}, A46:3198, 1992.

\bibitem{ysk93-pre}
Yu.~S. Kivshar.
\newblock Intrinsic localized modes as solitons with compact support.
\newblock {\em Phys. Rev. E}, 48:R43, 1993.

\bibitem{ysk93-pla}
Yu.~S. Kivshar.
\newblock Localized modes in a chain with nonlinear on-site potential.
\newblock {\em Phys. Lett.}, A173:172, 1993.

\bibitem{ysk93-prb}
Yu.~S. Kivshar.
\newblock Nonlinear impurity modes in a lattice.
\newblock {\em Phys. Rev. B}, 47:11167, 1993.

\bibitem{kc93}
Yu.~S. Kivshar and D.~K. Campbell.
\newblock Peierls-nabarro potential barrier for highly localized nonlinear
  modes.
\newblock {\em Phys. Rev. E}, 48:3077, 1993.

\bibitem{kt92}
Yu.~S. Kivshar and S.~K. Turitsyn.
\newblock Lattice solitons on a standing carrier wave.
\newblock {\em Phys. Lett. A}, 171:344, 1992.

\bibitem{kk74}
A.~M. Kosevich and A.~S. Kovalev.
\newblock Selflocalization of vibrations in a one-dimensional anharmonic chain.
\newblock {\em JETP}, 67:1793, 1974.

\bibitem{lks96}
E.~W. Laedke, O.~Kluth, and K.~H. Spatschek.
\newblock Existence of solitary solutions in nonlinear chains.
\newblock 
  {\em Phys. Rev. E}, 54:4299, 1996.

\bibitem{LLVII}
L.~D. Landau and E.~M. Lifshitz.
\newblock {\em Elastizit\"atstheorie, Lehrbuch der Theoretischen Physik VII}.
\newblock Akademie-Verlag Berlin, 1991.

\bibitem{LLIII}
L.~D. Landau and E.~M. Lifshitz.
\newblock {\em Quantenmechanik, Lehrbuch der Theoretischen Physik III}.
\newblock Akademie-Verlag Berlin, 1991.

\bibitem{lsp94}
K.~K. Lehman, G.~Scoles, and B.~H. Pate.
\newblock Intramolecular dynamics from eigenstate-resolved infrared spectra.
\newblock {\em Ann. Rev. Phys. Chem.}, 45:241, 1994.

\bibitem{slm97}
R.~Livi, R.~S.~MacKay and M.~Spicci.
\newblock Breathers on a diatomic FPU chain.
\newblock{\em Nonlinearity}, submitted, 1997.



\bibitem{ma94}
R.~S. MacKay and S.~Aubry.
\newblock Proof of existence of breathers for time-reversible or hamiltonian
  networks of weakly coupled oscillators.
\newblock {\em Nonlinearity}, 7:1623, 1994.

\bibitem{ms94}
R.~S. MacKay and J.~A. Sepulchre.
\newblock Multistability in networks of weakly coupled bistable units.
\newblock {\em Warwick preprint}, 44, 1994.

\bibitem{bam94}
B.~A. Malomed.
\newblock Damping and pumping of localized intrinsic modes in nonlinear
  dynamical lattices.
\newblock {\em Phys. Rev.}, B49:5962, 1994.

\bibitem{ma96}
J.~L. Marin and S.~Aubry.
\newblock Breathers in nonlinear lattices: numerical continuation from
the anticontinuous limit.
\newblock {\em Nonlinearity}, 9:1501, 1996.

\bibitem{mbr95}
P.~Marquie, J.~M. Bilbaut, and M.~Remoissenet.
\newblock Observation of nonlinear localized modes in an electrical lattice.
\newblock {\em Phys. Rev.}, E51:6127, 1995.

\bibitem{bml95}
J.~Carr and B.~McLeod.
\newblock Solitary waves on lattices.
\newblock {\em preprint}, 1997.

\bibitem{ahn93}
A.~H. Nayfeh.
\newblock {\em Introduction to Perturbation Techniques}.
\newblock John Wiley and Sons, New York, 1993.

\bibitem{nmf94}
A.~Neuper, F.~G. Mertens, and N.~Flytzanis.
\newblock Quasicontinuum approximation and iterative method for envelope
  solitons in anharmonic chains.
\newblock {\em Z. Phys.}, B95:397, 1994.

\bibitem{jbpe96}
J.~B. Pendry.
\newblock Calculating photonic band structure.
\newblock {\em J. Phys.: Condens. Matter}, 8:1085, 1996.

\bibitem{pp96}
P.~Perfetti.
\newblock An infinite-dimensional extension of a poincare's result concerning
  the continuation of periodic orbits.
\newblock {\em preprint}, 1996.

\bibitem{pbh89}
S.~R. Phillpot, A.~R. Bishop, and B.~Horovitz.
\newblock Amplitude breathers in conjugated polymers.
\newblock {\em Phys. Rev. B}, 40:1839, 1989.

\bibitem{pa96}
M. Abel, A.~Pikovsky and S. Flach.
\newblock {\em in preparation}, 1997.

\bibitem{numrec}
W.~H. Press, B.~P. Flannery, S.~A. Teukolsky, and W.~T. Vetterling.
\newblock {\em Numerical Recipes : the art of scientific computing}.
\newblock Cambridge University Press, 1986.

\bibitem{mr86}
M.~Remoissenet.
\newblock Low-amplitude breather and envelope solitons in quasi-one-dimensional
  physical models.
\newblock {\em Phys. Rev.}, B33:2386, 1986.

\bibitem{rp96}
T.~R\"ossler and J.~B. Page.
\newblock Driven intrinsic localized modes and their stability in anharmonic
  lattices with realistic potentials.
\newblock {\em Physica}, B219-220:387, 1996.

\bibitem{rp97}
T.~R\"ossler and J.~B.~Page.
\newblock Creation of intrinsic localized modes via control
of anharmonic lattices.
\newblock{\em Phys. Rev. Lett.}, 78:1287, 1997.


\bibitem{rzed96}
F.~M. Russell, Y.~Zolotaryuk, J.~C. Eilbeck, and T.~Dauxois.
\newblock Moving breathers in a chain of magnetic pendulums.
\newblock {\em Phys. Rev. B}, 55:6304, 1997.

\bibitem{suz88}
R.~S. Sagdeev, D.~A. Usikov, and G.~M. Zaslavski.
\newblock {\em Nonlinear Physics: from the Pendulum to Turbulence and Chaos}.
\newblock Harwood Academic Publishers, 1988.

\bibitem{ssbv92}
A.~Sanchez, R.~Scharf, A.~R. Bishop, and L.~Vasquez.
\newblock Sine-gordon breathers on spatially periodic potentials.
\newblock {\em Phys. Rev. A}, 45:6031, 1992.

\bibitem{sp94}
K.~W. Sandusky and J.~B. Page.
\newblock Interrelation between the stability of extended normal modes and the
  existence of intrinsic localized modes in nonlinear lattices with realistic
  potentials.
\newblock {\em Phys. Rev. B}, 50:866, 1994.

\bibitem{sps92}
K.~W. Sandusky, J.~B. Page, and K.~E. Schmidt.
\newblock Stability and motion of intrinsic localized modes in nonlinear
  periodic lattices.
\newblock {\em Phys. Rev.}, B46:6161, 1992.

\bibitem{sww96}
S.~A. Schofield, R.~E. Wyatt, and P.~G. Wolynes.
\newblock Computational study of many-dimensional quantum vibrational energy
  redistribution. i. statistics of the survival probability.
\newblock {\em J. Chem. Phys.}, 105:940, 1996.

\bibitem{seg94}
A.~C. Scott, J.~C. Eilbeck, and H.~Gilh$\o$j.
\newblock Quantum lattice solitons.
\newblock {\em Physica D}, 78:194, 1994.

\bibitem{sk87}
H.~Segur and M.~D. Kruskal.
\newblock Nonexistence of small-amplitude breather solutions in $\phi^4$
  theory.
\newblock {\em Phys. Rev. Lett.}, 58:747, 1987.

\bibitem{sh91}
Z.~Shi and G.~Huang.
\newblock Envelope-kink phonon localized modes in a homogeneous anharmonic
  chain.
\newblock {\em Phys. Rev.}, B44:12601, 1991.

\bibitem{sp95}
A.~J. Sievers and J.~B. Page.
\newblock Unusual anharmonic local mode systems.
\newblock In G.~K. Horton and A.~A. Maradudin, editors, {\em Dynamical
  Properties of Solids VII Phonon Physics The Cutting Edge}, page 137. Elsevier
  Amsterdam, 1995.

\bibitem{st88}
A.~J. Sievers and S.~Takeno.
\newblock Intrinsic localized modes in anharmonic crystals.
\newblock {\em Phys. Rev. Lett.}, 61:970, 1988.

\bibitem{smk95}
A.~W. Snyder, D.~J. Mitchell, and Yu.~S. Kivshar.
\newblock Unification of linear and nonlinear wave optics.
\newblock {\em Mod. Phys. Lett. B}, 9:1479, 1995.

\bibitem{sweb86}
P.~Stancioff, C.~R. Willis, M.~El-Batanouny, and S.~Burdick.
\newblock Sine-gordon kinks on a discrete lattice. ii. static properties.
\newblock {\em Phys. Rev.}, B33:1912, 1986.

\bibitem{sla76}
R.~L.~Swofford, M.~E.~Long and A.~C.~Albrecht.
\newblock C-H vibrational states of benzene, naphtalene, and
anthracene in the visible region by thermal lensing spectroscopy
and the local mode model.
\newblock{\em J. Chem. Phys.}, 65::179, 1976.


\bibitem{st90-jpsj-3}
S.~Takeno.
\newblock P-like stationary self-localized modes in pure one-dimensional
  lattice with quartic lattice anharmonicity.
\newblock {\em J. Phys. Soc. Japan}, 59:3861, 1990.

\bibitem{st90-jpsj-2}
S.~Takeno.
\newblock Quantum theory of vibron solitons - coherent states of a
  vibron-phonon system and self-localized modes -.
\newblock {\em J. Phys. Soc. Japan}, 59:3127, 1990.

\bibitem{st92-jpsj-2}
S.~Takeno.
\newblock Theory of stationary anharmonic localized modes in solids.
\newblock {\em J. Phys. Soc. Japan}, 61:2821, 1992.

\bibitem{thom90}
S.~Takeno and S.~Homma.
\newblock Self-localized anharmonic rotational modes of bases in dna.
\newblock {\em J. Phys. Soc. Japan}, 59:1890, 1990.

\bibitem{thom91}
S.~Takeno and S.~Homma.
\newblock Propagation of a soliton and a nonlinear self-localized state in a
  one-dimensional disordered nonlinear lattice.
\newblock {\em J. Phys. Soc. Japan}, 60:731, 1991.

\bibitem{th93}
S.~Takeno and S.~Homma.
\newblock Robust nature of dispersionless envelope lattice solitons and their
  propagation in one-dimensional disordered nonlinear lattices.
\newblock {\em J. Phys. Soc. Japan}, 62:835, 1993.

\bibitem{thor90}
S.~Takeno and K.~Hori.
\newblock A propagating self-localized mode in a one-dimensional lattice with
  quartic anharmonicity.
\newblock {\em J. Phys. Soc. Japan}, 59:3037, 1990.

\bibitem{thor91}
S.~Takeno and K.~Hori.
\newblock Self-localized modes in a pure one-dimensional lattice with cubic and
  quartic lattice anharmonicity.
\newblock {\em J. Phys. Soc. Japan}, 60:947, 1991.

\bibitem{tk92}
S.~Takeno and K.~Kawasaki.
\newblock Intrinsic self-localized magnons in one-dimensional antiferromagnets.
\newblock {\em Phys. Rev. B}, 45:5083, 1992.

\bibitem{tkt93}
S.~Takeno, K.~Kawasaki, and K.~Taniguchi.
\newblock Exact and approximate analytical solutions of stationary vortexlike
  modes in the d-dimensional anisotropic classical o(2) (xy) spin model.
\newblock {\em J. Phys. Soc. Japan}, 62:2192, 1993.

\bibitem{tks88}
S.~Takeno, K.~Kisoda, and A.~J. Sievers.
\newblock Intrinsic localized vibrational modes in anharmonic crystals.
\newblock {\em Prog. Theor. Phys. Suppl.}, 94:242, 1988.

\bibitem{tp96}
S.~Takeno and M.~Peyrard.
\newblock Nonlinear modes in coupled rotator models.
\newblock {\em Physica}, D92:140, 1996.

\bibitem{trp95}
J.~M. Tamga, M.~Remoissenet, and J.~Pouget.
\newblock Breathing solitary waves in a sine-gordon two-dimensional lattice.
\newblock {\em Phys. Rev. Lett.}, 75:357, 1995.

\bibitem{tip96}
J.~L. Ting and M.~Peyrard.
\newblock Effective breather trapping mechanism for dna transcription.
\newblock {\em Phys. Rev. E}, 53:1011, 1996.

\bibitem{mt89}
M.~Toda.
\newblock {\em Theory of Nonlinear Lattices}.
\newblock Springer Verlag Berlin, 1989.

\bibitem{rtm89}
R.~Trautman-Michalska.
\newblock Formation of an optical breather.
\newblock {\em J. Opt. Soc. America}, 6:36, 1989.

\bibitem{ta96}
G.~P.~Tsironis and S.~Aubry.
\newblock Slow relaxation phenomena induced by breathers
in nonlinear lattices.
\newblock{\em Phys. Rev. Lett.}, 77:5225, 1996.


\bibitem{at72}
A.~Tsurui.
\newblock Wave modulations in anharmonic lattices.
\newblock {\em Prog. Theor. Phys.}, 48:1196, 1972.

\bibitem{tu91}
T.~Uzer.
\newblock Theories of intramolecular vibrational energy transfer.
\newblock {\em Phys. Rep.}, 199:73, 1991.

\bibitem{vr94}
B.~G. Vekhter and M.~A. Ratner.
\newblock Spatial and temporal decay of localized electrons in solids:
  One-dimensional model.
\newblock {\em J. Chem. Phys.}, 101:9710, 1994.

\bibitem{vr95}
B.~G. Vekhter and M.~A. Ratner.
\newblock Energy and charge trapping by localized vibrations:
  Electron-vibrational coupling in anharmonic lattices.
\newblock {\em Phys. Rev.}, B51:3469, 1995.

\bibitem{wfb94}
R.~F. Wallis, A.~Franchini, and V.~Bortolani.
\newblock Localized modes in inhomogeneous one-dimensional anharmonic lattices.
\newblock {\em Phys. Rev.}, 50:9851, 1994.

\bibitem{wmb95}
R.~F. Wallis, D.~L. Mills, and A.~D. Boardman.
\newblock Intrinsic localized spin modes in ferromagnetic chains with on-site
  anisotropy.
\newblock {\em Phys. Rev.}, B52:R3828, 1995.

\bibitem{sw93}
S.~Wang.
\newblock Localized vibrational modes in an anharmonic chain.
\newblock {\em Phys. Lett. A}, 182:105, 1993.

\bibitem{wgbs96}
W.~Z. Wang, J.~Tinka Gammel, A.~R. Bishop, and M.~I. Salkola.
\newblock Quantum breathers in a nonlinear lattice.
\newblock {\em Phys. Rev. Lett.}, 76:3598, 1996.

\bibitem{wszo95}
S.~Watanabe, S.~H. Strogatz, H.~S.~J. van~der Zant, and T.~P. Orlando.
\newblock Whirling modes and parametric instabilities in the discrete
  sine-gordon equation: experimental tests in josephson rings.
\newblock {\em Phys. Rev. Lett.}, 74:23, 1995.

\bibitem{wb90}
C.~R. Willis and R.~Boesch.
\newblock Effect of lattice discreteness on the statistical mechanics of a
  dilute gas of kinks.
\newblock {\em Phys. Rev.}, B41:4570, 1990.

\bibitem{webs89}
C.~R. Willis, M.~El-Batanouny, R.~Boesch, and P.~Sodano.
\newblock Nonlinear internal-mode influence on the statistical mechanics of a
  dilute gas of kinks: The double-sine-gordon model.
\newblock {\em Phys. Rev. B}, 40:686, 1989.

\bibitem{wes86}
C.~R. Willis, M.~El-Batanouny, and P.~Stancioff.
\newblock Sine-gordon kinks on a discrete lattice. i. hamiltonian formalism.
\newblock {\em Phys. Rev.}, B33:1904, 1986.

\bibitem{gbw74}
G.~B. Witham.
\newblock {\em Linear and Nonlinear Waves}.
\newblock Wiley New York, 1974.

\bibitem{zwl93}
G.~S. Zavt, M.~Wagner, and A.~L\"utze.
\newblock Anderson localization and solitonic energy transport in
  one-dimensional oscillatory systems.
\newblock {\em Phys. Rev. E}, 47:4108, 1993.




\end{thebibliography}
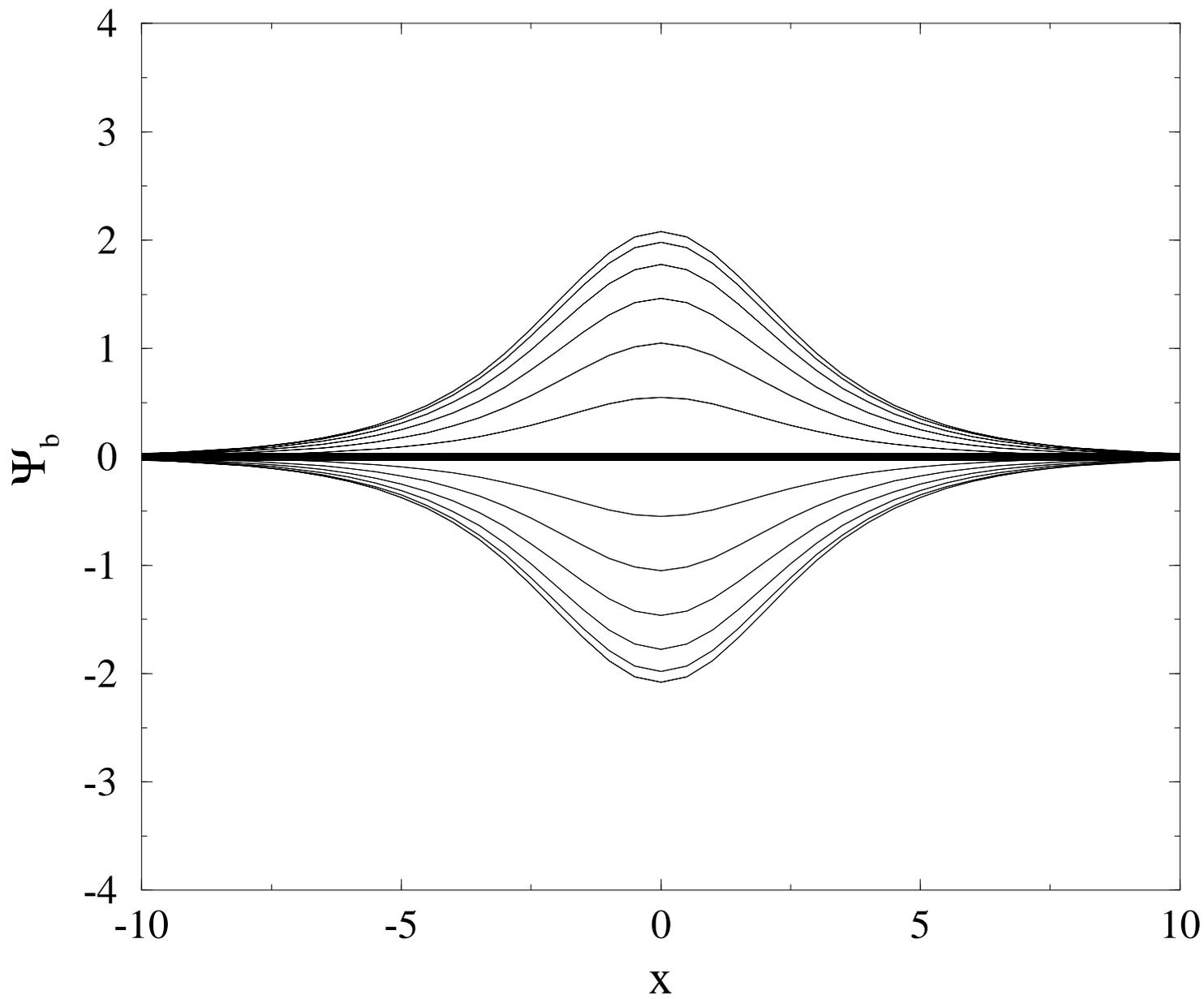
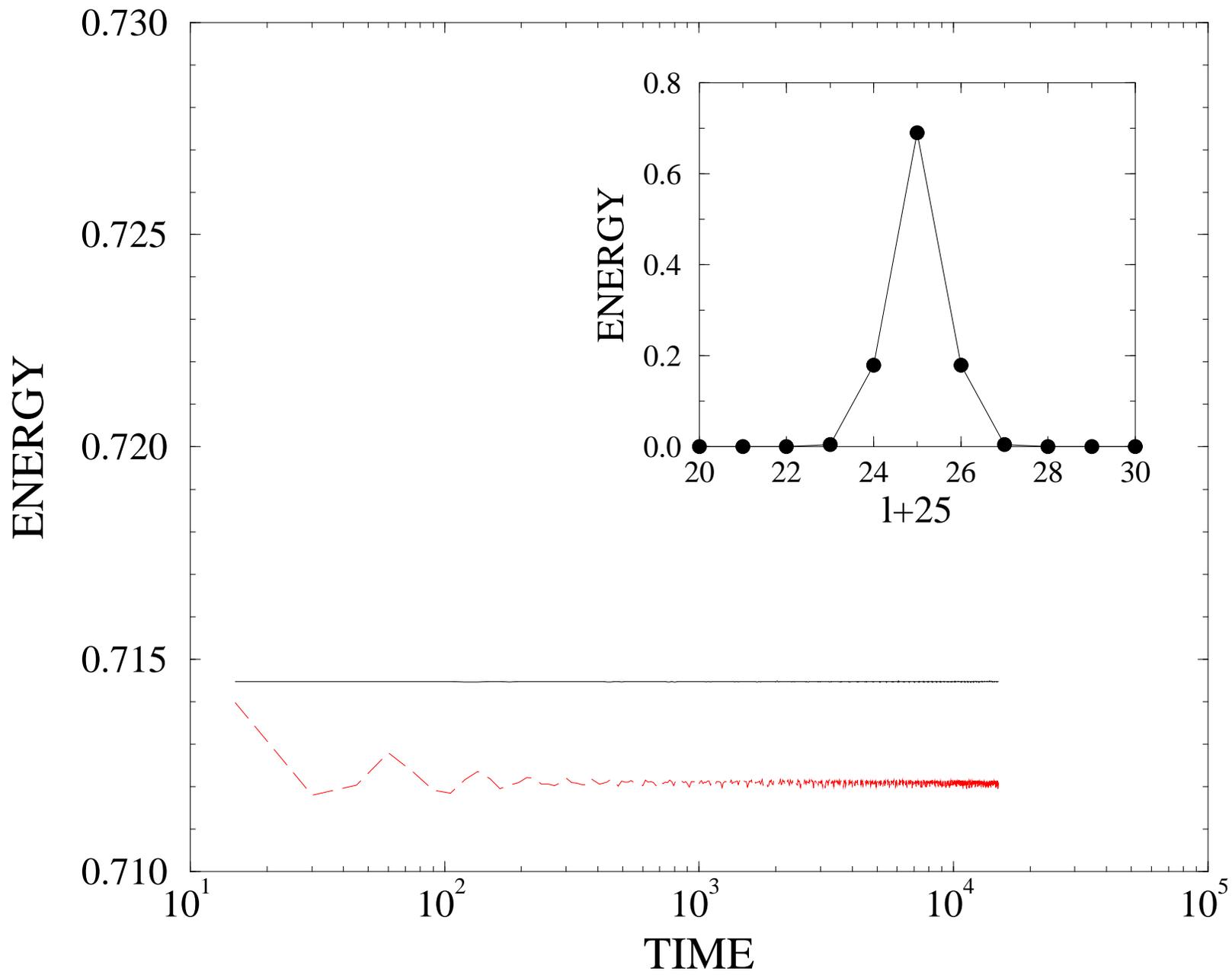
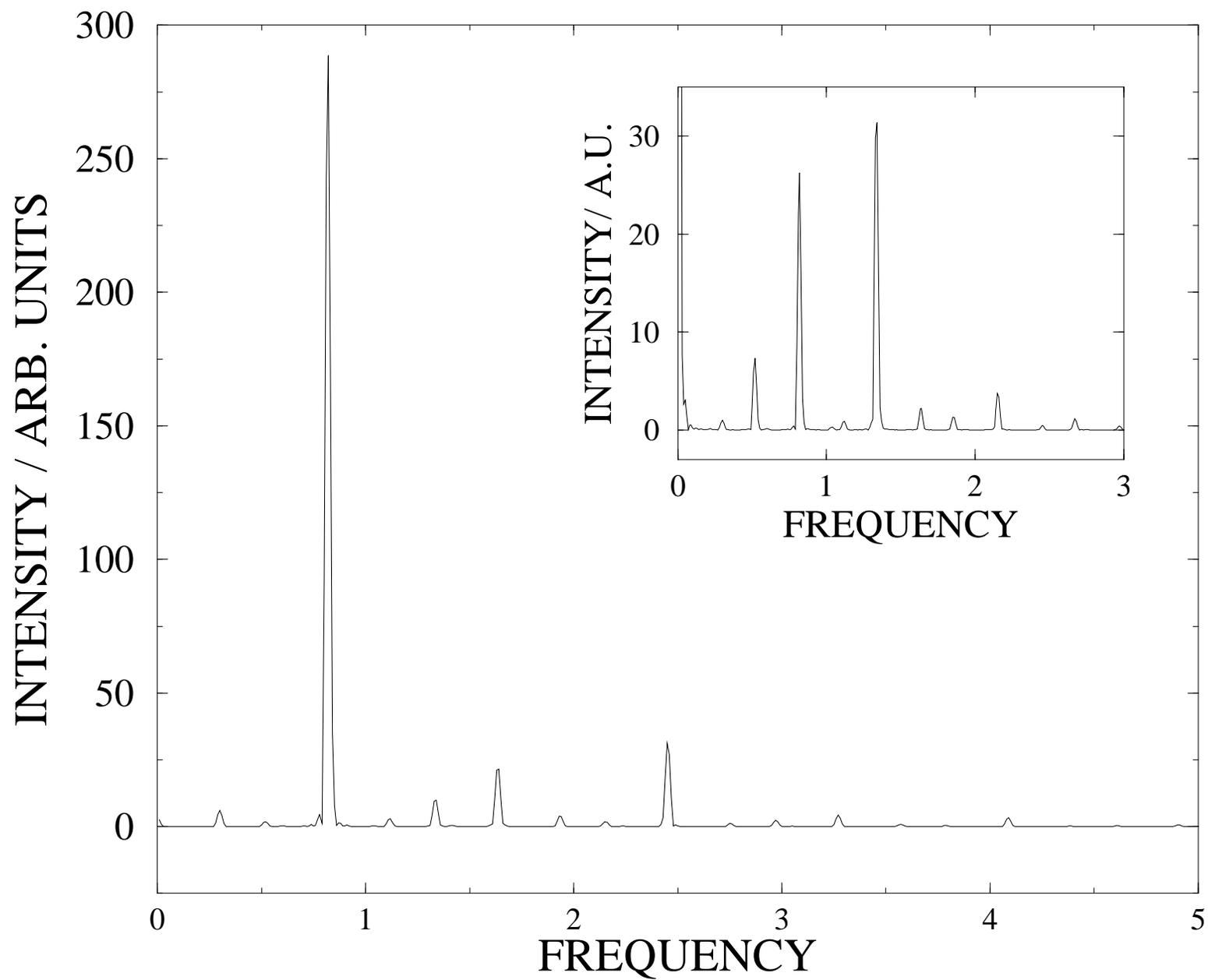
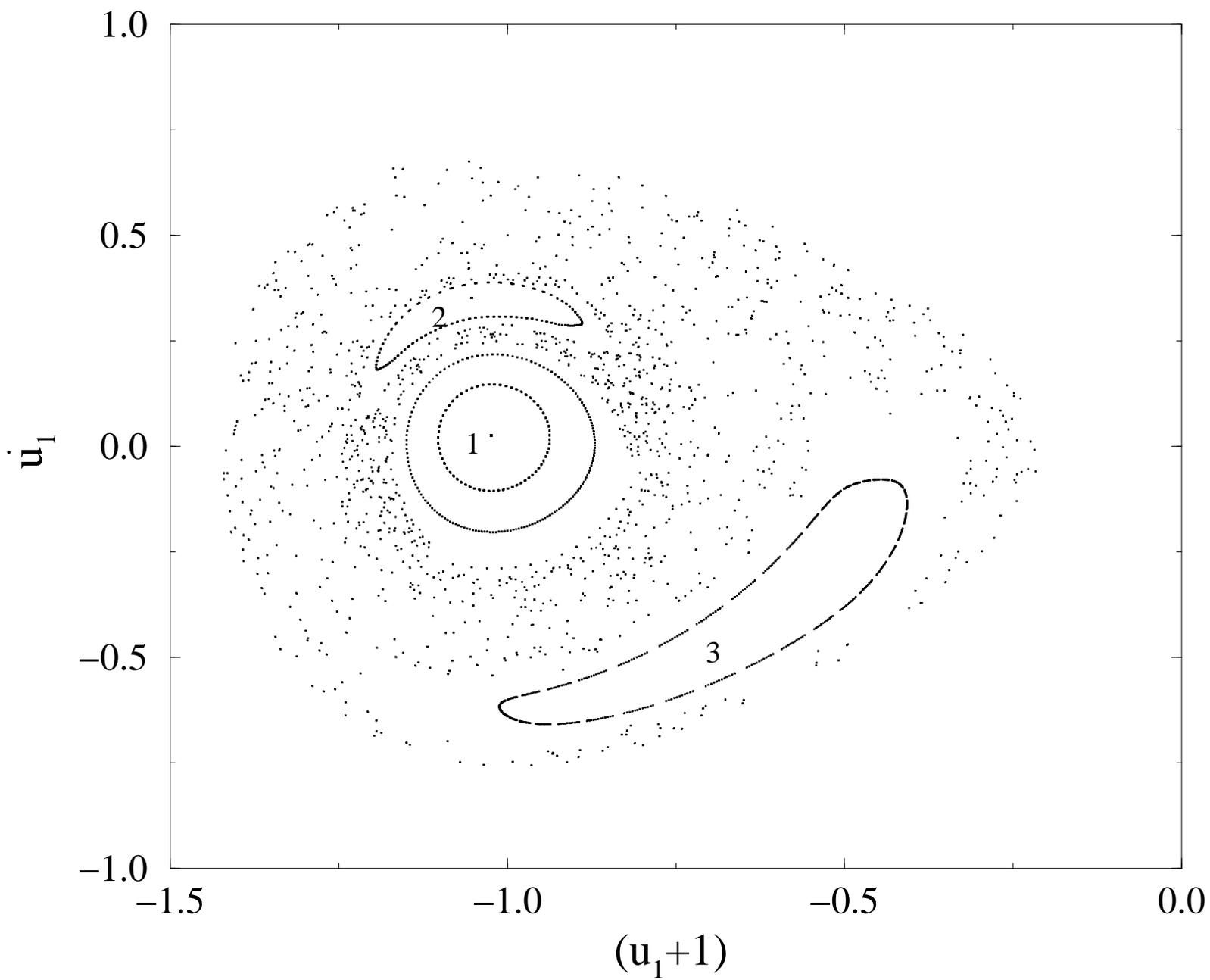
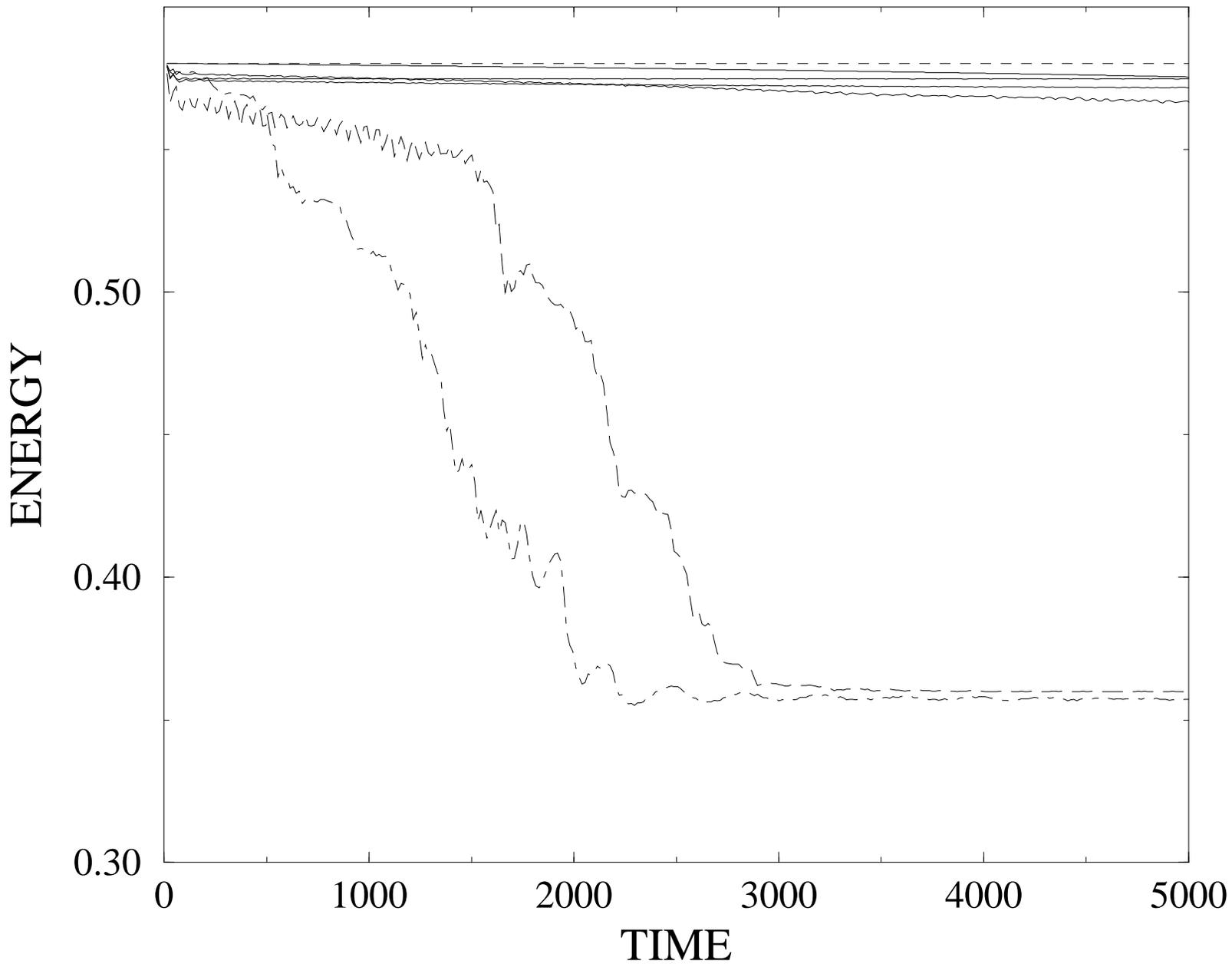
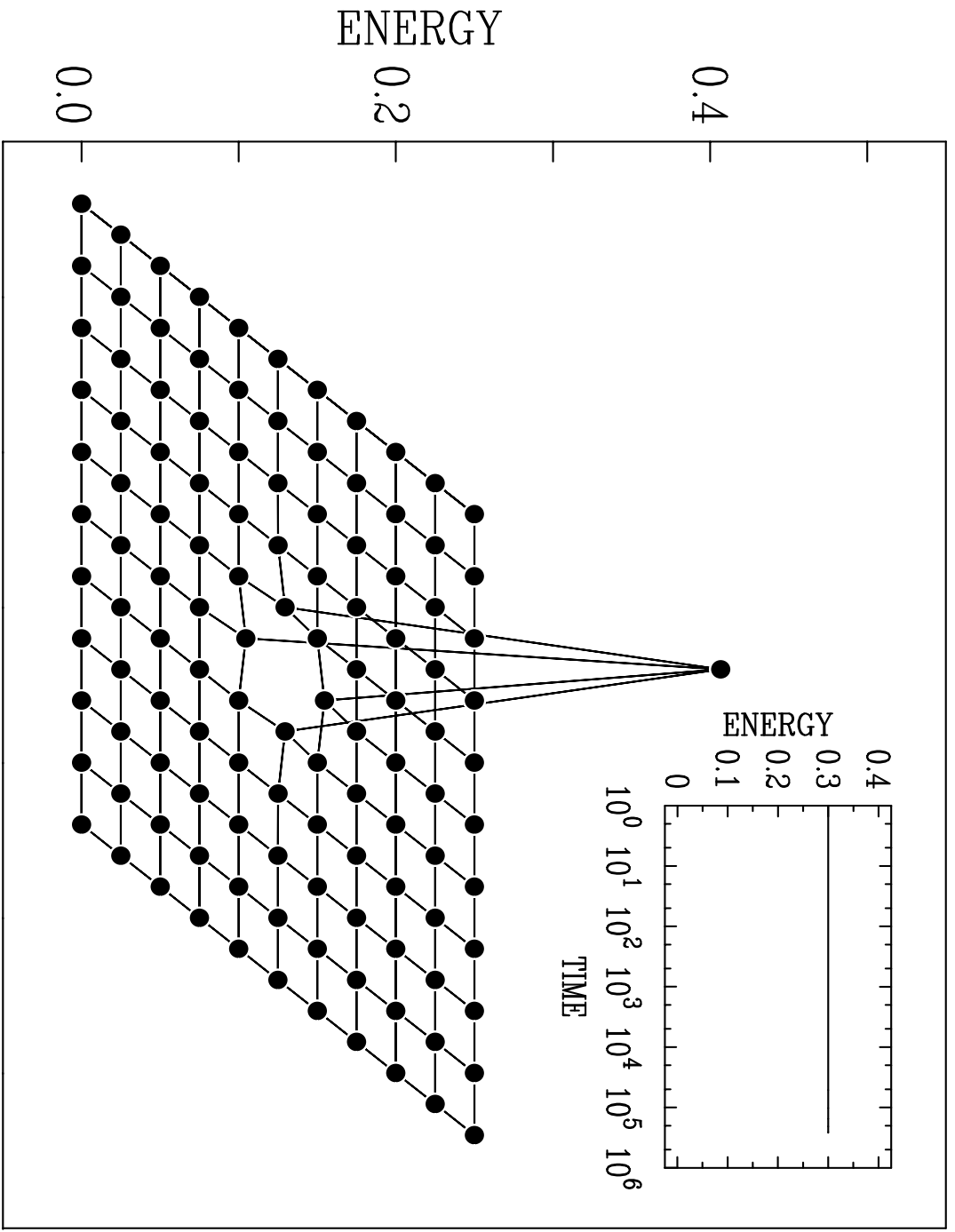
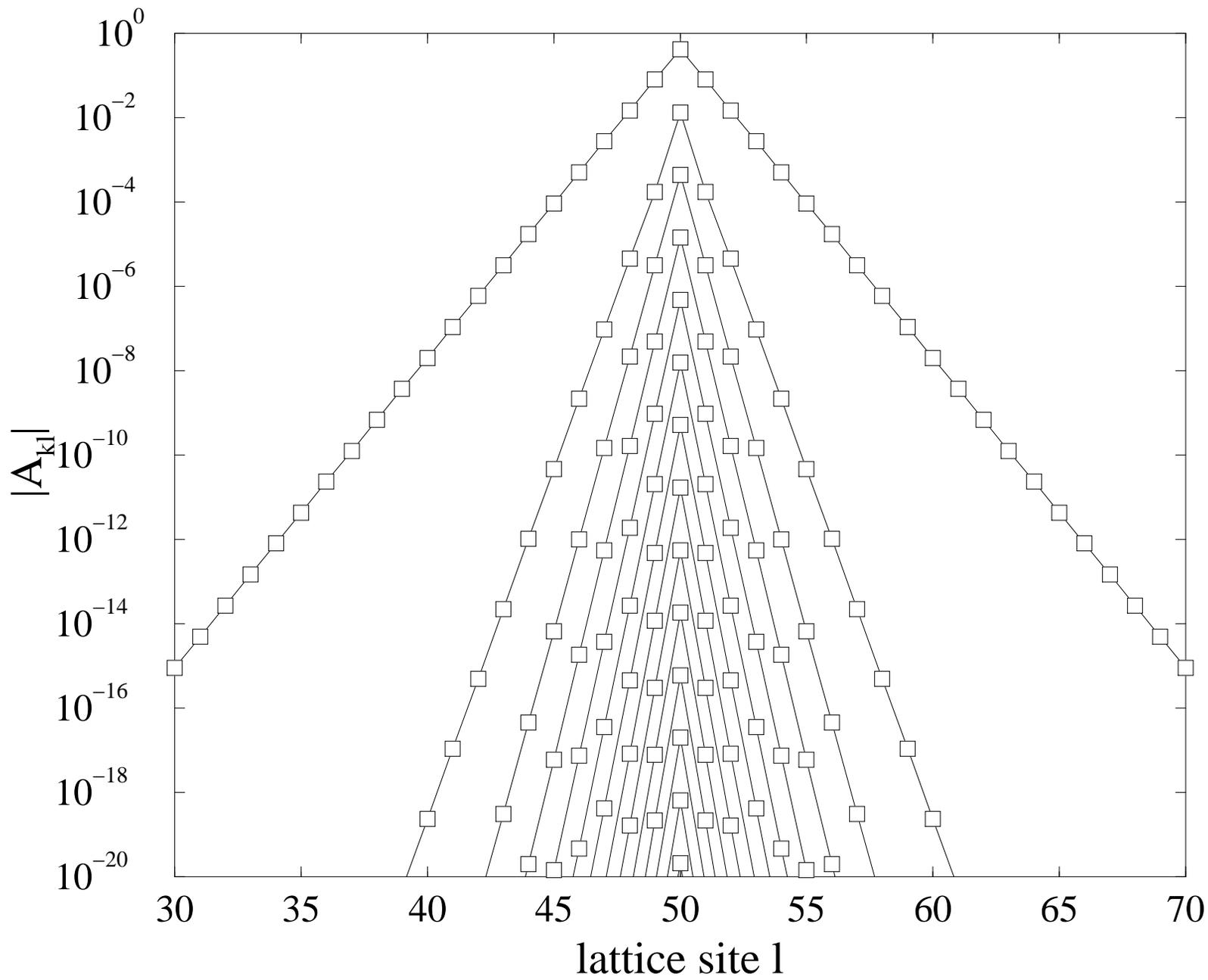
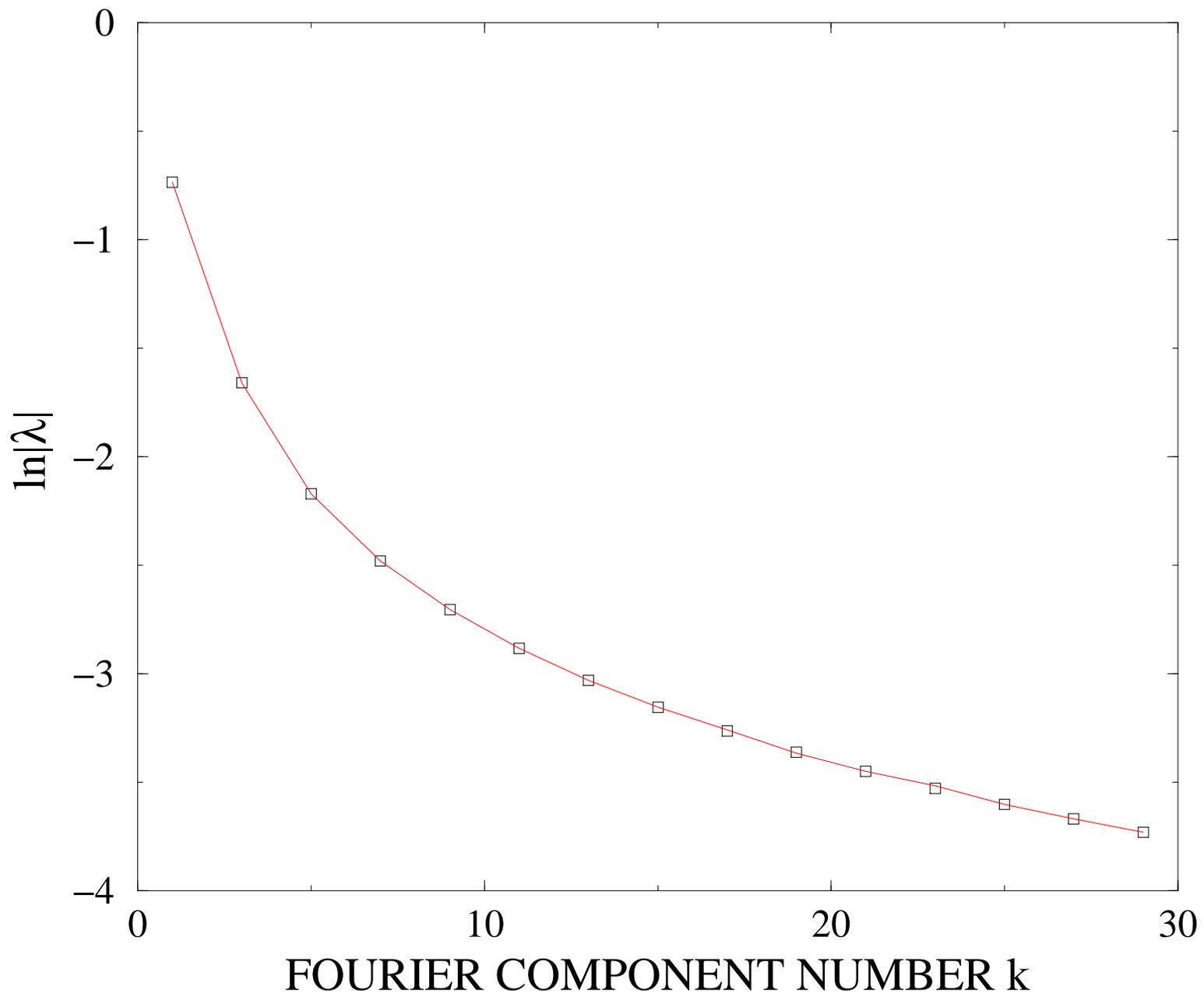
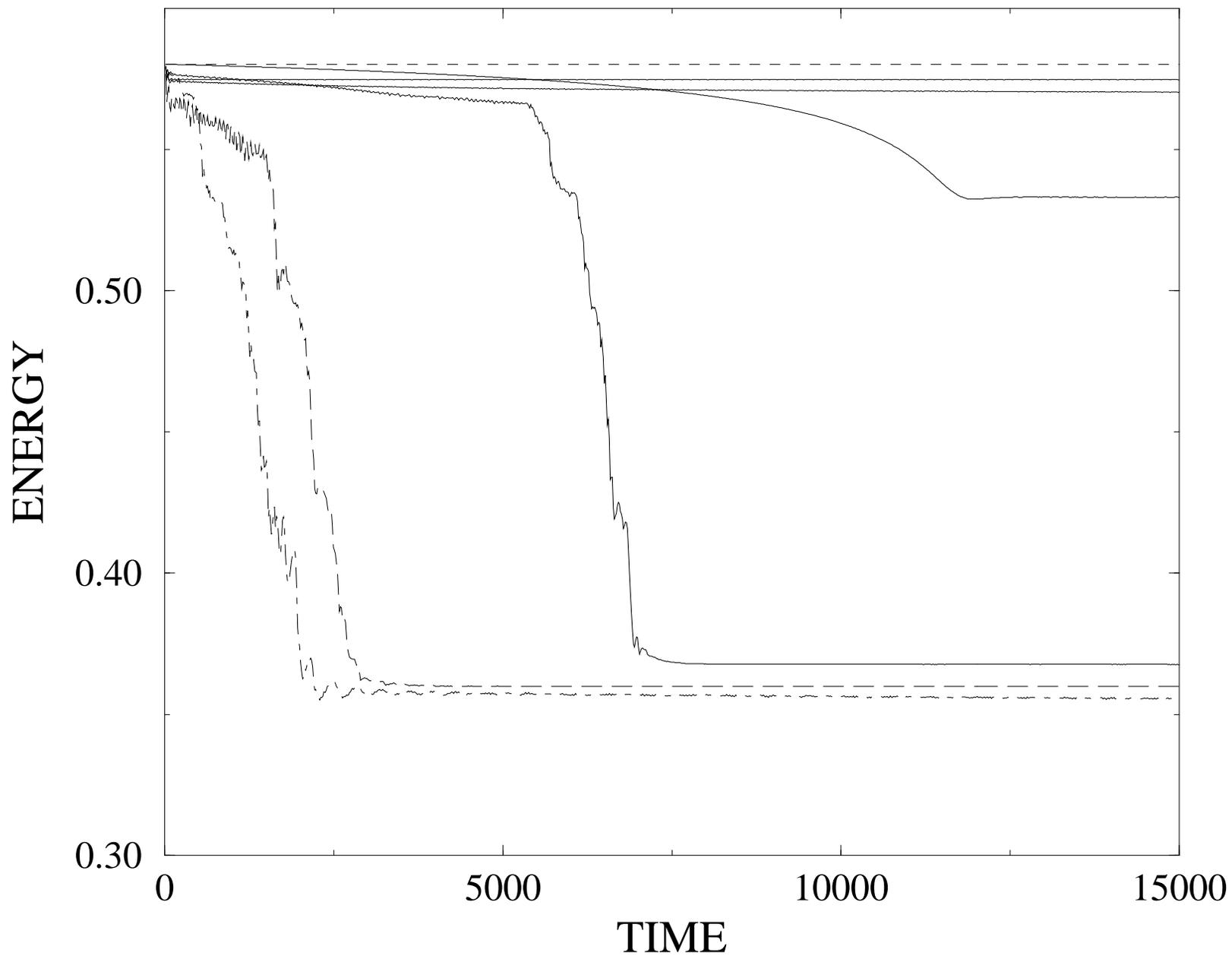
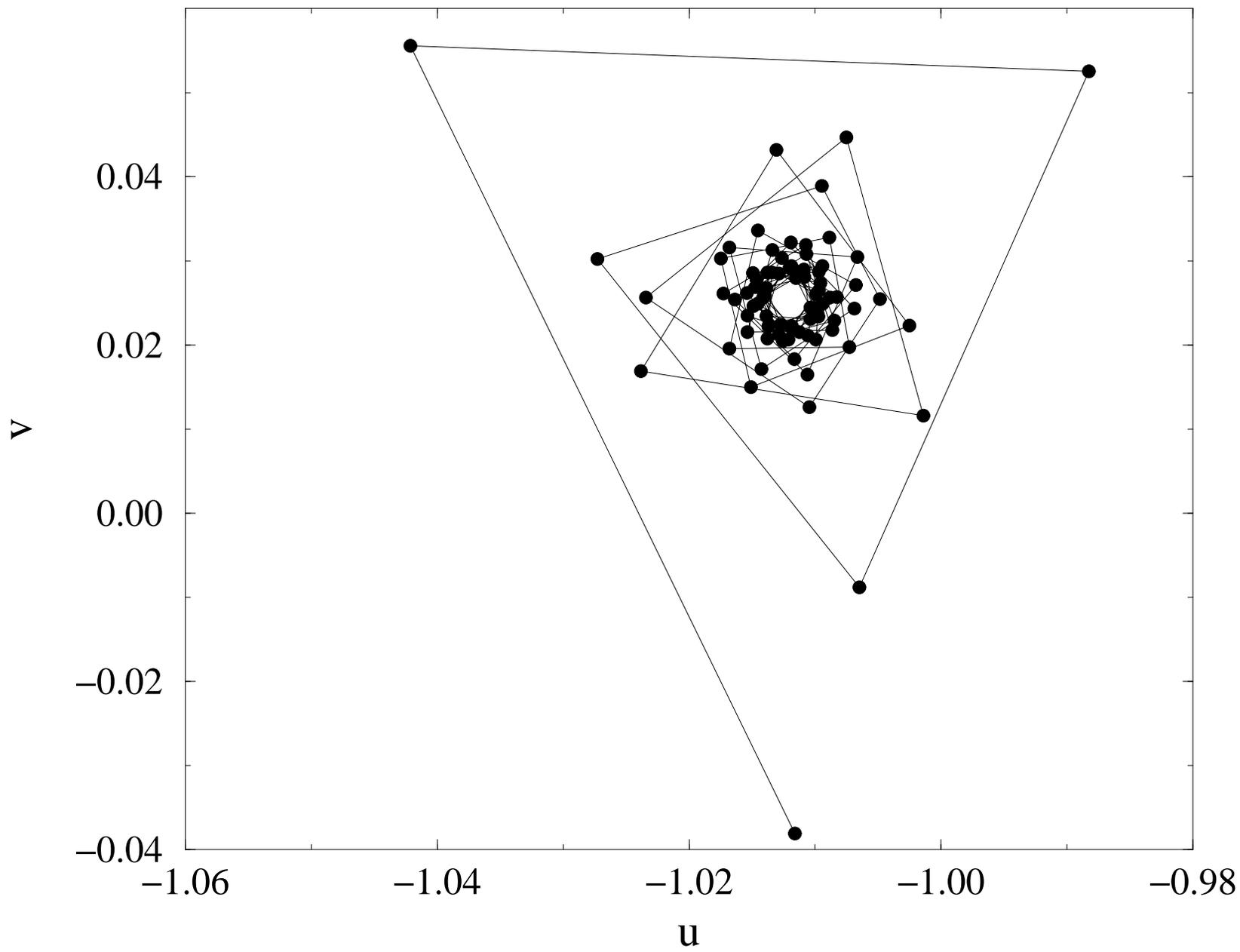
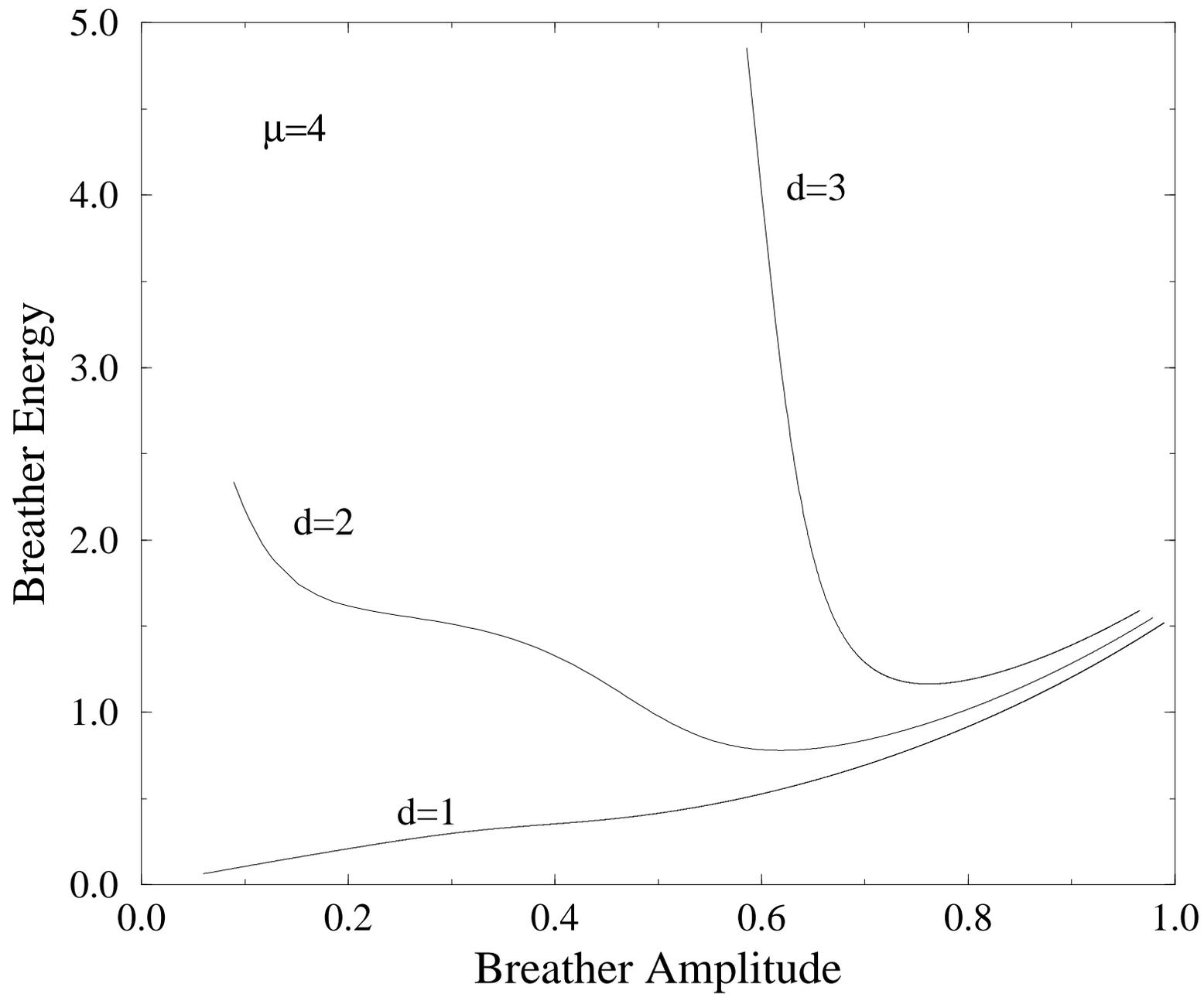
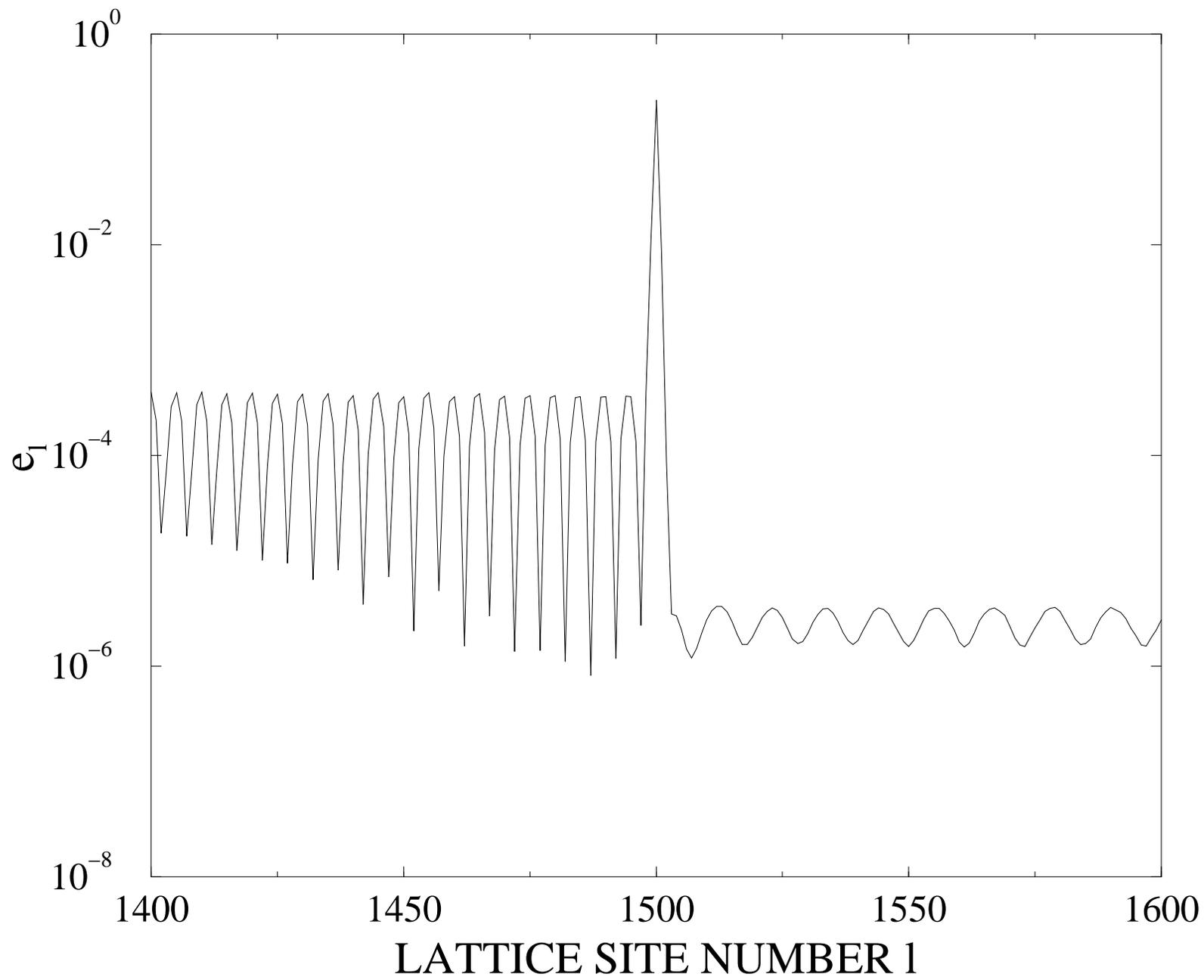
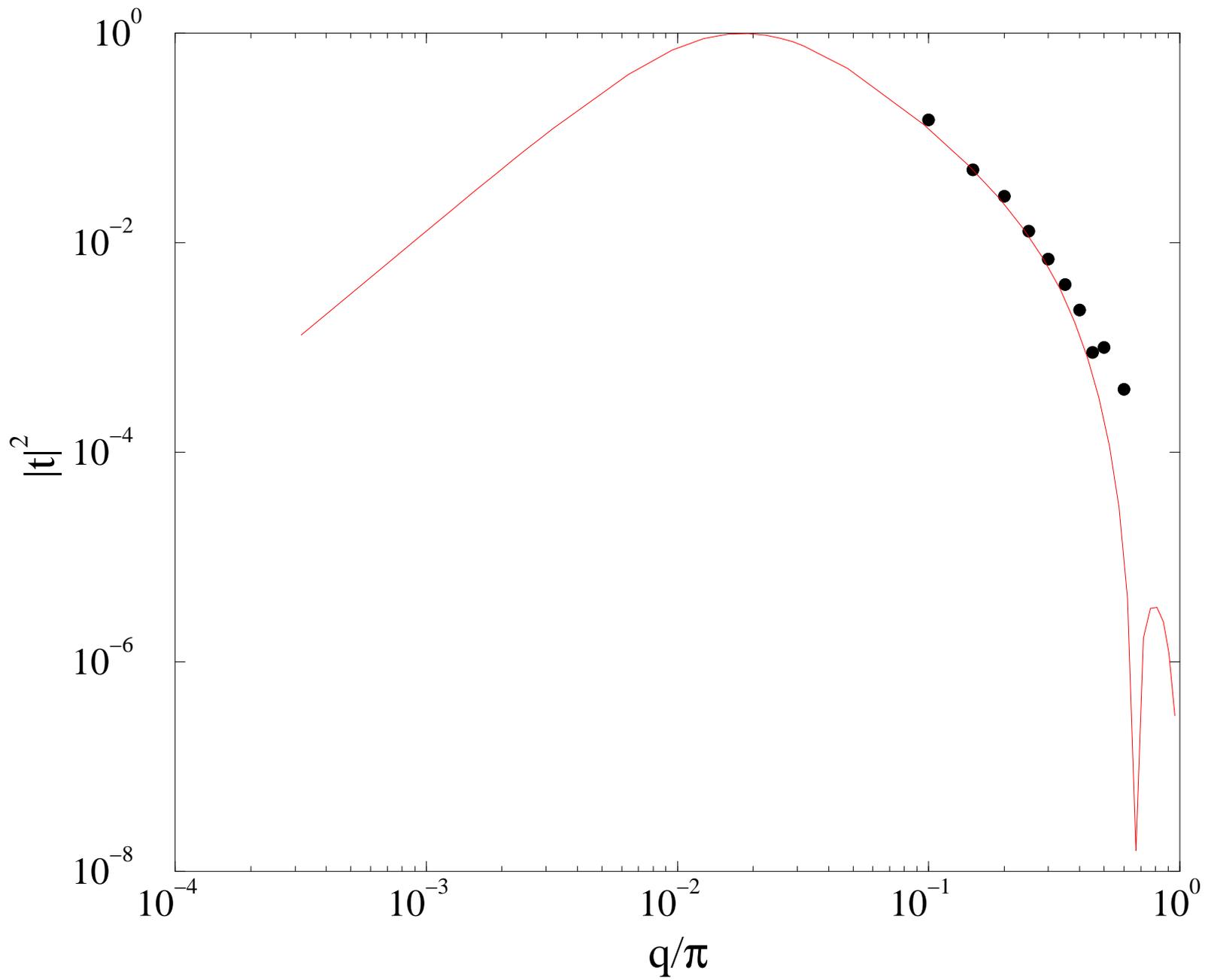
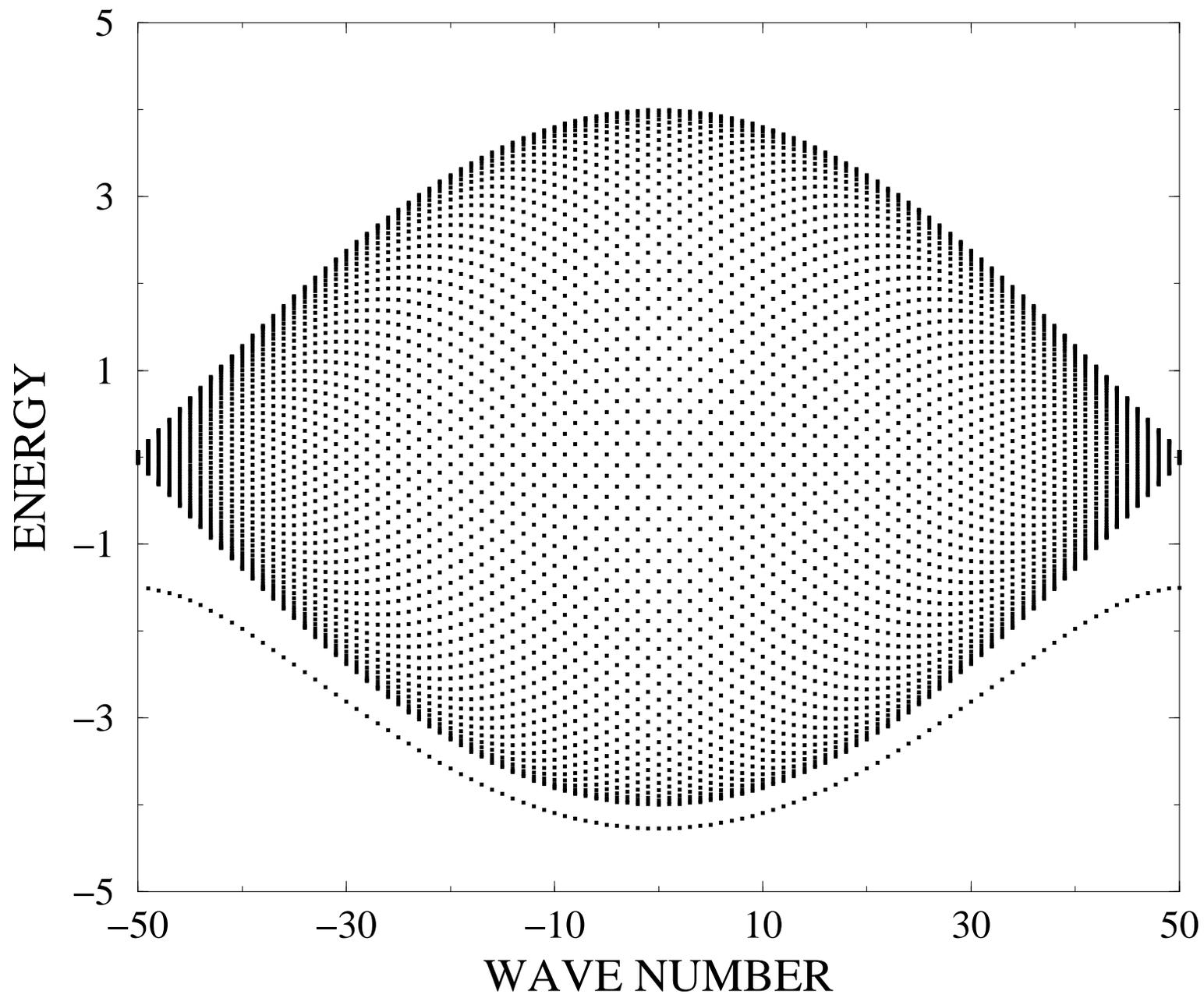
\end{document}